\documentstyle[12pt]{article}
\renewcommand{\thesubsection}{\arabic{section}.\arabic{subsection}}
\makeatletter
\def\section{\@startsection{section}{1}{\z@}{-3.5ex plus -1ex minus 
 -.2ex}{2.3ex plus .2ex}{\large}} 
\def\subsection{\@startsection{subsection}{2}{\z@}{-3.25ex plus -1ex minus 
 -.2ex}{1.5ex plus .2ex}{\normalsize\it}}
\makeatother

\renewcommand{\theequation}{\arabic{section}.\arabic{equation}}
\newcommand{\beq}{\begin{equation}}
\newcommand{\eeq}{\end{equation}}
\newcommand{\bea}{\begin{eqnarray}}
\newcommand{\eea}{\end{eqnarray}}
\newcommand{\beas}{\begin{eqnarray*}}
\newcommand{\eeas}{\end{eqnarray*}}
\newcommand{\tr}{\mbox{tr}}
\newcommand{\nn}{\nonumber}
\newcommand{\alp}{\alpha}
\newcommand{\lam}{\lambda}
\newcommand{\ep}{\varepsilon}
\newcommand{\sg}{\sigma}

\newcommand{\ze}{\zeta}
\newcommand{\dl}{\delta}
\newcommand{\prt}{\partial}
\newcommand{\cH}{{\cal H}}
\newcommand{\cM}{{\cal M}}
\newcommand{\cO}{{\cal O}}
\newcommand{\ket}{\rangle}
\newcommand{\bra}{\langle}
\newcommand{\oh}{\frac{1}{2}}
\newcommand{\limit}{\rightarrow}
\newcommand{\gst}{g_{\mbox{st}}}
\newcommand{\ointze}{\oint\frac{d\zeta}{2\pi i}}

\newcommand{\Jty}{\tilde{J}(y)}
\newcommand{\inty}{\int_{-i\infty}^{i\infty}\frac{dy}{2\pi i}}
\newcommand{\e}{{\rm e}}
\newcommand{\Jtyp}{\tilde{J}(y')}
\newcommand{\intyp}{\int_{-i\infty}^{i\infty}\frac{dy'}{2\pi i}}
\newcommand{\intyC}{\int_{C}\frac{dy}{2\pi i}}
\newcommand{\Lint}{\int_L {d\zeta \over 2\pi i}}
\newcommand{\Lintp}{\int_L {d\zeta' \over 2\pi i}}
\newcommand{\Z}{\mbox{\boldmath $Z$}}
\newcommand{\intAB}{\int d^{N^{2}}Ad^{N^{2}}B}
\newcommand{\Lintsg}{\int_L\frac{d\sg}{2\pi i}}
\newcommand{\ointsg}{\oint\frac{d\sigma}{2\pi i}}

\newcommand{\Lintzesgn}{\int_L\prod_{i=1}^{n}\frac{d\zeta_{i}}{2\pi i}
                               \frac{d\sigma_{i}}{2\pi i}}
\newcommand{\spJ}{\left(\frac{\dl}{\dl J}\vee\frac{\dl}{\dl J}\right)}
\newcommand{\mrJ}{\left(\wedge \frac{\dl}{\dl J}\right)}
\newcommand{\Jh}{\hat{J}}
\newcommand{\derJh}{\frac{\dl}{\dl\Jh}}
\newcommand{\Jt}{\tilde{J}}
\newcommand{\derJt}{\frac{\dl}{\dl\Jt}}
\newcommand{\intx}{\int_{-i\infty}^{i\infty}\frac{dx}{2\pi i}}

\newcommand{\intynxn}{\int_{-i\infty}^{i\infty}\prod_{i=1}^{n}
     \frac{dy_{i}}{2\pi i}\frac{dx_{i}}{2\pi i}}
\newcommand{\non}{\mbox{\scriptsize non}}
\newcommand{\mrJt}{\left(\wedge \frac{\dl}{\dl \Jt}\right)}
\newcommand{\oN}{\frac{1}{N}}

\newcommand{\TmatrixA}{T^{(\mbox{\scriptsize mat.})\, A}}
\newcommand{\TmatrixB}{T^{(\mbox{\scriptsize mat.})\, B}}
\newcommand{\intympxmp}{\int_{-i\infty}^{i\infty}\prod_{i=1}^{m}
     \frac{dy'_{i}}{2\pi i}\frac{dx'_{i}}{2\pi i}}
\newcommand{\MderJh}{\left(\cM\derJh\right)}

\begin{document}
\topmargin 0pt
\oddsidemargin 5mm
\headheight 0pt
\topskip 0mm
\begin{flushright}
UT-Komaba 95-8\\October, 1995
\end{flushright}
\vspace*{1cm}
\baselineskip=0.6cm
\Large
\begin{center}
Stochastic Hamiltonians for Non-Critical String Field Theories \\
from Double-Scaled Matrix Models

\vspace{2cm}

\normalsize
{\sc Fumihiko Sugino}\footnote{E-mail: sugino@hep1.c.u-tokyo.ac.jp}
and
{\sc Tamiaki Yoneya}\footnote{E-mail: tam@hep1.c.u-tokyo.ac.jp}

\vspace{0.5cm}
{\it Institute of Physics, University of Tokyo, Komaba,
Meguro-ku, Tokyo 153, Japan}

\vspace{2.5cm}
Abstract\\
\end{center}

\normalsize
\baselineskip=0.6cm
We present  detailed discussions on the stochastic 
Hamiltonians for non-critical string field theories on the 
basis of matrix models. 
Beginning from the 
simplest $c=0$ case, we 
derive the explicit forms of the Hamiltonians 
for the higher critical case $k=3$ (which corresponds to $c=-22/5$) 
 and for the case $c=1/2$, directly from the double-scaled 
matrix models.  
In particular, for the two-matrix case,
we do not put any restrictions on the 
spin configurations of the string fields. The properties of the 
resulting infinite algebras of 
Schwinger-Dyson operators associated with the Hamiltonians  
and the derivation of the Virasoro and $W_3$ algebras therefrom  
are also investigated. Our results suggest certain universal 
structure of the stochastic Hamiltonians, which might be 
useful for an attempt towards a background independent 
string field theory.

\setcounter{footnote}{0}

\newpage
\normalsize
\section{Introduction}
A common idea towards a non-perturbative formulation of 
string theory is to start from the concept of string fields. 
Just as the ordinary local fields describe the motion and  
interaction of particles in terms of  creation and annihilation 
operators, we can 
construct string field theories by appropriately slicing  
the world-sheets of strings and introducing 
the field operators to create and annihilate the 
strings. Clearly, there is continuously infinite amount 
of arbitrariness in choosing slicings. 
For instance, the light-cone string field theory 
\cite{Kaku-Kikkawa}
uses the 
light-like plane in the target spacetime to slice the world-sheet, 
while the covariant string field theories 
\cite{Witten}, in general, 
use different methods of slicing,
which are based on the geometry of 
the moduli space of Riemann surfaces. 
The arbitrariness of slicing may be interpreted as a sort of  
gauge freedom of the theory. At present, however, 
we have no satisfactory framework 
to formulate such a gauge 
structure in a systematic and general way. 

Recently, an interesting new way of slicing has been proposed
\cite{KKMW}, 
and the corresponding string field theories
\cite{IK,IIKMNS}
 have been 
suggested for the case of noncritical strings 
with $c=0$ and the case 
with minimal conformal matter $c=1-{6 \over m(m+1)}$. 
In this proposal, the world sheets are sliced by using a certain  
time parameter, 
 which is intrinsically defined on the world 
sheet as a measure of the distances 
 from the boundaries of 
the world sheets. In this paper, we will call the string 
field theories of this type as ``proper time" string field 
theories (PSFTs), in analogy with the familiar proper-time 
representation of propagators in ordinary field theories.   
In PSFTs, it seems less difficult to incorporate the higher-genus
(and hopefully, 
non-perturbative) effect than in the moduli space approach  
as employed in the ordinary covariant string field theories.  
Remarkably enough, there exists a single exact Hamilonian operator
which directly characterizes
all of the correlation functions in the system at once.

One of the many unsolved 
problems of present PSFTs, however, is that we do not know 
definite symmetry principles, if any, on the basis of 
 which one can more or less uniquely 
characterize  
the theories. 
Thus, most of previous attempts had to 
rely upon certain guess works, and 
one can only justify the theories by checking the 
agreement of amplitudes with known results obtained
from other methods,  such as the 
matrix models. In this situation, the observation 
\cite{JR}
 that the PSFT for 
$c=0$ can be interpreted as
the collective field theory of matrix models  
formulated in stochastic quantization 
seems very useful and suggestive.  In connection with this, 
we should recall an attractive idea
\cite{Banks-Martinec} 
 of relating the 
renormalization group formulation 
of string-field equations to stochastic quantization.

Another crucial question of PSFTs
for further developing the theory is   
whether or not this method of slicing is meaningful for 
constructing string field theories for $c>1$ and 
critical strings. The simplicity of proposed PSFTs for the 
case $c<1$ is of course due to the simplicity of the target 
spaces. For example, in the case where the target space 
is the  Ising model, 
one can deform the 
slicings such that the spin configuration on each string field 
is either all spin up or all spin down \cite{IK2}.
 If one goes to $c>1$, 
the slicing of this type would, however, be too singular 
to be tractable and 
one would have to introduce string fields 
without making any restrictions on possible matter configurations. 

{}From this view point, it seems important 
to treat even the cases  
$c<1$ without such restrictions and to study the structure of 
resulting PSFT, since we naturally expect that such a formulation 
should exhibit certain universal structure of the 
general PSFT which is common to PSFT for general critical strings.  
Since there is no known symmetry principle on the basis
of  which we can derive 
the theories, it is natural to directly derive such a formulation 
starting from the matrix models. That is what we shall present  in 
this paper. Our hope is to 
get some insight on the nature of the PSFTs 
by deriving the formalism  
from the matrix models as explicitly as possible.  
We will follow the suggestion of ref. \cite{JR}, 
using a slightly different approach,   
and construct the PSFT Hamiltonians directly by taking the 
double scaling limit of the matrix model Hamiltonians.\footnote
{
For previous works which
discuss the possibility of the PSFT with general 
matter configurations in the continuum formulation,
see \cite{IIKMNS,NS}.
}

In the next section, we will first review our method 
of deriving the stochastic Hamiltonian from the 
matrix models.  We illustrate the method by using a 
simple quantum mechanical model with two degrees of freedom 
and point out some crucial assumptions required for 
proper-time string field theories.  
In section 3, we treat the case of one-matrix model 
and derive the Hamiltonians for the cases of
$k=2 \, (c=0)$ critical point 
and, as a simplest example of higher critical models,  
$k=3$ critical point.  In section 4, we discuss the Virasoro algebra 
structure associated with the Hamiltonians. Using the example 
with $k=3$, we will 
clarify how the closed Virasoro algebra is
obtained for higher critical cases. 
In section 5, we proceed to discuss the two-matrix model. 
Technically, this case is much more complicated  than the 
case of one-matrix model and requires some new 
ingredients which have not shown up in
the case of the one-matrix model. 
We will exhibit  some interesting properties on the structure of the 
stochastic Hamiltonians, which may indeed be regarded 
as an example of the universal structure of the general PSFTs. 
In section 6, we will discuss the closure property of
  the infinite algebras 
associated with our Hamiltonians. Then, in section 
7, the $W_3$ algebra of the two matrix model 
will be derived starting from the infinite algebra. 
These two sections provide consistency checks for the 
results of section 5, by deriving the expected properties 
of the two-matrix model from the present formalism.   
In the final section, we will conclude the paper by discussing 
possible implications of our work and remaining
 problems. Throughout this paper, we had to perform 
a number of tedious computations 
for which we could not find any appropriate 
references. Most of such details will be 
described in the Appendix.

\vspace{1cm}
\section{The Hamiltonian of Stochastic Quantization}

In this section, we will briefly introduce our method for deriving 
the Hamiltonian of PSFT.  
For clarity, we take a simple example of zero-dimensional 
field theory with two degrees of freedom $x, y$ with 
action $S(x,y)$, 
\beq
Z=\int dxdy \, \e^{-S(x,y)}.
\eeq

  The idea of stochastic quantization
\cite{Parisi-Wu}
 can be summarized by 
introducing the following Hamiltonian, 
\beq
H= -{\partial \over \partial x}({\partial \over \partial x}
+ {\partial S \over \partial x})
-{\partial \over \partial y}
({\partial \over \partial y}+ {\partial S \over \partial y}),
\eeq
and the Fokker-Planck equation,  
\beq
{\partial \over \partial \tau}\Psi(x, y, \tau)=
-H\Psi(x, y, \tau).
\eeq
As is well known, the Fokker-Planck equation describes the 
statistical evolution of the probability distribution function 
$\Psi (x, y, \tau)$ for 
the system described by the stochastic equations of motion 
\bea
{d x \over d\tau}
&=&-{\partial S\over \partial x}+\eta_1,\\
{d y \over d\tau} 
&=&-{\partial S\over \partial y}+\eta_2,
\eea
where $\eta_1, \eta_2$ are Gaussian random noises. 
In the limit of $\tau \rightarrow \infty$, the  
solution of the Fokker-Planck
 equation reduces to the ground state 
\beq
\Psi \rightarrow  \e^{-S},
\label{groundstate}
\eeq
satisfying $H\Psi=0$ 
{\bf under} the assumption that $\e^{S/2} \Psi$ 
 rapidly decreases at infinity, 
corresponding to the positivity of the 
hermitian Laplace operator 
\beq
\e^{S/2}H \e^{-S/2}=D_1^{\dagger}D_1+D_2^{\dagger}D_2,
\eeq
with $D_1=\e^{-S/2}{\partial \over \partial x}\e^{S/2}
,D_2=\e^{-S/2}{\partial \over \partial y}\e^{S/2}$.
   
The Green function of an arbitrary observable ${\cal O}$ 
can be expressed as 
\beq
\bra {\cal O} \ket = \lim_{\tau \rightarrow \infty}
\int dxdy \, {\cal O}(x,y)\Psi(x, y, \tau). 
\eeq
When (\ref{groundstate}) is assumed to be the unique ground state 
of the Hamiltonian $H$, we are entitled to suppose that the 
entire Schwinger-Dyson equation is replaced by
a single ground-state condition
given as 
\beq
\lim_{\tau \rightarrow \infty} \int dxdy \,
{\cal O}(x,y)H\Psi(x,y,\tau) =0, 
\eeq
which is, after partial integrations, equivalent to 
\beq
\int dxdy \, [{\partial \over \partial x}\e^{-S}
{\partial \over \partial x}
+{\partial \over \partial y}\e^{-S}
{\partial \over \partial y}]{\cal O} =0.
\eeq
Using the generating functional with ${\cal O}=\e^{J_1 x+J_2 y}$, 
this is rewritten as 
\beq
H(J, {\partial \over \partial J})Z[J]=0,
\label{hamiltonconstraint}
\eeq
\beq
Z[J]=\int dxdy \, \e^{-S(x,y)+J_1 x+J_2y},
\eeq
with
\beq
H(J, {\partial \over \partial J})=J_1T_1(J, {\partial \over \partial J})
+J_2T_2(J, {\partial \over \partial J}),
\eeq
\bea
T_1 &=&-S_x({\partial \over \partial J_1}, \, 
{\partial \over \partial J_2}) +J_1, \\
T_2 &=&-S_y({\partial \over \partial J_1}, \,
{\partial \over \partial J_2}) +J_2.
\eea
{}From the assumption of the uniqueness of the solution for the 
ground state condition (\ref{hamiltonconstraint})
(that we call Hamilton constraint),
we can impose   
\beq
T_1Z[J]=T_2Z[J]=0. 
\label{SchwingerDysonxy}
\eeq
which are nothing but the general form of the Schwinger-Dyson 
equation for our system, 
\bea
T_1Z[J]&=&\int dxdy \,
{\partial \over \partial x}\e^{-S(x,y)+J_1 x+J_2 y}, \\
T_2Z[J]&=&\int dxdy \,
{\partial \over \partial y}\e^{-S(x,y)+J_1 x+J_2 y}.
\eea
Note that the integrability condition is automatically 
satisfied,
\beq 
[T_1, T_2]=S_{xy}(J, {\partial \over \partial J})
-S_{yx}(J, {\partial \over \partial J})=0. 
\eeq

In general, by introducing more general 
source terms $\sum_i J_if_i(x,y)$, the stochastic Hamiltonian 
takes the following form with a general set of operators $T_i$ 
\beq
H=\sum_i J_iT_i.
\label{generalsourcehamil}
\eeq
Then, the assumption of uniqueness of the ground state 
implies that the partition function 
satisfies
\beq
T_iZ[J]=0
\eeq
where
\beq
T_iZ[J]=\int dxdy \,({\partial \over 
\partial x}{\partial f_i\over \partial x}+
{\partial \over \partial y}{\partial f_i\over \partial y}
)\e^{-S+\sum_iJ_if_i}
\eeq
which can be, for appropriate choice of 
the source terms, expressed as functional 
differential operators in terms of $J_i$'s and 
are equivalent to the Schwinger-Dyson equations of the system. 
This should be regarded as
 a fundamental assumption of the method of 
stochastic quantization. We note that in general the 
algebra of the Schwinger-Dyson operators $T_i$ is 
non-Abelian.  

Here, we add an important remark which partly 
underlies our later discussions. Namely, 
by introducing general source terms, together with 
this assumption,  we can make the formalism  background independent. 
After making a shift of the source function 
$J_i \rightarrow J_i +\delta_{i, S}$, the Hamiltonian equation
(\ref{generalsourcehamil}) 
\beq
HZ\equiv \int dxdy [{\partial \over \partial x}\e^{-S}
{\partial \over \partial x}
+{\partial \over \partial y}\e^{-S}
{\partial \over \partial y}]\e^{S+\sum_i J_if_i} =0
\eeq
is then recast into the following form 
\beq
T^0_S Z
+H^0 Z=0,
\label{shiftedhamilton}
\eeq
where
\beq
H^0Z \equiv \int dxdy  ({\partial^2 \over \partial x^2} +
{\partial^2 \over \partial y^2})\e^{\sum_i J_i f_i} =(\sum_iJ_iT^0_i)Z.
\eeq
Thus, under the 
assumption that the Hamilton equation is equivalent to 
the Schwinger-Dyson equations $T^0_iZ=0$, the first term of
(\ref{shiftedhamilton}) 
vanishes and   
the Hamilton equation is 
reduced to  $H^0 Z=0$, a form which is formally independent of 
the starting action $S$.  Here, $T^0_i$'s are
the Schwinger-Dyson operators 
with the shifted source $J_i +\delta_{i, S}$,
or in other words, with no bare 
action.

Now, to gain a more concrete understanding 
on the above assumption, 
let us consider a simple example with the bare action
\beq
S(x,y)=V(x)+V(y)+cxy, \quad V(x)={x^2\over 2}+g{x^3\over 3}.
\eeq
The operators $T_1, T_2$ are given as 
\bea
T_1&=&-{\partial \over \partial J_1}
-g{\partial^2 \over \partial J_1^2}
-c{\partial \over \partial J_2}+J_1,\\
T_2&=&-{\partial \over \partial J_2}
-g{\partial^2 \over \partial J_2^2}
-c{\partial \over \partial J_1}+J_2.
\eea
On the other hand, from the view point of the Schwinger-Dyson 
equations, 
it is easy to check  that the following 
set of equations 
gives a closed recursion equation for  
$\bra x^n\ket$:
\bea
0&=&\int dxdy {\partial \over \partial x} x^n \e^{-S},  
\label{sd1} \\
0&=&\int dxdy {\partial \over \partial y} x^n \e^{-S}, 
\label{sd2}  \\
0&=&\int dxdy {\partial \over \partial x} x^n y \e^{-S}, 
\label{sd3}
\eea
which are obtained from the $T_1, T_2$ constraints 
by making a power series expansion in $J_1, J_2$ as 
\bea
{\partial^n \over \partial J_1^n}T_1Z[J] \biggr|_{J=0}&=&0, \label{t1}\\
{\partial^n \over \partial J_1^n}T_2Z[J]\biggr|_{J=0}&=&0, \label{t2}\\
{\partial^{n+1} \over \partial J_1^n\partial J_2}
T_1Z[J]\biggr|_{J=0}&=&0, \label{t3}
\eea
respectively. 
The closed recursion equation for
$\bra x^n\ket$ is obtained by expressing 
the  correlators of the form $\bra x^ny\ket, \bra x^ny^2\ket$ 
using the last two equations 
in terms of $\bra x^n\ket$  
and by substituting the results into the first equation.\footnote{
This procedure is essentially the same as the one employed 
in ref. \cite{St}
to derive a closed subset of the Schwinger-Dyson equations 
for the two-matrix model.} 

However, it is not difficult to see
 that we cannot derive all of these conditions 
(\ref{sd1})$\sim$(\ref{sd3}) 
directly by taking finite order derivatives with respect to 
$J_1, J_2$ from the single Hamiltonian constraint,
\beq
(J_1T_1+J_2T_2)Z[J] =0.
\label{hamiltonxy}
\eeq
For example, by taking a derivative 
${\partial^{n+1} \over \partial J_1^n\partial J_2}$ 
of (\ref{hamiltonxy}), we obtain the sum of (\ref{t2}) and (\ref{t3}), but 
can never obtain (\ref{t2}) or (\ref{t3}), separately, by taking any 
derivatives of finite order. 

Thus, the equivalence of the Hamiltonian constraint (\ref{hamiltonxy}) 
with the 
Schwinger-Dyson equations (\ref{SchwingerDysonxy}) 
is based on the 
assumption of the uniqueness of the solution for (\ref{hamiltonxy}) 
which requires that $\e^{-S/2}$ rapidly decreases at 
infinity $x, y \rightarrow \pm \infty$. 
This uniqueness assumption amounts to setting certain conditions 
on the partition function $Z[J]$ which {\bf cannot} be expressed in any 
{\bf finite} order of the expansion with respect to the source functions 
$J_i $.  
If the appropriate global conditions for the uniqueness are  
not satisfied, the stochastic Hamiltonian would fail\footnote{
This is obvious for the case of usual laplacian $
\triangle=\sum_i\partial_i \partial_i$. 
There are an infinite number of polynomial solutions 
for $\triangle f=0$.  The uniqueness can be guaranteed under 
the requirement, say, of $L^2$ normalizability condition. 
}
to give a unique ground state in the limit $\tau\rightarrow \infty$, and 
the limit would, in general, depend on the choice of the initial state. 
   
In the case of simple quantum mechanical models, it is 
relatively easy to identify the necessary global conditions. 
However, in  more complex systems such as the 
double-scaling limit of matrix models, 
it is quite nontrivial to state such conditions, 
and, in fact, there has been no known result replying 
this question.

It is clear that the PSFT proposed in ref. \cite{IK} 
is based on the tacit assumptions of similar nature as above. 
Unless the Hamiltonian is able to define a more or less unique 
ground state under the same constraint as 
for the Schwinger-Dyson equations, 
the concept of the PSFT Hamiltonians 
would become less significant, since in that case we have to 
recourse to the Schwinger-Dyson equations themselves 
for the definition of the theory.

    In the following sections, we will discuss the Hamiltonians of PSFT 
for one- and two-matrix models using the above methods, 
keeping  those assumptions in mind.
We here mention that our method can be
translated into the language of ref. \cite{IK}
by making a functional transformation
\beq
Z[J]\rightarrow 
\bra Z| =\bra 0| \exp\bigl(\sum \psi_i{\partial \over \partial J_i}\bigr)
Z[J]\Big\vert_{J=0}.
\label{functransform}
\eeq
The source functions and their derivatives are replaced by  the 
annihilation- and creation-string fields
\beq
J_i  \leftrightarrow \psi_i, \, {\partial \over \partial J_i}
\leftrightarrow \psi_i^{\dagger},
\eeq
respectively. Then, the correlators are given as 
\beq
\bra Z|\psi_1^{\dagger}\psi_2^{\dagger}\cdots \psi_n^{\dagger}|0\ket
=\Big(\prod_{i=1}^n {\partial\over \partial J_i}\Big)Z[J]\Big\vert_{J=0}.
\eeq
Note that under the
above transformation
the normal ordering $J\cdots{\partial\over\partial J}\cdots$ 
for $J, {\partial \over \partial J}$ is automatically transformed into the one
$\psi^{\dagger}\cdots\psi\cdots$ for $\psi, \psi^{\dagger}$.
The Hamilton constraint is thus
 $\bra Z\vert :H(\psi, \psi^{\dagger}):=0$,
and the state $\bra Z\vert$ is obtained as
$\bra Z\vert =\lim_{\tau \rightarrow \infty}\bra 0\vert \e^{-\tau :H:}$.

\vspace{1cm}
\section{PSFTs from One-Matrix Model}
\setcounter{equation}{0}
 
In this section,  we derive the stochastic Hamiltonians 
from the one-matrix 
model at the $k=2$ and $k=3$ critical points which correspond 
to the matter central 
charges $c=0$ and $c=-22/5$, respectively.   

\vspace{0.5cm}
\subsection{Stochastic Hamiltonian at $c=0$}

   We first treat the case of $c=0$.  
Although this case has already been discussed in ref. \cite{JR} 
within the framework of the collective field method,   
we present some details for the purpose of  
explaining our method which is slightly different from ref. \cite{JR}.  

  The generating functional of the $c=0$ one-matrix model is defined by 
\begin{eqnarray}
 Z[J] & = & \frac{1}{Z} \int d^{N^{2}}M~
\e^{-N\tr V(M)} \e^{J\cdot \Phi},\nn \\
 Z & = & \int d^{N^{2}}M~\e^{-N\tr V(M)},   \label{3-1-0} \\
 V(M) & = & \oh M^{2}-\frac{g}{3}M^{3},   \label{c=0pot} \\
J\cdot \Phi &  = & \Lint J(\ze) \Phi(\ze),     
\end{eqnarray}
where 
\beq
\Phi(\ze) =\frac{1}{N} \tr \frac{1}{\ze-M}  \label{loopoperator}
\eeq
is a loop operator and the contour of $\ze$-integral $L$ is 
chosen to be parallel 
to the imaginary axis 
such that in the region of the right of $L$ there are no poles 
of $\Phi(\ze)$. 
The source function $J(\ze)$ can take
an arbitrary form as a function on $L$.  
The variable $\ze$ can be regarded as being conjugate to 
the length of the loop in the sense of Laplace transform. 

We start with the ground state condition of the stochastic Hamiltonian 
\beq
0=-\frac{1}{Z}\int d^{N^{2}}M~\sum_{\alp=1}^{N^{2}}
\frac{\prt}{\prt M_{\alp}}
         \left( \e^{-N \tr V(M)}\frac{\prt}{\prt M_{\alp}}\e^{J\cdot \Phi}
        \right)                     \label{3-1-1}
\eeq
where $M$ is expanded by the basis of $N\times N$ hermitian matrices 
$\{ t^{\alp}\}$:
$$
M=\sum_{\alp=1}^{N^{2}}M_{\alp}~t^{\alp}.  
$$
Using the identities 
\beas
\sum_{\alp}\tr(At^{\alp}Bt^{\alp}) & = & \tr A \, \tr B,  \\
\sum_{\alp}\tr(At^{\alp})\tr(Bt^{\alp}) & = & \tr AB, 
\eeas
we obtain formulas such as 
\beq
\sum_{\alpha}{\partial \over \partial M_{\alpha}}\Phi(\ze)
{\partial \over \partial M_{\alpha}}\Phi(\ze' )
= -\frac{1}{N} \prt_{\ze}\prt_{\ze'}D_z(\ze,\ze')\Phi(z),
\eeq
\beq
\sum_{\alpha}{\partial^2 \over \partial M_{\alpha}^2}\Phi(\ze)
=-N{\partial \over \partial \ze}\Phi(\ze)^2,
\eeq
\beq
{1\over N}\sum_{\alp}\tr(M^nt^{\alp})\tr(t^{\alp}{1\over (\ze-M)^2})
=-{d\over d\ze} (\ze^n\Phi(\ze)-
\sum_{k=0}^{n-1}\ze^k {1\over N}\tr(M^{n-1-k})),
\label{formula3}
\eeq
where the symbol $D_z$ is the so-called combinatorial derivative,
 defined as  
\beq
D_z(\ze,\ze') f(z) \equiv \frac{f(\ze)-f(\ze')}{\ze-\ze'}, 
\eeq
which appears when two loops merge 
into a new loop. 

 We can then reduce eq. (\ref{3-1-1}) to a functional differential 
equation with respect to the source $J(\ze)$,  
\bea
0 & = & \cH Z[J]                   \label{3-2-1} \\
\cH & =& \Lint  J(\ze)\prt_{\ze}\left[ \left( \frac{\dl}{\dl J(\ze)}-
        \oh (\ze-g \ze^{2})\right)^{2}
-\frac{1}{4}(\ze-g \ze^{2})^{2}-g \ze
        \right]       \nn \\
   &  & +\frac{1}{N^{2}}\Lint J(\ze)\Lintp J(\ze')\prt_{\ze}\prt_{\ze'}
               D_z(\ze,\ze') \frac{\dl}{\dl J(z)}.         \label{3-2-2}
\eea
The functional derivative $\frac{\dl}{\dl J(\ze)}$ is defined for $\ze$ 
on the contour and acts on the source $J(\ze')$ as a delta function 
\beq
\frac{\dl J(\ze')}{\dl J(\ze)}=2\pi i \delta(\ze-\ze').             
\eeq
when both $\ze$ and  $\ze'$ resides on the same contour. 
$\cH$ is the exact stochastic Hamiltonian for the $c=0$ PSFT 
before taking the double-scaling limit. 
The first term represents the splitting process
of a loop, while the second represents the merging process 
of two loops.   
Note that the expression 
(\ref{3-2-2}) is normal ordered in the sense that the differential 
operators ${\delta \over \delta J(\ze)}$ always sit right of $J(\ze)$.


 Introducing a lattice spacing $a$, we now take the continuum limit (the 
double scaling limit) $a\limit 0$ by defining the scaling variables as 
$$
\ze=\ze_{*}(1+a y),~~~g=g_{*}(1-a^{2}t),~~~\frac{1}{N}=a^{5/2} \gst 
$$
where the critical points are 
$$
\ze_{*}=(\sqrt{3}+1)\cdot 3^{1/4},~~~g_{*}=\frac{3^{1/4}}{6}.
$$
The meanings of the variables $y$, $t$ and $\gst$ are 
the Laplace conjugate of loop length, 
the cosmological constant, and the string coupling constant,
 respectively. From the result of the disk amplitude \cite{BIPZ}, 
it can be seen that the contribution
of  the poles of $\Phi(\ze)$ accumulates  
to a cut of the interval $[-(\sqrt{3}-1)\cdot 3^{3/4}+O(a^2),
{}~\ze_{*}-4\cdot 3^{-1/4}a \sqrt{t}+O(a^2)]$.
The region $\mbox{Re}\,\ze\geq\ze_{*}$
contains no singularities of $\Phi(\ze)$.
So, in the scaling limit we can choose as the contour $L$  the line 
$[\ze_{*}-i\infty, \ze_{*}+i\infty]$, 
which is mapped to the imaginary axis in $y$-plane.
 
     In order to obtain the correct continuum limit, we have to subtract a 
non-universal part from the correlation functions. In the present case, 
it is required only for the one-point disk amplitude. 
Namely, the connected $K$-point function 
$$
W(\ze_{1},\cdots,\ze_{K})=\bra \Phi(\ze_{1})\cdots\Phi(\ze_{K})\ket_{c} 
$$
is written as
\beas
W(\ze) & = & \oh (\ze-g \ze^{2})+a^{3/2} w(y)+O(a^{2})     \\
W(\ze_{1},\cdots,\ze_{K}) & = & a^{3K/2}w(y_{1},\cdots,y_{K})
                    +O(a^{(3K+1)/2})~~~(K\geq 2)    
\eeas
where $\oh (\ze-g \ze^{2})$ is the non-universal part of the disk amplitude,  
and $w$ is the universal part giving the correct continuum limit. 

  Thus we  redefine the source $\Jty$ and the functional derivative by    
\bea
\frac{\dl}{\dl J(\ze)}&=&\oh (\ze-g\ze ^{2})+ a^{3/2}\frac{\dl}{\dl \Jty},   
\label{chderizeta}  \\
J(\ze)&=&\ze_{*}^{-1}a^{-5/2}\Jty.  
\eea
The shift (\ref{chderizeta}) corresponds to the rescaling of the 
partition function as 
 $Z[J]=\exp(\Lint J(\ze)(\ze -g\ze^2)/2)Z[\tilde J(y)]$. 
Then, $\cH$ becomes 
\begin{eqnarray*}
\cH & = &
a^{1/2}\ze_{*}^{-1}\left[\int_{-i\infty}^{i\infty} {dy\over  2\pi i} \Jty
\prt_{y}
 \left(\frac{\dl^{2}}{\dl\Jty^{2}}-C (y^{3}-\frac{3}{4}T y)
            +O(a^{1})\right)  \right.    \\
 &  & \left. + \gst^{2}\ze_{*}^{-2}
   \int_{-i\infty}^{i\infty}{dy\over2\pi i}\Jty  
  \int_{-i\infty}^{i\infty}{dy' \over2\pi i}\tilde{J}(y')
  \prt_{y}\prt_{y'} D_z(y,y')\frac{\dl}{\dl  \tilde{J}(z)}\right],
\end{eqnarray*}
where 
$$
T=\frac{16}{3 (1+\sqrt{3})^{2}} t,~~~C=\frac{\sqrt{3}}{12}(1+\sqrt{3})^{3}.
$$
Note that in the merging interaction  (namely,
the term of the form $JJ\frac{\dl}{\dl J}$) the shift of 
the functional derivative does not contribute because 
\beq
\prt_{\ze}\prt_{\ze'} D_z(\ze,\ze') \oh (z-gz^2)=0.
\eeq

  After finite rescalings
$$
\Jty\limit\Jty C^{-1/2},~~
\frac{\dl}{\dl\Jty}\limit\frac{\dl}{\dl\Jty}C^{1/2},~~
\gst\limit\gst\ze_{*}C^{1/2},
$$
 we have the stochastic Hamiltonian in the 
continuum theory 
\bea
\cH & = & \int_{-i\infty}^{i\infty} {dy\over  2\pi i} \Jty \prt_{y}
 \frac{\dl^{2}}{\dl\Jty^{2}}-\inty  \Jty\tilde{\rho}(y)    \nn   \\
 &  & + \gst^{2}
   \int_{-i\infty}^{i\infty}{dy\over2\pi i}\Jty  
  \int_{-i\infty}^{i\infty}{dy' \over2\pi i}\tilde{J}(y')
  \prt_{y}\prt_{y'} D_z(y,y')\frac{\dl}{\dl  \tilde{J}(z)},
                \label{3-6-1}      \\
\tilde{\rho}(y) & = & 3y^2-\frac{3}{4}T ,     \label{k=2rho}
\eea
where the overall factor $a^{1/2}\ze_{*}^{-1}C^{1/2}$ was absorbed by a 
redefinition of the fictitious time.
This result, which has been already known 
from ref. \cite{JR}, essentially coincides 
with the form of the $c=0$ non-critical string field 
theory\footnote
{
For a derivation of the $c=0$ Hamiltonian directly from 
dynamical triangulation, see ref. \cite{Watabiki}. 
}
proposed by Ishibashi and Kawai \cite{IK},
 if one uses the 
Laplace-transformed string fields instead of
their loop-length representation.

\subsection{PSFT for a Higher Critical One-Matrix Model}

  We next treat a case of higher critical point ($k=3 (c=-22/5)$).\footnote{
Extension of the formalism of ref. \cite{KKMW} to higher critical 
cases has been given 
in \cite{Kleb}.} 
This problem is interesting since a naive extension of the 
$c=0$ hamiltonian leads to an apparent contradiction 
as discussed in ref. \cite{IK3}.

   The $k=3$ critical theory is realized by the potential of 
the fourth-degree polynomial 
\beq
V(M)=\frac{\beta}{N} (\frac{g_2}{2}M^2
+\frac{g_3}{3}M^3+\frac{1}{20}M^4),
\label{k=3pot}
\eeq
where $g_3$ is a real solution of the cubic equation 
$$
25g_3^3-30g_3+32=0,
$$
or explicitly
\beq
g_3=-\left(\frac{2}{25}\right)^{1/3}((8+3\sqrt{6})^{1/3}+(8-3\sqrt{6})^{1/3}),
\label{g3}
\eeq
and 
\beq
g_2=\frac{5g_3^2-2}{3}.
\label{g2}
\eeq
Note that we do not use the well known even critical potential (of 
sixth order) 
at $k=3$ critical point, in order to avoid a complication  
caused by the $Z_2$ symmetry.\footnote
{
We do not know any previous work discussing the $k=3$ critical 
point using the quartic potential. 
For a brief explanation of the derivation of the quartic
 critical potential, see  the Appendix A. 
}

   Now, let us derive the Hamiltonian
 for the $k=3$ critical case. Considering the 
generating functional (\ref{3-1-0}) with the potential (\ref{k=3pot}), 
we have the Hamiltonian before taking the scaling limit: 
\bea
\cH & =& \Lint  J(\ze)\prt_{\ze}\left[ \left( \frac{\dl}{\dl J(\ze)}-
        \oh V'(\ze)\right)^{2}-\frac{1}{4}V'(\ze)^{2}\right.          \nn \\
   &  & \left. +\frac{\beta}{N}(g_3\ze+\frac{1}{5}\ze^2)+\frac{\beta}{N}
               \frac{1}{5}\ze \oint\frac{d\ze'}{2\pi i}\ze'\frac{\dl}{\dl
J(\ze')} 
                \right] \nn \\
   &  & +\frac{1}{N^{2}}\Lint J(\ze) \Lintp J(\ze') \prt_{\ze}\prt_{\ze'}
              D_z(\ze, \ze') \frac{\dl}{\dl J(z)},
                           \label{k=3LH}
\eea
where $\oint$ is the integral over the contour encircling the poles of 
$\Phi(\ze)$.
Note that $\oint\frac{d\ze'}{2\pi i}\ze' \frac{\dl}{\dl J(\ze')}$ 
corresponds to the insertion of a microscopic loop
 represented by the operator 
$\frac{1}{N}\tr M$. In the $c=0$ case, 
no such term appears if one uses
the third order potential of (\ref{c=0pot}), 
because of the formula (\ref{formula3}).


Next,  we 
need to identify the non-universal parts of loop operators 
in the scaling limit $a\limit 0$ defined by 
\beq
\frac{N}{\beta}=1-a^3t,            \label{N/B}
\eeq
\beq
\ze=\ze_*(1+ay), ~~~\ze_*=\frac{-5g_3+2}{3}.
\label{zeta}
\eeq   
{}From the results of the Appendix A, we have 
\bea
\bra \Phi(\ze) \ket_0 & = & \oh V'(\ze)+a^{5/2}w(y)+O(a^{7/2}),  
\label{k=3sdiskmatrix}\\
\left \bra \frac{1}{N} \tr M \right \ket_0 & = & -\frac{32+25g_3}{15}
     -a^3\frac{4}{5}t+a^4\frac{3}{4}t^{4/3}+O(a^5), 
\label{k=3microdisk}
\eea
where, in the disk amplitude corresponding to a  
macroscopic loop, the first term $\oh V'(\ze)$ is the non-universal part, 
while the second term $w(y)$ denotes the universal one:
\beq
w(y)=-\frac{1}{5}\ze_*^{5/2}(y^2-\oh T^{1/3}y+\frac{3}{8}T^{2/3})
  \sqrt{y+T^{1/3}} 
\label{k=3diskmatrix}
\eeq
with $T=(2\ze_*^{-1})^3t$. 
For the microscopic disk amplitude (\ref{k=3microdisk}), 
the first two terms represents the non-universal part, and the third term 
is universal. 



   Using the above results, we see that the source $\Jty$ and the 
microscopic loop operator $\cO_0$ in the continuum theory 
should be defined by 
\beq
\frac{\dl}{\dl J(\ze)}=\oh V'(\ze)+ a^{5/2}\frac{\dl}{\dl \Jty},~~~
         J(\ze)= a^{-7/2}\ze_*^{-1}\Jty,
\eeq
\beq
\oint\frac{d\ze'}{2\pi i}\ze'\frac{\dl}{\dl J(\ze')} = 
   -\frac{32+25g_3}{15}-a^3\frac{4}{5}t+a^{4}\cO_{0}.
\eeq
The string coupling constant is introduced as 
\beq
\frac{1}{N}=a^{7/2}\gst.
\eeq

  Substituting (\ref{N/B}), (\ref{zeta}) and these 
rescaled expressions into the lattice Hamiltonian 
(\ref{k=3LH}), we obtain  
\bea
\cH & = &
a^{3/2}\ze_*^{-1}\inty \Jty \prt_y\left[\frac{\dl^{2}}{\dl \Jty^{2}}
     -\frac{1}{25}\ze_*^5(y^5+\frac{5}{8}T y^2)
+\frac{1}{5}\ze_* y\cO_0\right]   \nn \\
 & & +a^{3/2}\ze_*^{-3}
\gst^2\inty\Jty\intyp\Jtyp\prt_y\prt_{y'}D_z(y,y')
           \frac{\dl}{\dl\tilde{J}(z)}    \nn \\
 & & +O(a^2),
\eea
where as in the $c=0$ case  we chose the line 
$[\ze_*-i\infty,~\ze_*+i\infty]$ as the contour $L$. 
Note that in the merging term the shift of the 
derivative $\frac{\dl}{\dl J}$ 
produces a quadratic term with respect to $\tilde{J}$, but it 
does not contribute to the leading term
of $\cH$ as $a\rightarrow 0$, 
because 
\bea
\lefteqn{\frac{1}{N}\Lint J(\ze)\Lintp J(\ze')\prt_{\ze}\prt_{\ze'}
   D_z(\ze,\ze')\oh V'(\ze)}  \nn \\
 & = & a^2\gst ^2 \inty\Jty\intyp\Jtyp\frac{1}{10}. 
\eea

    After making the rescalings again,  
$$
\Jty\limit (\frac{1}{5}\ze_*^{5/2})^{-1}\Jty,~~~
\frac{\dl}{\dl \Jty}\limit \frac{1}{5}\ze_*^{5/2}\frac{\dl}{\dl \Jty},  
$$
\beq
\cO_0\limit -\frac{1}{5}\ze_*^4\cO_0,~~~
\gst\limit\frac{1}{5}\ze_*^{7/2}\gst, 
\label{k=3rescale}
\eeq
we finally obtain the continuum Hamiltonian 
\bea
\cH & = & \inty \Jty \prt_y\frac{\dl^{2}}{\dl \Jty^{2}}
     -\inty\Jty\tilde{\rho}(y)  \nn \\
 & & +\gst^2\inty\Jty\intyp\Jtyp\prt_y\prt_{y'}D_z(y,y')
           \frac{\dl}{\dl\tilde{J}(z)} , 
\label{k=3H}       \\
\tilde{\rho}(y) & = & 5y^4+\frac{5}{4}T y+\cO_0,      \label{k=3rho}
\eea
where the overall factor
$a^{3/2}\frac{1}{5}\ze_*^{3/2}$ was absorbed into 
the fictitious time. 

We emphasize that in this Hamiltonian, the tadpole term 
$-\inty\Jty\tilde{\rho}(y)$, is not a pure c-number, but 
contains the $y$-independent  
operator $\cO_0$.  This is in contrast to the $c=0$ case 
where the tadpole term consists 
only of the c-number function. 
Actually, the ``operator" part of the tadpole is a misnomer. 
We should rather call it a kinetic term. 
The Hamiltonian description of the 
higher critical point requires a kinetic term for 
infinitesimally small loop, in addition to the genuine tadpole 
corresponding to the c-number part of $\tilde{\rho}$. 
  
In the Appendix B, we will determine the operator 
$\cO_0$ using the Schwinger-Dyson equations. And 
in the next section, using this result, we confirm that 
it is just necessary for ensuring the closure  
of the algebra of the Schwinger-Dyson operators 
appearing in the Hamiltonian.  
As is discussed in ref. \cite{IK3} 
in trying the extension of the $c=0$ Hamiltonian 
to the higher critical case, the integrability 
condition would not be satisfied if one had naively replaced the 
$\tilde{\rho }(y)$ of the $c=0$ case with  c-number polynomials of   
higher degree.  
The authors in ref. \cite{IK3} proposed a possible way out, which 
is, however, different from ours. 


Before concluding this section,  
we derive the disk amplitude from the PSFT Hamiltonian 
(\ref{k=3H}) and compare with the matrix  model  result as a 
consistency check of our result. 
  In the sphere approximation, the condition  
\beq
\left.\frac{\dl}{\dl\Jty}\cH Z[J]\right|_{J=0}=0 , 
\eeq
is reduced to the equation for the one-point function $w(y)$, 
\beq
\prt_y [w(y)^2-(y^5+\frac{5}{8}T y^2+y\bra \cO_0\ket_0)] =0. 
\eeq
By demanding that the cut of $w(y)$ resides only on the real axis, 
as is required from the original definition of the loop operator 
(\ref{loopoperator}), 
both of  
$w(y)$ and the expectation value of $\cO_0$ are uniquely determined 
as 
 \bea
w(y) & = & (y^{2}-\oh T^{1/3}y+\frac{3}{8}T^{2/3})\sqrt{y+T^{1/3}},  
                                           \label{k=3diskPSFT} \\
\bra \cO_{0}\ket_{0} & = & -\frac{15}{64}T^{4/3},   
\eea
which coincide with the results
(\ref{k=3diskmatrix}) and (\ref{k=3microdisk})
obtained directly without using the
Hamiltonian,   after taking account
 the rescalings (\ref{k=3rescale}).

\vspace{1cm}

\section{Derivation of the Virasoro Constraints}
\setcounter{equation}{0}

  In this section, we examine the integrability condition of the 
Schwinger-Dyson operators associated with the 
Hamiltonian equation,   
\beq
 \cH Z[J]=0    \label{4HC}
\eeq
for $k=2,3$ cases.
As a warmup exercise, let us begin from the simplest case of 
$c=0$. 

\subsection{The Virasoro algebra at $k=2$}

  Recalling the discussions of section 2,  we rewrite the 
Hamiltonian $\cH$ in the form
\bea
  \cH & = & -\inty \Jty\prt_{y} T(y),               \\
  \prt_y T(y) & = &\prt_y  T_{0}(y)+\tilde{\rho} (y),                \\
  T_{0}(y) & = & -\frac{\dl^{2}}{\dl \Jty^{2}}
      -\gst^{2}\intyp \Jtyp\prt_{y'} D_z(y,y') \frac{\dl}{\dl\tilde{J}(z)}.
\eea
Thus the Schwinger-Dyson equation associated with the $c=0$ 
Hamiltonian is 
\beq
   \prt_{y}T(y) Z[J]=0.                      \label{k=2SD}
\eeq
After a straightforward  calculation 
using the functional derivative 
\beq
\frac{\dl\Jty}{\dl\Jtyp}=2\pi i \dl(y-y'), 
\eeq
we find a closed algebra for $T_0$ operators
\beq
 [\prt_{y_{1}}T_{0}(y_{1}), \prt_{y_{2}}T_{0}(y_{2})]=-\gst^{2} 
   \prt_{y_{1}}\prt_{y_{2}}(\prt_{y_{1}}-\prt_{y_{2}})\frac{1}{y_{1}-y_{2}}
   (T_{0}(y_{1})-T_{0}(y_{2})),                \label{algT0}
\eeq
The algebra of $\prt_y T(y)$ is obtained by substituting 
$$ 
T_0(y)=T(y)-\int^ydy' \tilde{\rho}(y')
$$
into  (\ref{algT0}). Then, using the explicit 
expression of  $ \tilde{\rho}$ (\ref{k=2rho})  we find that the 
effect of $\tilde{\rho}$ vanishes 
\beq
\prt_{y_1}\prt_{y_2} (\prt_{y_1}-\prt_{y_2})D_z(y_1,y_2)
   \int^zdy\tilde{\rho}(y)=0.
\eeq
Thus, $\prt_y T(y)$ forms the same closed algebra as $\prt_y T_0(y)$ 
\beq
 [\prt_{y_{1}}T(y_{1}), \prt_{y_{2}}T(y_{2})]=-\gst^{2} 
   \prt_{y_{1}}\prt_{y_{2}}(\prt_{y_{1}}-\prt_{y_{2}})\frac{1}{y_{1}-y_{2}}
   (T(y_{1})-T(y_{2})).              \label{k=2algT}
\eeq
This agrees with the Laplace-transformed version
of the result in ref. \cite{IK2}.

\vspace{0.5cm}

\subsection{The Schwinger-Dyson operators at $k=3$}

    The $k=3$ case is less trivial.
The only difference from the $k=2 \, (c=0)$
case lies in 
$\tilde{\rho}$ (\ref{k=3rho}). Namely, the c-number part of the 
tadpole term  
$5y^4$ in $\tilde{\rho}$  gives a non-vanishing effect 
\beq
\prt_{y_1}\prt_{y_2} (\prt_{y_1}-\prt_{y_2})D_z(y_1,y_2)
   \int^zdy\tilde{\rho}(y)=2(y_1-y_2), 
\eeq
which is the reason why the naive extension violates the integrability 
condition. 
However,  
$\tilde{\rho}$ is not a pure c-number and contains the operator 
part 
$\cO_0$. 
Then, on making the substitution as in the 
$c=0$ case, the algebra of $\prt_y T_0(y)$ becomes 
\bea
 [\prt_{y_{1}}T(y_{1}), \prt_{y_{2}}T(y_{2})] & = & -\gst^{2} 
   \prt_{y_{1}}\prt_{y_{2}}(\prt_{y_{1}}
-\prt_{y_{2}})\frac{1}{y_{1}-y_{2}}
   (T(y_{1})-T(y_{2})) \nn \\
 &  & +2\gst^{2}(y_{1}-y_{2}) \nn \\
 &  & +\prt_{y_{1}}[T_{0}(y_{1}),\cO_{0}]
-\prt_{y_{2}}[T_{0}(y_{2}),\cO_{0}].
                                         \label{k=3algT1}
\eea
Since $\cO_0$ inserts a 
microscopic loop, it can be expressed by some local operator, 
obtained as  
 some coefficient of large $y$ expansion of the loop operator 
$\frac{\dl}{\dl\Jty}$. 
This is done in the Appendix B. 
The final results are  
\bea
\cO_{0} & = & 2\intyC y^{1/2}\frac{\dl}{\dl\Jty}  \label{O01}\\
                & = & 2\lim_{\ep\limit +0}\inty \e^{\ep
y}y^{1/2}\frac{\dl}{\dl\Jty}, 
\label{O02}
\eea 
where the contour $C$ surrounds the negative real axis, 
and the contour $[-i\infty, i\infty]$ is understood to avoid the singularity 
at the origin to the right. Also, the integral 
is defined by the analytic continuation
using the Beta function and  the limit $\ep\limit +0$
must be taken {\bf after} the integration. 

Now in calculating the commutator 
$[T_0(y),\cO_0]$, we can use ({\ref{O02}) rather than (\ref{O01}),  
because the functional derivative 
$$
\frac{\dl\Jty}{\dl\Jtyp}=2\pi i\dl(y-y')
$$
is defined for $y, y'$ on the imaginary axis. 
    As a result of the straightforward calculation  performing 
partial integration once, 
 the commutator $[T_0(y),\cO_0]$ 
becomes 
\beq
[T_{0}(y),\cO_{0}] = -\gst^{2}\int_{C}\frac{dy'}{2\pi i}y'^{-1/2}
       \frac{1}{y'-y}\left(\frac{\dl}{\dl\Jtyp}-\frac{\dl}{\dl\Jty}\right). 
\eeq
Substituting (\ref{4-11-0}), and using the formulas in the Appendix B,
we see that 
only the term 
$$
-\gst^2y^{5/2}\int_C \frac{dy'}{2\pi i}y'^{-1/2}\frac{1}{y-y'}
$$
survives after some cancellations. Thus, we obtain 
\beq
[T_{0}(y),\cO_{0}]=-\gst^{2}y^{2}, 
\eeq
which makes the algebra (\ref{k=3algT1}) for $T(y)$ closed: 
\beq
 [\prt_{y_{1}}T(y_{1}), \prt_{y_{2}}T(y_{2})]=-\gst^{2} 
   \prt_{y_{1}}\prt_{y_{2}}(\prt_{y_{1}}
-\prt_{y_{2}})\frac{1}{y_{1}-y_{2}}
   (T(y_{1})-T(y_{2})).               \label{k=3algT}
\eeq

We note that this algebra is of course identical with the usual 
Virasoro algebra (\ref{k=3Virasoro}), 
after taking into account the contribution from
 the transformation (\ref{k=3rescaleZ}). All what we have done is merely 
a check of self consistency. It  clarifies, however,  
how the closure of the Schwinger-Dyson operators associated 
with the Hamiltonian is satisfied for higher critical points, owing to 
the presence of the operator part of $\tilde{\rho}(y)$.

\vspace{1cm}
\section{$c=1/2$ PSFT from Two-Matrix Model}
\setcounter{equation}{0}

 We now apply our method to the two-matrix model and derive 
the $c=1/2$ PSFT without making any restrictions 
on the spin configurations of the string fields. 
As we emphasized in the Introduction, 
such a treatment will hopefully  reveal 
certain universal properties of the PSFT 
which are basically independent of the 
structure of target spaces. 
This is our motivation for performing 
this analysis in spite of its technical difficulties.

\subsection{Stochastic Hamiltonian of  the Two-Matrix Model}

  The generating functional of the two-matrix model is defined by 
\bea
Z[J] & = & \frac{1}{Z}\intAB\, \e^{-S}\e^{J\cdot \Phi}, \nn \\
Z & = & \intAB\, \e^{-S}, \nn  \\
S & = & N\tr (V(A)+V(B)-cAB),~~~V(A)=\oh A^{2}-\frac{g}{3}A^{3},  
\eea
\bea
\lefteqn{J\cdot\Phi = \Lint J_{A}(\ze)\Phi_{A}(\ze)
      +\Lintsg J_{B}(\sg)\Phi_{B}(\sg) }  \nn \\
  &   &  +\sum_{n=1}^{\infty}\Lintzesgn J_{n}(\ze_{1},\sg_{1},\cdots,
          \ze_{n},\sg_{n})\Phi_{n}(\ze_{1},\sg_{1},\cdots,\ze_{n},\sg_{n}), 
                    \label{5-1-1}    
\eea
where the components of the string field $\Phi$ 
are defined in terms of the matrix variables as 
\beas
\Phi_{A}(\ze) & = & \frac{1}{N}\tr\frac{1}{\ze-A},\\
\Phi_{B}(\sg) & = & \frac{1}{N}\tr\frac{1}{\sg-B},\\
\Phi_{n}(\ze_{1},\sg_{1},\cdots,\ze_{n},\sg_{n}) & = & \frac{1}{N}
   \tr\left(\frac{1}{\ze_{1}-A}\frac{1}{\sg_{1}-B}\cdots\frac{1}{\ze_{n}-A}
   \frac{1}{\sg_{n}-B}\right)    \\
   &    &   ~~~~~~~~~~~~~~~~~~~~~~~~~~~~~~~~(n=1,2,\cdots).
\eeas
The component $\Phi_{A}$ ($\Phi_{B}$) represents a loop on which 
 only  a 
single spin $A$ ($B$) is put, and $\Phi_{n}$ represents a loop on 
which two domains of $A$- and $B$-spins alternatively appear $n$ times.
The variables $\ze_i, \sg_i \quad (i=1,2,\ldots, n)$ 
can be regarded as the 
conjugate variables corresponding to 
lengths on the loop for spinstates $A, B$, respectively.

As before, the stochastic Hamiltonian for this system is 
derived from the 
identity
\beq
0=-\intAB \sum_{\alp=1}^{N^{2}}\left[\frac{\prt}{\prt A_{\alp}}\e^{-S}
                                    \frac{\prt}{\prt A_{\alp}}+
   \frac{\prt}{\prt B_{\alp}}\e^{-S}\frac{\prt}{\prt B_{\alp}}\right]
       \e^{J\cdot\Phi}.   \label{5-2-1}
\eeq
By extending the formulas for the one-matrix model, 
we can arrange 
eq. (\ref{5-2-1})  in the following form:
\bea
0 & = & \cH Z[J] \nn \\
\cH & = & -J\cdot \left(K\frac{\dl}{\dl J}\right)
      -J\cdot\left(\frac{\dl}{\dl J}\vee\frac{\dl}{\dl J}\right)
       -\frac{1}{N^{2}}J\cdot\left(J\cdot\left(\wedge\frac{\dl}{\dl J}
\right)\right)
       -J\cdot T.
       \label{5-3-1}
\eea
Here, reflecting  the cyclic symmetry of pairs $(\ze_{i},\sg_{i})$, 
the derivative $\frac{\dl}{\dl J_{n}}$ is defined by 
\bea
\lefteqn{\frac{\dl J_{m}(\ze '_{1},\sg '_{1},\cdots,\ze '_{m},\sg '_{m})}
{\dl J_{n}(\ze_{1},\sg_{1},\cdots,\ze_{n},\sg_{n})}
 =\dl_{m,n}\frac{1}{n}(2\pi i)^{2n}} \nn \\
 & \times & \sum_{c:\mbox{\scriptsize cyclic permutation}}
\dl(\ze_{1}-\ze '_{c(1)}) \dl(\sg_{1}-\sg '_{c(1)})\cdots
\dl(\ze_{n}-\ze '_{c(n)})\dl(\sg_{n}-\sg '_{c(n)}).
                                                 \label{5-3-2}
\eea

\vspace{0.3cm}
Let us explain the meaning of each term. 

\noindent
i) 
The first term, ``kinetic term", $J\cdot 
\left(K\frac{\dl}{\dl J}\right)$, coming from a part of the product of 
the derivatives of the classical action and 
the source term,  symbolizes the contributions which 
preserve the number of string fields.  
The first few components of the kinetic operator $K$ 
  are as follows. 
\beas
\left(K\frac{\dl}{\dl J}\right)_{A}(\ze) & = &
\vec{\prt_{\ze}}(\ze-g \ze^{2})
   \frac{\dl}{\dl J_{A}(\ze)}-c\prt_{\ze} 
   \ointsg\sg\frac{\dl}{\dl J_{1}(\ze,\sg)}, \\
\left(K\frac{\dl}{\dl J}\right)_{B}(\sg) & = &
\vec{\prt_{\sg}}(\sg-g \sg^{2})
   \frac{\dl}{\dl J_{B}(\sg)}-c\prt_{\sg} 
   \ointze\ze\frac{\dl}{\dl J_{1}(\ze,\sg)}, 
\eeas
\beas
\lefteqn{\left(K\frac{\dl}{\dl J}\right)_{1}(\ze_{1},\sg_{1}) = 
 (\vec{\prt_{\ze_{1}}}(\ze_{1}-g \ze_{1}^{2})
 +\vec{\prt_{\sg_{1}}}(\sg_{1}-g \sg_{1}^{2}))
             \frac{\dl}{\dl J_{1}(\ze_{1},\sg_{1})} }\\
  &  & +c\ointsg\sg\frac{\dl}{\dl J_{2}(\ze_{1},\sg,\ze_{1},\sg_{1})}
   +c\ointze\ze\frac{\dl}{\dl J_{2}(\ze_{1},\sg_{1},\ze,\sg_{1})} \\
  &  & +g\left(\frac{\dl}{\dl J_{A}(\ze_{1})}
+\frac{\dl}{\dl J_{B}(\sg_{1})}
  \right),
\eeas
\beas
\lefteqn{\left(K\frac{\dl}{\dl J}\right)_{2}
(\ze_{1},\sg_{1},\ze_{2},\sg_{2})
 =  \sum_{j=1}^{2}\{ \vec{\prt_{\ze_{j}}}(\ze_{j}-g \ze_{j}^{2})+
    \vec{\prt_{\sg_{j}}}(\sg_{j}-g \sg_{j}^{2})\} \frac{\dl}{\dl 
     J_{2}(\ze_{1},\sg_{1},\ze_{2},\sg_{2})} }\\
 & & +c\ointsg\sg\left(\frac{\dl}
     {\dl J_{3}(\ze_{1},\sg,\ze_{1},\sg_{1},\ze_{2},\sg_{2})}
      +\frac{\dl}
     {\dl J_{3}(\ze_{1},\sg_{1},\ze_{2},\sg,\ze_{2},\sg_{2})}\right) \\
 & & +c\ointze\ze\left(\frac{\dl}
     {\dl J_{3}(\ze_{1},\sg_{1},\ze,\sg_{1},\ze_{2},\sg_{2})}
      +\frac{\dl}
     {\dl J_{3}(\ze_{1},\sg_{1},\ze_{2},\sg_{2},\ze,\sg_{2})}\right) \\
 & & -gD_{\sg}(\sg_{1},\sg_{2})\left(\frac{\dl}{\dl J_{1}(\ze_{1},\sg)}+
             \frac{\dl}{\dl J_{1}(\ze_{2},\sg)}\right) \\
 & & -gD_{\ze}(\ze_{1},\ze_{2})\left(\frac{\dl}{\dl J_{1}(\ze,\sg_{1})}+
             \frac{\dl}{\dl J_{1}(\ze,\sg_{2})}\right),
\eeas
\beq
  \cdots,                                    \label{5-3-3}
\eeq
where the arrow over $\prt$ indicates that
 the derivative acts on the whole 
functions that follow it. 

The structure of the higher components can be inferred from 
these expressions. Basically,  
 each component represents one of  the following 
two processes. The first is the 
propagation of string preserving a spin configuration on a loop, with 
the loop length being either kept fixed or 
decreased by one-lattice unit. The second is a process 
in which only a single spin is flipped
and the loop length is preserved. 
For example, 
let us consider 
$\left(K\frac{\dl}{\dl J}\right)_{1}(\ze_{1},\sg_{1})$. The 
first and last terms express the former process. Note that, 
as a special case when a domain consists of only 
one spin, the process 
can annihilate the domain. 
The last term represents this. On the other hand, the second 
and third terms express the latter process, 
with a single spin flip preserving the loop length. 

In the case of $\left(K\frac{\dl}{\dl J}\right)_{2}$, it is noted 
that 
$-D_{\sg}(\sg_{1},\sg_{2})\frac{\dl}{\dl J_{1}(\ze_{1},\sg)}$ in 
$\left(K\frac{\dl}{\dl J}\right)_{2}(\ze_{1},\sg_{1},\ze_{2},\sg_{2})$ 
represents the following loop when it acts on the partition function, 
$$
-D_{\sg}(\sg_{1},\sg_{2})\frac{\dl}{\dl J_{1}(\ze_{1},\sg)}=
\frac{1}{N}\tr\left(\frac{1}{\ze_{1}-A}\frac{1}{\sg_{1}-B}\frac{1}{\sg_{2}-B}
    \right). 
$$
Namely, the $\ze_{2}$-domain has disappeared in 
$\Phi_{2}(\ze_{1},\sg_{1},\ze_{2},\sg_{2})$.

\vspace{0.3cm}
\noindent
ii) The second term 
$J\cdot\left(\frac{\dl}{\dl J}\vee\frac{\dl}{\dl J}\right)$, 
coming from the second derivative of the 
source term, represents processes where a string splits into two. 
The symbol $\spJ_I$ represents the result of splitting of a string 
with the spin configuration $I$. The first few components 
are 
\beas
\spJ_{A}(\ze) & = & -\prt_{\ze}\frac{\dl^{2}}{\dl J_{A}(\ze)^{2}},  \\
\spJ_{B}(\sg) & = & -\prt_{\sg}\frac{\dl^{2}}{\dl J_{B}(\sg)^{2}},  \\
\spJ_{1}(\ze_{1},\sg_{1}) & = & 
   -2\left(\frac{\dl}{\dl J_{A}(\ze_{1})}\prt_{\ze_{1}}
   +\frac{\dl}{\dl J_{B}(\sg_{1})}\prt_{\sg_{1}}\right)
   \frac{\dl}{\dl J_{1}(\ze_{1},\sg_{1})}, \\
\spJ_{2}(\ze_{1},\sg_{1},\ze_{2},\sg_{2}) & = & 
  -2\sum_{j=1}^{2}\left(\frac{\dl}{\dl J_{A}(\ze_{j})}\prt_{\ze_{j}}
  +\frac{\dl}{\dl J_{B}(\sg_{j})}\prt_{\sg_{j}}\right)
  \frac{\dl}{\dl J_{2}(\ze_{1},\sg_{1},\ze_{2},\sg_{2})}  \\
 &  & +2D_{\ze}(\ze_{1},\ze_{2})\frac{\dl}{\dl J_{1}(\ze,\sg_{1})}
        D_{\ze}(\ze_{1},\ze_{2})\frac{\dl}{\dl J_{1}(\ze,\sg_{2})}  \\
 &  & +2D_{\sg}(\sg_{1},\sg_{2})\frac{\dl}{\dl J_{1}(\ze_{1},\sg)}
        D_{\sg}(\sg_{1},\sg_{2})\frac{\dl}{\dl J_{1}(\ze_{2},\sg)}, 
\eeas
\beq
  \cdots.                \label{5-5-1}
\eeq
We hope the structure of the higher terms
containing is self-explanatory 
from these examples. 

\noindent
iii) The third term
$\frac{1}{N^{2}}J\cdot\left(J\cdot\left(
\wedge\frac{\dl}{\dl J}\right)\right)$, 
coming from the square of the first derivative 
of the source term, symbolizes processes in which  two strings 
merge into a single string. The symbol $\mrJ_{IJ}$ 
expresses the result of  the two strings 
with the spin configurations $I, J$ merging into a single string   
: 
\beas
\mrJ_{A,A}(\ze; \ze') & = &
-\prt_{\ze}\prt_{\ze'}D_z(\ze,\ze')\frac{\dl}{\dl J_{A}(z)},  \\
\mrJ_{B,B}(\sg; \sg') & = &
 -\prt_{\sg}\prt_{\sg'}D_s(\sg,\sg')\frac{\dl}{\dl J_{B}(s)},  \\
\mrJ_{A,B}(\ze; \sg) & = & 0,
\eeas
\beas
\lefteqn{\mrJ_{n,A}(\ze_1,\sg_1,\cdots,\ze_n, \sg_n; \ze') =
\mrJ_{A,n}(\ze'; \ze_1,\sg_1,\cdots, \ze_n,\sg_n)} \\
 & = & -\prt_{\ze'}\sum_{j=1}^{n}\prt_{\ze_j}D_z(\ze_j, \ze')
             \frac{\dl}{\dl J_n(\ze_1,\sg_1,\cdots, z,\sg_j,
\cdots, \ze_n,\sg_n)}, 
\eeas
\beas
\lefteqn{\mrJ_{n,B}(\ze_1,\sg_1,\cdots,\ze_n, \sg_n; \sg') =
\mrJ_{B,n}(\sg'; \ze_1,\sg_1,\cdots, \ze_n,\sg_n)} \\
 & = & -\prt_{\sg'}\sum_{j=1}^{n}\prt_{\sg_j}D_s(\sg_j, \sg')
             \frac{\dl}{\dl J_n(\ze_1,\sg_1,\cdots,
\ze_j,s, \cdots, \ze_n,\sg_n)}, 
\eeas
\beas
\lefteqn{\mrJ_{n,1}
(\ze_1,\sg_1,\cdots,\ze_n,\sg_n ;\ze'_1, \sg'_1) =
\mrJ_{1,n}(\ze'_1,\sg'_1; \ze_1,\sg_1,\cdots, \ze_n.\sg_n)} \\ 
 & = & \sum_{j=1}^n   D_z(\ze_j,\ze'_1)D_w(\ze_j,\ze'_1)
\frac{\dl}{\dl J_{n+1}(\ze_1,\sg_1,\cdots,\ze_{j-1},\sg_{j-1},
                                       z, \sg'_1,w,\sg_j, \cdots,\ze_n,\sg_n)}
\\
 &+ & \sum_{j=1}^n D_s(\sg_j,\sg'_1)D_t(\sg_j,\sg'_1)
\frac{\dl}{\dl J_{n+1}(\ze_1,\sg_1,\cdots, \ze_j,s,\ze'_1,t,  
                                     \ze_{j+1},\sg_{j+1},\cdots,\ze_n,\sg_n)},
\\
\eeas
\beq
  \cdots. ~~~~~~~~~~~~~~~~~~~~~~~~~~~~~~(n=1,2,\cdots)
        \label{5-5-2}
\eeq
We again hope that the structure of the generic term is 
self-explanatory from these examples. 

\noindent 
iv) The last term (the tadpole term), 
which together with the kinetic term is originated from the 
product of the derivatives of 
the classical action and the source term, 
shows the processes of the annihilation of a string into nothing:
$$
J\cdot T=\Lint J_{A}(\ze)g+\Lintsg J_{B}(\sg)g.
$$

We here would like to emphasize  an important property  
which characterizes all of the above formulas and plays
an essential role later in studying the scaling limit.  Namely, 
all the processes occur  
{\bf locally}  with 
respect to  the spin 
domains.  Because of locality, 
more than two domains never be created or 
annihilated at the same 
time.  As a consequence of this rule, 
 only the strings consisting of one domain, 
$\Phi_{A}$ and $\Phi_{B}$, 
can be annihilated into nothing. 
Also, only a single 
pair of domains can participate in the splitting or merging processes, 
and other domains are left intact. 

\subsection{Hamiltonian in Continuum Theory}

Let us now consider the continuum limit of the Hamiltonian 
(\ref{5-3-1}). As in the one-matrix cases, 
the first task in  carrying out this is to identify 
and to subtract the non-universal 
parts of the disk amplitudes.  At this point, 
a new feature arises.  Namely, as we discuss  in the Appendix C, 
the non-universal part of a disk amplitude 
with a given spin configuration contains, in general,   
the universal parts of the disk amplitudes with simpler spin 
configurations, in addition to the non-universal c-number function
which has already appeared in the one-matrix model. 
 
  By introducing the connected $k$-point
correlator in the $J=0$ background as
\beq
G^{(k)}_{I_{1},\cdots,I_{k}} = \bra\Phi_{I_{1}}\cdots\Phi_{I_{k}}\ket,
                          ~~~~~~~~  I_{1},\cdots,I_{k}=A,B,1,2,\cdots,
                                 \label{5-7-1}
\eeq
the generating functional is written as
\beq
Z[J]=\exp\left[J\cdot G^{(1)}+\frac{1}{2!}J\cdot (J\cdot G^{(2)})+
  \frac{1}{3!}J\cdot(J\cdot (J\cdot G^{(3)}))+\cdots\right].
\label{5-7-2}
\eeq
Now,  the investigation of disk amplitudes
in the Appendix C shows that 
the universal part $\hat{\Phi}_{I}$ 
of the operator $\Phi_{I}$
is obtained by a linear transformation of the following form:
\beq
\Phi_{I}=\sum_{J}\cM_{IJ}\hat{\Phi}_{J}+\phi_{I},   \label{5-8-1}
\eeq
where $\cM_{IJ}$ is a mixing matrix of the universal parts, which is 
the new feature  noticed above, and $\phi_{I}$ is 
the non-universal c-number function. 
The mixing matrix $\cM_{IJ}$ is upper-triangular, i.e.
$\cM_{IJ}=0$ for $I<J$ and is invertible.   
 The first few components of (\ref{5-8-1}) are 
\beas
\Phi_{A}(\ze) & = & \hat{\Phi}_{A}(\ze)+\phi_{A}(\ze),  \\
\Phi_{B}(\sg) & = & \hat{\Phi}_{B}(\sg)+\phi_{B}(\sg),   \\
\Phi_{1}(\ze,\sg) & = & \sqrt{10c}(\hat{\Phi}_{A}(\ze)+
   \hat{\Phi}_{B}(\sg)) +\hat{\Phi}_{1}(\ze,\sg)+\phi_{1}(\ze,\sg),
                                                                   \\
\Phi_{2}(\ze_{1},\sg_{1},\ze_{2},\sg_{2}) & = & -10c
                   (D_{\ze}(\ze_{1},\ze_{2})\hat{\Phi}_{A}(\ze)
                     +D_{\sg}(\sg_{1},\sg_{2})\hat{\Phi}_{B}(\sg))      \\
  &  & -\sqrt{10c}[D_{\ze}(\ze_{1},\ze_{2})
          (\hat{\Phi}_{1}(\ze,\sg_{1})+\hat{\Phi}_{1}(\ze,\sg_{2}))\nn \\
  &  & \mbox{  }+D_{\sg}(\sg_{1},\sg_{2}) (\hat{\Phi}_{1}(\ze_{1},\sg)+
           \hat{\Phi}_{1}(\ze_{2},\sg))] \nn \\
  &  & +\hat{\Phi}_{2}(\ze_{1},\sg_{1},\ze_{2},\sg_{2})+ 
           \phi_{2}(\ze_{1},\sg_{1},\ze_{2},\sg_{2}),  
\eeas
\beq
      \cdots,                \label{5-8-2}
\eeq
where
\beas
  \phi_{A}(\ze) & = & -\frac{c}{3g}+\frac{2}{3}(\ze-g\ze^{2}),  \\
  \phi_{B}(\sg) & = & -\frac{c}{3g}+\frac{2}{3}(\sg-g\sg^{2}),  \\
\phi_{1}(\ze,\sg) & = & c(1+2s-\sqrt{10c}(\ze+\sg)), \\
\phi_{2}(\ze_{1},\sg_{1},\ze_{2},\sg_{2}) & = & 10c^{2},  
\eeas
\beq
                 \cdots,                         
\eeq
and $c$ takes its critical value: $c=\frac{-1+2\sqrt{7}}{27}$,
and $s=2+\sqrt{7}$. 
Then the connected correlators are transformed as 
\beas
G^{(1)}_{I} & = & \sum_{I}\cM_{IJ}\hat{G}^{(1)}_{J}+\phi_{I},  \\
G^{(k)}_{I_{1},\cdots,I_{k}} & = & \sum_{J_{1},\cdots,J_{k}}
  \cM_{I_{1}J_{1}}\cdots\cM_{I_{k}J_{k}}
\hat{G}^{(k)}_{J_{1},\cdots,J_{k}} ~~~
    (k\geq 2),
\eeas
where $\hat{G}^{(k)}_{I_{1},\cdots,I_{k}}$
stands for the universal part of 
$G^{(k)}_{I_{1},\cdots,I_{k}}$.

   Thus, by introducing the transformed source 
\beq
J_{I}=\sum_{K}\hat{J}_{K}(\cM^{-1})_{KI},              \label{5-10-1}
\eeq
the generating functional $\hat{Z}[\hat{J}]$ in the continuum theory  
is obtained by  the rescaling $
Z[J]= \e^{J\cdot\phi}\hat{Z}[\hat{J}]$,  and takes the form 
\beq
\hat{Z}[\hat{J}]=
\exp\left[\hat{J}\cdot \hat{G}^{(1)}+\frac{1}{2!}\hat{J}\cdot 
  (\hat{J}\cdot \hat{G}^{(2)})+
  \frac{1}{3!}\hat{J}\cdot(\hat{J}\cdot (\hat{J}\cdot \hat{G}^{(3)}))
  +\cdots\right]. 
\eeq
The Hamiltonian acting on $\hat{Z}[\hat{J}]$ now becomes 
\bea
0 & = & \cH \hat{Z}[\hat{J}],   \label{5-10-2} \\
\cH & = & -(\Jh\cM^{-1})\cdot \left(K(\cM\frac{\dl}{\dl \Jh}+\phi)\right)
                   -(\Jh\cM^{-1})\cdot T            \nn \\
 &  & -(\Jh\cM^{-1})\cdot \left( (\cM\frac{\dl}{\dl \Jh}+\phi)\vee
         (\cM\frac{\dl}{\dl \Jh}+\phi)\right) \nn \\
 &  & -\frac{1}{N^{2}}(\Jh\cM^{-1})\cdot \left( (\Jh\cM^{-1})\cdot
         \left(\wedge(\cM\frac{\dl}{\dl \Jh}+\phi)\right)\right).
                  \label{5-10-3}
\eea

    Next, we arrange eq. (\ref{5-10-3}) to a simpler form in which 
the mixing matrix disappears.  
We claim the validity of the following 
equations:
\bea
\lefteqn{-(\Jh\cM^{-1})\cdot \left(K(\cM\derJh+\phi)\right)
                  -(\Jh\cM^{-1})\cdot T } \nn \\
     &  &   -(\Jh\cM^{-1})\cdot\left((\cM\derJh+\phi)\vee
      (\cM\derJh+\phi)\right) \nn \\
 & = & -\Jh\cdot \left(F\derJh\right)
-(\Jh\cM^{-1})\cdot \left(\left(\cM\derJh\right)
              \vee\left(\cM\derJh\right)\right),           \label{5-10-4} 
\eea
\bea
(\Jh\cM^{-1})\cdot \left(\left(\cM\derJh\right)
\vee\left(\cM\derJh\right)
\right) & = & \Jh\cdot\left(\derJh
\vee\derJh\right),   \label{5-11-1} \\
(\Jh\cM^{-1})\cdot\left((\Jh\cM^{-1})\cdot\left(
\wedge\cM\derJh\right) \right)& = &
                      \Jh\cdot\left(\Jh\cdot\left(
\wedge\derJh\right)\right),         \label{5-11-2}\\
(\Jh\cM^{-1})\cdot( (\Jh\cM^{-1})\cdot(
\wedge\phi ))& = & 0,     \label{5-11-3}
\eea
where $F$ is a part of the kinetic operator $K$,
representing only the spin flip 
process.   These equations make possible to rewrite the 
Hamiltonian as  
\beq
\cH=
-\Jh\cdot \left(F\derJh\right)-\Jh\cdot\left
(\derJh\vee\derJh\right)
 -\frac{1}{N^{2}}
\Jh\cdot\left(\Jh\cdot\left(\wedge\derJh\right)\right).  
\label{5-11-4}
\eeq

\paragraph{Justification  of  the equations
(\ref{5-10-4})$\sim$(\ref{5-11-3}) }:
 
We now present the arguments for establishing the above equations.    
Firstly, we consider the spin flip process in the continuum theory. 
For the universal parts of the disk 
amplitudes, the loop containing 
a microscopic domain which consists only of a single flipped spin  
is  obtained from the loop containing only macroscopic 
domains by the following rule:
\bea
-\prt_{\ze}\hat{W}_{1}(\ze)
   & = & -s^{-1}\hat{\oint}\frac{d\sg}{2\pi i}
\sg\prt_{\ze}\hat{W}^{(2)}(\ze,\sg),
                           \label{5-12-1} \\
\hat{W}_{1}(\ze_{1};\ze_{2},\sg_{1}) & = & 
s^{-1}\hat{\oint}\frac{d\sg}{2\pi i}\sg\hat{W}^{(4)}
        (\ze_{1},\sg,\ze_{2},\sg_{1}),            \label{5-12-2} \\
       &    &   \cdots ,   \nn
\eea
where the domain $\sg$ has been shrunk into the microscopic domain 
by integration. 
Here we borrow the notations of ref. \cite{St} for the disk
amplitudes. For their definitions, see the Appendix C.
The integral symbol $\hat{\oint}\frac{d\sg}{2\pi i} \sg$ 
is used in the sense of the integral of 
the variable $x$ in the continuum theory ($\sg=P_{*}(1+ax)$) 
\beq
\hat{\oint}\frac{d\sg}{2\pi i}\sg=P_{*}^{2}a\int_{C}\frac{dx}{2\pi i}, 
        \label{5-12-3}
\eeq
where the contour $C$ encircles 
around the negative real axis 
and the singularities of the left half plane. 
$P_{*}$ stands for the critical value 
$P_{*}=s(10c)^{-1/2}$, and $a$ is a lattice spacing. 
Note that there is a sort of finite {\it renormalization} 
represented by the factor $s^{-1}$. 
The derivation of the above formulas is given in the Appendix D. 

These relations reflect the property
that the spin flip process occurs  locally 
with respect to domains; 
in  (\ref{5-12-1}), (\ref{5-12-2}) only the $\sg$ domain 
is concerned and the other domains do not change at all. 
This implies that the relation such as (\ref{5-12-1}) and (\ref{5-12-2}) 
should hold for any amplitudes with arbitrary number of handles 
with generic spin 
configurations.  The complete general  proof of this 
important property would be, however, technically 
formidable, since it would 
require explicit identification of the universal part for 
general loops with arbitrary spin configurations.   
We can now rewrite  the spin flip process in 
$ \left(K\left(\cM\frac{\dl}{\dl \Jh}+\phi\right)\right)_{A}(\ze)$ as, 

\bea
\lefteqn{-\prt_{\ze}\ointsg\sg[\left(\cM\frac{\dl}{\dl \Jh}\right)_{1}
 (\ze,\sg)+\phi_{1}(\ze,\sg)]}  \nn \\
 & = &  
 -s^{-1}\hat{\oint}\frac{d\sg}{2\pi i}
\sg\prt_{\ze}\frac{\dl}{\dl\Jh_{1}(\ze,\sg)}-
\prt_{\ze}\left(\left(\frac{2}{3g}-\frac{1}{3c}(\ze-g\ze^{2})\right)
 \frac{\dl}{\dl\Jh_{A}(\ze)}\right)  \nn \\
 &  &  -\frac{1}{c}\prt_{\ze}[g\ze+\frac{c}{9g}(\ze-g\ze^{2})+
       \frac{2}{9}(\ze-g\ze^{2})^{2}],  \label{5-14-1}
\eea
where the first term is the universal part, and the others are 
the non-universal parts which are obtained 
by replacing $\hat{W}(\ze)$ with 
$\frac{\dl}{\dl \hat{J}_{A}(\ze)}$ in (\ref{A-10-3}).

Similarly, we can derive the expressions for  
$$
-\prt_{\sg}\ointze\ze[\left(\cM\frac{\dl}{\dl \Jh}\right)_{1}(\ze,\sg)
      +\phi_{1}(\ze,\sg)] ~~~ \mbox{in}~~
    \left(K\left(\cM\frac{\dl}{\dl \Jh}+\phi\right)\right)_{B}(\sg),
$$
and 
\beas
\ointsg\sg[\left(\cM\frac{\dl}{\dl\Jh}\right)_{2}
 (\ze_{1},\sg,\ze_{1},\sg_{1}) +\phi_{2}],      &    &              \\
\ointze\ze[\left(\cM\frac{\dl}{\dl \Jh}\right)_{2}
(\ze_{1},\sg_{1},\ze,\sg_{1})+\phi_{2}] & \mbox{in}&    
    \left(K\left(\cM\frac{\dl}{\dl \Jh}+\phi\right)\right)_{1}
                                 (\ze_{1},\sg_{1}). 
\eeas 

   Using these results,  we arrive at 
eq. (\ref{5-10-4}) with 
\bea
\Jh\cdot \left(F\derJh\right)
 & = & 
\Lint\Jh_{A}(\ze)cs^{-1}(-\prt_{\ze})\hat{\oint}
\frac{d\sg}{2\pi i}\sg\frac{\dl}
         {\dl\Jh_{1}(\ze,\sg)} \nn \\
 &  & 
+\Lintsg\Jh_{B}(\sg)cs^{-1}(-\prt_{\sg})\hat{\oint}
\frac{d\ze}{2\pi i}\ze\frac{\dl}
         {\dl\Jh_{1}(\ze,\sg)} \nn \\
 &  & 
+\int_L\frac{d\ze_{1}}{2\pi i}\frac{d\sg_{1}}{2\pi i}
      \Jh_{1}(\ze_{1},\sg_{1})cs^{-1}\left[\hat{\oint}
\frac{d\sg}{2\pi i}\sg\frac{\dl}{\dl\Jh_{2}
         (\ze_{1},\sg,\ze_{1},\sg_{1})}\right.     \nn \\
 &  &  
{}~~~~~~~~~~~~~~~+\left. \hat{\oint}
\frac{d\ze}{2\pi i}\ze\frac{\dl}{\dl\Jh_{2}
         (\ze_{1},\sg_{1},\ze,\sg_{1})}\right] \nn \\
 &  & +\cdots,  \label{5-15-1}
\eea
where the ellipsis stands for the terms containing 
the higher components $\Jh_{n}$ ($n\geq 2$). 

  We here note that the tadpole term is cancelled 
with a contribution of the same form from the kinetic term. 
This can be regarded as a consequence of
our definition of the disk universal part  of
the single-spin flip amplitude $\hat{W}_{1}(\ze)$: Roughly,
the kinetic term contains the product of the
form $J_A \times {\delta \over \delta J_A^{{\rm one-spin \, flip}}}
+(A\rightarrow B)$, 
and hence there is always a freedom of absorbing the
tadpole terms, which are polynomials times the sources $J_A\, (J_B)$,
by appropriately defining the universal parts of the 
spin-flip disk amplitudes and making 
the shifts for ${\delta \over \delta
J_A^{{\rm one-spin \, flip}}}
({\delta \over \delta
J_B^{{\rm one-spin \, flip}}})$.
In the case of one-matrix model, the spin-flip process is
absent 
 so that the tadpole term is directly responsible for determining the
disk amplitude.

Although we do not elaborate further on determining
the explicit continuum 
limits for higher components, it is natural, because of the 
local nature of the spin-flipping process, to suppose that 
the above expression already indicates the generic structure of 
the kinetic term, namely the flipping of a single spin with 
general spin configurations, the absence of the tadpole term and 
the kinetic term without spin flipping.

   Secondly, we consider 
the splitting and merging processes, eqs.  (\ref{5-11-1}) and 
(\ref{5-11-2}). 
In the Appendix E, it is explicitly checked that
\beq
\left(\left(\cM\derJh\right)\vee\left(\cM\derJh\right)\right)_{I}=
\left(\cM\left(\derJh\vee\derJh\right)\right)_{I}  
\label{splitmix}
\eeq
for $I=A,B,1,2$, and that 
\beq
\left(\wedge\left(\cM\derJh\right)\right)_{I,J}=
\sum_{K,L}\cM_{IK}\cM_{JL}\left(\wedge\derJh\right)_{K,L}
\label{mergemix}
\eeq
for $(I,J)=(A,A), (B,B), (A,1), (B,1), (A,2), (B,2), (1,1).$ 

 As seen from eqs. (\ref{5-8-1}) and (\ref{5-8-2}),
$\cM$ mixes the operators with the configurations 
which are smaller parts of the configuration $I$, 
 when it acts on  
the loop operator $\frac{\dl}{\dl\Jh_{I}}$. 
Eq. (\ref{splitmix}) (eq. (\ref{mergemix})) means that the mixing  
is commutative  with the splitting (merging) 
processes of loops.  
This result can again be regarded as a consequence  
of locality of the splitting and 
merging processes.  It is then 
reasonable to assume that the splitting (merging ) commutes 
with mixing matrix for arbitrary 
spin configurations. 
This assertion is nothing but the claims
(\ref{5-11-1}), (\ref{5-11-2}). 
Proving this for completely general case is
prohibitively difficult in our present technical 
machinery for treating the double scaling limit. 
We have to be satisfied with the explicit confirmation 
of this property for several nontrivial cases
as given in the Appendix E.

   Thirdly,  in regard of eq. (\ref{5-11-3}),
by direct calculation we can check that 
\beq
(\wedge\phi)_{I,J}=0
\label{wedgephi}
\eeq
for $(I,J)=(A,A), (B,B), (A,1), (B,1), (A,2), (B,2), (1,1).$ 
In this case, we can give a general proof of this equation as 
follows. 
First we show that $\phi_k$ must be polynomial for general $k$. 
Suppose that this is valid up to some $k-1$. Then, 
from  Staudacher's recursion equations, as described
by (\ref{W(2k)Staudacher}), which  
relates $W^{(2k)}$ with amplitudes with lower values 
of $k$, we can see that the part of 
the numerator for $W^{(2k)}$ 
 consisting  only of $\phi$ is  
a polynomial, because in general the
combinatorial derivative of a polynomial 
is also a polynomial. The denominator, on the other hand,
 behaves in the scaling limit as 
$$
P_1-gP_1^2-cQ_1-W(P_1)=-a\frac{c^{1/2}s}{\sqrt{10}}(y_1+x_1)
+O(a^{4/3}). 
$$
Thus, by using the scaled variables
 $P_i=\ze_*(1+ay_i), Q_i=\ze_*(1+ax_i),$ 
the form of $\phi_k$ can be written as 
$$
\phi_k=\frac{\mbox{Polynomial of }(y_1, x_1,\cdots, y_k, x_k)}
                 {y_1+x_1}.  
$$
However, from the cyclic symmetry of  ($P_i, Q_i$) in 
$W^{(2k)}(P_1, Q_1, \cdots, P_k, Q_k)$,
the denominator $y_1+x_1$ must be 
cancelled with the numerator, and thus $\phi_k$ should have the form: 
$$
\phi_k=\mbox{Polynomial of } (y_1, x_1, \cdots, y_k, x_k), 
$$
where the polynomial has the same symmetry as $W^{(2k)}$. 
Thus, by induction, the $\phi_k$ is a polynomial for general $k$. 

     Now from the scaling of the universal part $\hat{W}^{(2k)}$ 
$$
\hat{W}^{(2k)}(P_1, Q_1, \cdots, P_k, Q_k) =
a^{\frac{7}{3}-\frac{2}{3}k} w^{(2k)}(y_1,x_1, \cdots, y_k, x_k), 
$$
which is presented in the Appendix C,
we expect that the relevant part of $\phi_k$ 
takes the form 
\beq
\phi_k=\left\{\begin{array}{ll}\mbox{const.} &  k=3 \\
                                                               
0 & k\geq 4 .     \end{array}  \right. 
\label{phik}
\eeq
Since every component of $\wedge\phi$ contains the derivative or 
the combinatorial derivative, eq. (\ref{phik}) leads to 
\beq
(\wedge\phi)_{I,J}=0  ~~~~\mbox{for the general components},
\eeq
as is claimed.

\paragraph{Continuum Limit} : 

After these preparations, we are now 
ready to take the continuum limit of the stochastic 
Hamiltonian (\ref{5-11-4}). 
{}From the scaling behaviors of disk amplitudes
presented in the Appendix C, 
the scaling of various variables are fixed as follows:
\bea
g & = & g_{*}(1-a^{2}\frac{s^{2}}{20}T),~~~\ze=P_{*}(1+ay),
          ~~\sg=P_{*}(1+ax),\nn \\
\frac{1}{N} & = & a^{7/3}\gst,\nn \\
\frac{\dl}{\dl\Jh_{A}(\ze)} & = &
a^{4/3}P_{*}^{-1}\frac{\dl}{\dl\Jt_{A}(y)}, 
{}~~~\Jh_{A}(\ze)=a^{-7/3}\Jt_{A}(y),\nn \\
\frac{\dl}{\dl\Jh_{B}(\sg)} & = &
a^{4/3}P_{*}^{-1}\frac{\dl}{\dl\Jt_{B}(x)}, 
{}~~~\Jh_{B}(\sg)=a^{-7/3}\Jt_{B}(x),\nn \\
\frac{\dl}{\dl\Jh_{1}(\ze,\sg)} & = &
a^{5/3}P_{*}^{-2}\frac{\dl}
 {\dl\Jt_{1}(y,x)},~~~\Jh_{1}(\ze,\sg)=
a^{-11/3}\Jt_{1}(y,x),\nn \\
\frac{\dl}{\dl\Jh_{2}(\ze_{1},\sg_{1},\ze_{2},\sg_{2})}
& = & a^{1}P_{*}^{-4}
 \frac{\dl}{\dl\Jt_{2}(y_{1},x_{1},y_{2},x_{2})}, \nn \\
 &  & \Jh_{2}(\ze_{1},\sg_{1},\ze_{2},\sg_{2})=a^{-5}\Jt_{2}
       (y_{1},x_{1},y_{2},x_{2}), \nn \\
 &  &\cdots,          \label{5-18-1}
\eea
where the critical values are \cite{BK2,GN}
$$
g_{*}=\sqrt{10c^{3}},~~~P_{*}=s(10c)^{-1/2}.
$$
Indeed in the limit $a\limit 0$ all the universal 
contributions in the Hamiltonian start with $O(a^{1/3})$, 
as it should be for the correct continuum limit.
After finite rescaling as 
$$
\Jt_{I}\limit P_{*}^{-2}\Jt_{I},
{}~~~\frac{\dl}{\dl\Jt_{I}}\limit P_{*}^{2}\frac{\dl}{\dl\Jt_{I}},
{}~~~\gst^{2}\limit P_{*}^{4}\gst^{2},
$$
and absorbing the overall factor $a^{1/3}$ into the fictitious time, 
we obtain the continuum Hamiltonian in the form 
\beq
\cH=-\Jt\cdot\left(F\derJt\right)-\Jt\cdot\left(\derJt\vee\derJt\right)-
   \gst^{2}\Jt\cdot\left(\Jt\cdot\left(\wedge\derJt\right)\right),      
  \label{5-19-1}
\eeq
where the inner product is defined by 
\bea
\lefteqn{f\cdot g=\inty f_{A}(y)g_{A}(y)+\intx f_{B}(x)g_{B}(x)} \nn \\
 &  & +\sum_{n=1}^{\infty}\intynxn f_{n}(y_{1},x_{1},\cdots,y_{n},x_{n})
            g_{n}(y_{1},x_{1},\cdots,y_{n},x_{n}).
\eea
As already emphasized above, 
only the spin flip process survives in the kinetic term: 
\beas
\left(F\derJt\right)_{A}(y) & = & cs^{-1}(-\prt_{y})\int_{C}\frac{dx}{2\pi i}
       \frac{\dl}{\dl\Jt_{1}(y,x)},  \\
\left(F\derJt\right)_{B}(x) & = & cs^{-1}(-\prt_{x})\int_{C}\frac{dy}{2\pi i}
       \frac{\dl}{\dl\Jt_{1}(y,x)}, \\
\left(F\derJt\right)_{1}(y_{1},x_{1}) & = & cs^{-1}\left[
 \int_{C}\frac{dx}{2\pi i}\frac{\dl}{\dl\Jt_{2}(y_{1},x,y_{1},x_{1})}+
 \int_{C}\frac{dy}{2\pi i}\frac{\dl}{\dl\Jt_{2}(y_{1},x_{1},y,x_{1})}\right],
\eeas
\bea
\lefteqn{\left(F\derJt\right)_{2}(y_{1},x_{1},y_{2},x_{2})} \nn \\
& = & cs^{-1}\left[\int_{C}\frac{dx}{2\pi i}\left(
\frac{\dl}{\dl\Jt_{3}(y_{1},x,y_{1},x_{1},y_{2},x_{2})}+
\frac{\dl}{\dl\Jt_{3}(y_{1},x_{1},y_{2},x,y_{2},x_{2})}\right)\right. \nn \\
&  & +\int_{C}\frac{dy}{2\pi i}\left.\left(
\frac{\dl}{\dl\Jt_{3}(y_{1},x_{1},y,x_{1},y_{2},x_{2})}+
\frac{\dl}{\dl\Jt_{3}(y_{1},x_{1},y_{2},x_{2},y,x_{2})}\right)\right], \nn \\
&  &  \cdots.   \label{5-19-2}
\eea
The forms of the splitting $\derJt\vee\derJt$ and the merging 
$\wedge\derJt$ are the same as in the lattice theory, 
because of the commutativity of the mixing matrix with the processes: 
 The first few components of the splitting  term are 
\bea
\left(\derJt\vee\derJt\right)_{A}(y) & = & -\prt_{y}\frac{\dl^{2}}
   {\dl \Jt_{A}(y)^{2}},  \nn \\
\left(\derJt\vee\derJt\right)_{B}(x) & = & -\prt_{x}\frac{\dl^{2}}
   {\dl \Jt_{B}(x)^{2}},  \nn \\
\left(\derJt\vee\derJt\right)_{1}(y,x) & = & -2\left(
    \frac{\dl}{\dl\Jt_{A}(y)}\prt_{y}+\frac{\dl}{\dl\Jt_{B}(x)}\prt_{x}\right)
    \frac{\dl}{\dl\Jt_{1}(y,x)},  \nn \\  
\left(\derJt\vee\derJt\right)_{2}(y_{1},x_{1},y_{2},x_{2})
 & = & -2\sum_{j=1}^{2}\left(\frac{\dl}{\dl\Jt_{A}(y_{j})}\prt_{y_{j}}+
          \frac{\dl}{\dl\Jt_{B}(x_{j})}\prt_{x_{j}}\right)
          \frac{\dl}{\dl\Jt_{2}(y_{1},x_{1},y_{2},x_{2})}  \nn \\
 &  & +2\left(D_{y}(y_{1},y_{2})\frac{\dl}{\dl\Jt_{1}(y,x_{1})}\right)
        \left(D_{y}(y_{1},y_{2})\frac{\dl}{\dl\Jt_{1}(y,x_{2})}\right)  \nn \\
 &  & +2\left(D_{x}(x_{1},x_{2})\frac{\dl}{\dl\Jt_{1}(y_{1},x)}\right)
        \left(D_{x}(x_{1},x_{2})\frac{\dl}{\dl\Jt_{1}(y_{2},x)}\right), \nn \\
 &  &  \cdots.                           \label{5-20-1}
\eea
Similarly, the merging processes  are given as  
\beas
\mrJt_{A,A}(y; y') & = &
 -\prt_{y}\prt_{y'}D_z(y,y')\frac{\dl}{\dl \Jt_{A}(z)},  \\
\mrJt_{B,B}(x; x') & = &
 -\prt_{x}\prt_{x'}D_s(x,x')\frac{\dl}{\dl \Jt_{B}(s)},  \\
\mrJ_{A,B}(y; x) & = & 0,
\eeas
\beas
\lefteqn{\mrJt_{n,A}(y_1,x_1,\cdots,y_n, x_n; y') =
\mrJt_{A,n}(y'; y_1,x_1,\cdots, y_n,x_n)} \\
 & = & -\prt_{y'}\sum_{j=1}^{n}\prt_{y_j}D_z(y_j, y')
             \frac{\dl}{\dl \Jt_n(y_1,x_1,\cdots, z,x_j, \cdots, y_n,x_n)}, 
\eeas
\beas
\lefteqn{\mrJt_{n,B}(y_1,x_1,\cdots,y_n, x_n; x') =
\mrJt_{B,n}(x'; y_1,x_1,\cdots, y_n,x_n)} \\
 & = & -\prt_{x'}\sum_{j=1}^{n}\prt_{x_j}D_s(x_j, x')
             \frac{\dl}{\dl \Jt_n(y_1,x_1,\cdots, y_j,s, \cdots, y_n,x_n)}, 
\eeas
\beas
\lefteqn{\mrJt_{n,1}(y_1,x_1,\cdots,y_n,x_n ;y'_1, x'_1) =
\mrJt_{1,n}(y'_1,x'_1; y_1,x_1,\cdots, y_n,x_n)} \\ 
 & = & \sum_{j=1}^n   D_z(y_j,y'_1)D_w(y_j,y'_1)
\frac{\dl}{\dl \Jt_{n+1}(y_1,x_1,\cdots,y_{j-1},x_{j-1},
                                       z, x'_1,w,x_j, \cdots,y_n,x_n)} \\
 &+ & \sum_{j=1}^n D_s(x_j,x'_1)D_t(x_j,x'_1)
\frac{\dl}{\dl \Jt_{n+1}(y_1,x_1,\cdots, y_j,s,y'_1,t,  
                                     y_{j+1},x_{j+1},\cdots,y_n,x_n)}, \\
\eeas
\beq
  \cdots. ~~~~~~~~~~~~~~~~~~~~~~~~~~~~~~(n=1,2,\cdots)
        \label{5-20-2}
\eeq

We remark that the final Hamiltonian has no tadpole term 
and no dependence on the cosmological constant $T$. 
Thus in the two-matrix model case,
 the cosmological constant should be 
regarded as an integration constant. 
This   
 contrasts with what one would naively expect from the 
result for the one-matrix cases. 
In  section 7, to check consistency of the above results, 
we will discuss how to obtain a closed set of Schwinger-Dyson 
equations, leading to the $W_3$ constraints, from this Hamiltonian. 

\vspace{1cm}

\section{Closure of the Schwinger-Dyson Algebra}
\setcounter{equation}{0}

   Now we proceed to discuss the algebra of
 the Schwinger-Dyson operators associated 
with the Hamiltonian (\ref{5-19-1}).  Comparing with the case of 
one-matrix models, this requires a much more intricate analysis, 
since there are  
 an infinite number of components for  the $T$ operators: 
\bea
\lefteqn{\cH = \inty\Jt_A(y)(-\prt_yT_A^A(y))
+\intx\Jt_B(x)(-\prt_xT_B^B(x))}\nn\\
 & + & \sum_{n=1}^{\infty}\intynxn \Jt_n(y_1,x_1,\cdots,y_n,x_n)
             T_n(y_1,x_1,\cdots, y_n,x_n), 
\label{Hamiltonian6}
\eea
\bea
-\prt_yT_A^A(y) & = & -\left(F\derJt\right)_A(y)
-\left(\derJt\vee\derJt\right)_A(y)\nn\\
& & -\gst^2\left(\Jt\cdot\mrJt\right)_A(y),  \\
-\prt_xT_B^B(x) & = & -\left(F\derJt\right)_B(x)
-\left(\derJt\vee\derJt\right)_B(x)\nn\\
& &-\gst^2\left(\Jt\cdot\mrJt\right)_B(x), 
\eea
\bea
\lefteqn{T_n(y_1,x_1,\cdots,y_n,x_n) =
-\left(F\derJt\right)_n(y_1,x_1,\cdots,y_n,x_n)}
\nn \\
 & -&\left(\derJt\vee\derJt\right)_n(y_1,x_1,\cdots,y_n,x_n) \nn \\
 & - & \gst^2\left(\Jt\cdot\mrJt\right)_n(y_1,x_1,\cdots,y_n,x_n).
  ~~~(n=1,2,\cdots)
\eea
The  $T_n$ operators appearing 
in the Hamiltonian can be regarded as the
  symmetrized versions of the general Schwinger-Dyson 
operators $T_n^A, T_n^B$: 
\bea
T_1(y_1,x_1) & = &
T_1^A(y_1;y_1,x_1)+T_1^B(y_1,x_1;x_1), \nn\\
T_2(y_1,x_1,y_2,x_2) & = & T_2^A(y_1;y_1,x_1,y_2,x_2)+
T_2^A(y_1,x_1,y_2;y_2,x_2) \nn\\
 &  &  +T_2^B(y_1,x_1;x_1,y_2,x_2)
+T_2^B(y_1,x_1,y_2,x_2;x_2), \nn\\
  &  &  \cdots ,        
\eea
where the semicolon in the argument in the right hand side 
denotes the  point where  the deformation of a loop occurs  
in constructing the Schwinger-Dyson equation. For example, 
$T_1^A(y_1;y_1,x_1) \, (T_1^B(y_1,x_1;x_1))$ represents 
a deformation of the loop 
with one pair of domains by attaching on it any loops which have 
at least one  
$A(B)$-domain.   
The explicit forms of the general Schwinger-Dyson operators 
$T_n^A, T_n^B$ are 
\bea
\lefteqn{T_n^A(y_{n+1};y_1,x_1,\cdots,y_n,x_n)} \nn \\
 & = & \left(\frac{\dl}{\dl\Jt_A(y_1)}
+\frac{\dl}{\dl\Jt_A(y_{n+1})}\right)
              D_z(y_1,y_{n+1})
\frac{\dl}{\dl\Jt_n(z,x_1,\cdots,y_n,x_n)} \nn \\
 & - & \sum_{j=1}^{n-1} D_z(y_1,y_{j+1})
\frac{\dl}{\dl\Jt_{j}(z,x_1,\cdots,y_{j},x_{j})}
            D_w(y_{j+1},y_{n+1})
\frac{\dl}{\dl\Jt_{n-j}(w,x_{j+1},\cdots,y_n,x_n)} \nn \\
& -&cs^{-1}\int_C\frac{dx'}{2\pi i}
\frac{\dl}{\dl\Jt_{n+1}(y_{n+1},x',y_1,x_1,\cdots,y_n,x_n)}
              \nn \\
& +& \gst^2\int_{-i\infty}^{i\infty}
\frac{dy'}{2\pi i}\Jt_A(y')\prt_{y'}
        D_z(y_{n+1},y')D_w(z,y_1)
\frac{\dl}{\dl\Jt_n(w,x_1,\cdots,y_n,x_n)}\nn \\
 & -& \gst^2\sum_{m=1}^{\infty}\intympxmp
\Jt_m(y'_1,x'_1,\cdots,y'_m,x'_m)
        \sum_{j=1}^m D_z(y'_j,y_1)D_w(y'_j,y_{n+1})\nn \\
& &  \times\frac{\dl}{\dl\Jt_{n+m}
(y'_1,x'_1,\cdots,y'_{j-1},x'_{j-1},z,x_1,\cdots,y_n,x_n,
             w,x'_j, y'_{j+1},x'_{j+1},\cdots,y'_m,x'_m)}, \nn\\
& & 
\eea
\bea
\lefteqn{T_n^B(y_1,x_1,\cdots,y_n,x_n;x_{n+1})} \nn \\
 & = & \left(\frac{\dl}{\dl\Jt_B(x_n)}
+\frac{\dl}{\dl\Jt_B(x_{n+1})}\right)
              D_z(x_n,y_{n+1})
\frac{\dl}{\dl\Jt_n(y_1,x_1,\cdots,y_n,z)} \nn \\
 & - & \sum_{j=1}^{n -1}D_z(x_{n+1},x_{j})
\frac{\dl}{\dl\Jt_{j}(y_1,x_1,\cdots,y_{j},z)}
            D_w(x_{j},x_n)
\frac{\dl}{\dl\Jt_{n-j}(y_{j+1},x_{j+1},\cdots,y_n,w)} \nn \\
& -&cs^{-1}\int_C\frac{dy'}{2\pi i}
\frac{\dl}{\dl\Jt_{n+1}(y_1,x_1,\cdots,y_n,x_n,y',x_{n+1})}
              \nn \\
& +& \gst^2\int_{-i\infty}^{i\infty}
\frac{dx'}{2\pi i}\Jt_B(x')\prt_{x'}
        D_z(x_{n},x')D_w(z,x_{n+1})
\frac{\dl}{\dl\Jt_n(y_1,x_1,\cdots,y_n,w)}\nn \\
 & -& \gst^2\sum_{m=1}^{\infty}\intympxmp
\Jt_m(y'_1,x'_1,\cdots,y'_m,x'_m)
        \sum_{j=1}^m D_z(x'_j,x_{n+1})D_w(x'_j,x_{n})\nn \\
& &  \times\frac{\dl}{\dl\Jt_{n+m}
(y'_1,x'_1,\cdots,y'_{j-1},x'_{j-1},y'_j,z,y_1,x_1,\cdots,y_n,w,
              y'_{j+1},x'_{j+1},\cdots,y'_m,x'_m)}. \nn \\
& & 
\eea
It should be noted that the 
Hamiltonian contains the particular set of the above 
operators $T^A_n, T^B_n$, with partial 
identification of the variables as $y_{n+1}=y_1$, $x_n=x_{n+1}$, 
respectively. 

We also remark that, as is expected from our 
construction of the Hamiltonian, the general 
Schwinger-Dyson operators introduced above 
are the continuum versions of the Schwinger-Dyson operators 
of the original two-matrix model, appearing in  
\beq
0=\intAB ~T^{(\mbox{\scriptsize mat.})}
 (\ze, \cdots)\e^{-S+J\cdot\Phi}.
\eeq
The explicit forms of the matrix model operators are  
\beas
\TmatrixA_A(\ze) & = & -\frac{1}{N^2}
\sum_{\alp=1}^{N^2}\vec{\frac{\prt}{\prt A_{\alp}}}
\tr\left(\frac{1}{\ze-A}t^{\alp}\right), \\
\TmatrixB_B(\sg)&  =&  -\frac{1}{N^2}
\sum_{\alp=1}^{N^2}\vec{\frac{\prt}{\prt B_{\alp}}}
\tr\left(\frac{1}{\sg-B}t^{\alp}\right), 
\eeas
\beas
\lefteqn{\TmatrixA_n
(\ze_{n+1};\ze_1,\sg_1,\cdots,\ze_n,\sg_n) } \\
& = &
-\frac{1}{N^2}\sum_{\alp=1}^{N^2}\vec{\frac{\prt}{\prt A_{\alp}}}
\tr\left(\frac{1}{\ze_{n+1}-A}t^{\alp}
\frac{1}{\ze_1-A}\frac{1}{\sg_1-B}\cdots
\frac{1}{\ze_n-A}\frac{1}{\sg_n-B}\right), 
\eeas
\bea
\lefteqn{\TmatrixB_n
(\ze_1,\sg_1,\cdots,\ze_n,\sg_n;\sg_{n+1}) }\nn \\
& = &
 -\frac{1}{N^2}\sum_{\alp=1}^{N^2}\vec{\frac{\prt}{\prt B_{\alp}}}
\tr\left(\frac{1}{\ze_1-A}\frac{1}{\sg_1-B}\cdots\frac{1}{\ze_n-A}
\frac{1}{\sg_n-B}t^{\alp}\frac{1}{\sg_{n+1}-B}\right). 
\label{MMoperators}
\eea

We now compute the algebra of the Schwinger-Dyson operators, 
as in section 4, by changing 
the contour $C$ to the imaginary axis. 
For the commutators of the operators 
with the same superscript, 
the result is found to coincide with 
that of  the 
corresponding matrix model operators,
after making the identification: 
$$
\gst\leftrightarrow\frac{1}{N}, ~~~y_i\leftrightarrow\ze_i,~~~
x_i\leftrightarrow\sg_i. 
$$
They are   
\beq
[\prt_{y}T_A^A(y), \prt_{y'}T_A^A(y')] = -\gst^{2} 
   \prt_{y}\prt_{y'}(\prt_{y}-\prt_{y'})\frac{1}{y-y'}
   (T_A^A(y)-T_A^A(y')), 
\label{algebraAAAA}
\eeq
\bea
\lefteqn{[\prt_yT_A^A(y), T_1^A(y';y',x')]=
-\gst^2\prt_y\left\{\frac{-2}{(y-y')^2}
   (T_1^A(y;y',x')+T_1^A(y';y,x'))\right.}\nn\\
 & + & \left.\frac{1}{(y-y')^2}(T_1^A(y;y,x')+3T_1^A(y';y',x'))
     +\frac{1}{y-y'}\prt_{y'}T_1^A(y';y',x')\right\}, 
\label{algebraAAA1}
\eea
$$ \cdots. $$
Similar result is obtained for the operators with the superscript $B$. 

  For commutators between the operators with different  
superscripts, 
 the situation is not so simple, 
except  for the trivial case  
\beq
[\prt_y T_A^A(y), \prt_x T_B^B(x)] = 0.
\label{algebraAABB}
\eeq 
For example, for the commutator $[\prt_y T_A^A(y), T_1^B(y',x';x')]$, 
the closed algebra  
\beq
[\prt_y T_A^A(y), T_1^B(y',x';x')]
=\gst^2\prt_y\prt_{y'}\frac{1}{y-y'}(T_1^B(y,x';x')-T_1^B(y',x';x')),
\label{algebraAAB1}
\eeq
which coincides with the result of the matrix model operators, 
can be obtained only when 
the formula 
\beq
\prt_y\int_C\frac{dx}{2\pi i}D_z(x,x')D_w(x,x')
\frac{\dl}{\dl\Jt_2(y,z,y',w)}=0
\label{formulaJ2}
\eeq
holds. 
To check this formula, we need to 
know a more detailed property of the functional derivatives 
$\frac{\dl}{\dl\Jt_n}, \,(n\geq 2)$. 

Here we try to justify this formula by using the 
short-domain expansion of the functional derivative operators. 
It will be useful also for the 
derivation of the $W_3$ constraints discussed in the next section. 
The short-domain expansions, in general,
depend on the choice of a 
  background. 
Since a specific background was already
assumed in taking the continuum limit 
of the two-matrix model, 
use of the short-domain expansion should be allowed here, 
as we have already done in the one-matrix cases  
in studying the closure of the Schwinger-Dyson 
algebra as a consistency 
check of the continuum results.  
The difference between the present case and the 
one-matrix cases is that 
the tadpole term is absorbed into the 
definition of the spin-flip amplitudes. 
Because of this peculiarity, the Hamiltonian 
(\ref{5-19-1}) itself apparently contains no terms which fix
 the dimensions of the loop operators. 
In order to fix the dimensions, however, it is sufficient to give the 
dimension of  the Hamiltonian, since it then
 determines the dimensions of the loop operators and 
the string coupling $\gst$ uniquely.  
Following our result  of a direct derivation of the  
Hamiltonian from the two-matrix model, 
we assume the dimension of $\cH$ to be $[\cH]=[y]^{1/3}$. 
 Also, it is noted that a new dimensional parameter  
can enter as the integration constant of the Schwinger-Dyson 
equation $\partial_y T_A^A(y)Z[J]=0$ or
$\partial_x T_B^B(x)Z[J]=0$.  
We can relate this parameter to the 
cosmological constant with dimension two ($[T]=[y]^2$), 
characterizing the $c=1/2$ background. 
See the next section, in particular, eq.  (\ref{7-8-1}), for more 
details. 

Let us now write down the short-domain expansions.  
 Firstly, for the operators with a single domain, $\frac{\dl}{\dl\Jt_A}$, 
$\frac{\dl}{\dl\Jt_B}$, 
by considering the dimensions of 
the loop operators and the cosmological constant, 
we can assume the expansions analogous to the one-matrix case 
(\ref{4-11-0})
\bea
\frac{\dl}{\dl\Jt_A(y)} & = &
a^A_{4/3}y^{4/3}+a^A_{-2/3}Ty^{-2/3}
+\sum_u A_uy^{-u-1} ~~~(y:\mbox{large}), 
\label{expansionJA}  \\
\frac{\dl}{\dl\Jt_B(x)} & = &
a^B_{4/3}x^{4/3}+a^B_{-2/3}Tx^{-2/3}
+\sum_u B_ux^{-u-1} ~~~(x:\mbox{large}), 
\label{expansionJB} 
\eea
where $a$'s stand for dimensionless
constants, and $u$ runs over positive integers 
divided by 3. The first two terms come from the
disk singular terms (thus, 
$a^A_{4/3}=a^B_{4/3}$, $a^A_{-2/3}=a^B_{-2/3}$),
and the remaining terms 
represent both the contributions of the large $y$
or $x$ expansion of the cylinder singular parts 
and those of the local operators. 

   Secondly, referring to the explicit form
of $w^{(2)}(y,x)$ in (\ref{w(2)}), 
we can assume that the operator
$\frac{\dl}{\dl\Jt_1(y,x)}$ has
 the following large $x$ 
expansion:
\bea
\frac{\dl}{\dl\Jt_1(y,x)} & = &
a^1_{5/3}x^{5/3}+a^1_{2/3}x^{2/3}y
+x^{-1/3}(a^1_{-1/3}y^2+a'^1_{-1/3}T)\nn\\
 & &+ x^{-4/3}(a^1_{-4/3}y^3+a'^1_{-4/3}Ty
+a''^1_{-4/3}B_{2/3})\nn\\
 & & +x^{-7/3}(a^1_{-7/3}y^4+a'^1_{-7/3}Ty^2
+a''^1_{-7/3}B_{2/3}y
          +a'''^1_{-7/3}B_{5/3}) \nn\\
 & & +a^1_{1/3}x^{1/3}\frac{\dl}{\dl\Jt_A(y)}
+a^1_{-2/3}x^{-2/3}y\frac{\dl}{\dl\Jt_A(y)}\nn\\
 & &+x^{-5/3}((a^1_{-5/3}y^2+a'^1_{-5/3}T)
\frac{\dl}{\dl\Jt_A(y)}+a''^1_{-5/3}B_1)\nn\\
 & & +x^{-7/3}a^{ ({\rm iv})1}_{-7/3}B_{1/3}
\frac{\dl}{\dl\Jt_A(y)}\nn\\
 & & +x^{-1}(a^1_{-1}\cO_1 (\frac{\dl}{\dl\Jt})(y)
+a'^1_{-1}B_{1/3})\nn\\
 & & +x^{-2}(a^1_{-2}y\cO_1 (\frac{\dl}{\dl\Jt})(y)
+a'^1_{-2}B_{1/3}y+a''^1_{-2}B_{4/3})\nn\\
 & & +x^{-7/3}a^{({\rm v})1}_{-7/3}
\cO_2 (\frac{\dl}{\dl\Jt})(y) +O(x^{-8/3}), 
\label{expansionJ1}
\eea
where $\cO_m (\frac{\dl}{\dl\Jt})(y)$ represents
the loop operator with a microscopic $m$-spin flipped 
domain being 
added to the domain $y$, and the remaining
notations are the same as before. 
Here, we assume that the nonpolynomial
powers of $y$ all come from the loop operators 
with the macroscopic $y$ domain,
and that the fractional powers of $T$ appear only through 
the $B$'s. For every disk amplitude whose explicit  form 
is derived in the Appendix C, we can confirm these properties. 
The above form (\ref{expansionJ1}) 
is then a consequence of the dimensional analysis.
Note that we do not have to worry
about the ordering of $B$'s and loop operators, because 
$[B_u,\frac{\dl}{\dl\Jt_I}]=0$ due to
$[\frac{\dl}{\dl\Jt_B}, \frac{\dl}{\dl\Jt_I}]=0$.

  Similarly, the large $x$ expansions of
$\frac{\dl}{\dl\Jt_2(y,x,y',x')}$, 
$\frac{\dl}{\dl\Jt_2(y,x,y',x)}$
are assumed to be, respectively, 
\bea
\frac{\dl}{\dl\Jt_2(y,x,y',x')} & = & a^2_1 x+
a^2_{2/3}x^{2/3}D_z(y,y')\frac{\dl}{\dl\Jt_A(z)}\nn\\
 & &+a^2_{1/3}x^{1/3}D_z(y,y')\frac{\dl}{\dl\Jt_1(z,x')}\nn\\
& & +a^2_0y+a'^2_0y'+a''^2_0x' +O(x^{-1/3}), 
\label{expansionJ2-1} \\
\frac{\dl}{\dl\Jt_2(y,x,y',x)} & = & \tilde{a}^2_1x+
\tilde{a}^2_{2/3}x^{2/3}D_z(y,y')\frac{\dl}{\dl\Jt_A(z)}\nn\\
& & +\tilde{a}^2_0 y+\tilde{a}'^2_0 y' +O(x^{-1/3}). 
\label{expansionJ2-2}
\eea

   We can fix the normalizations of
the spin flipped loop operators, referring to the disk results 
in the Appendix D. For example, $a^1_{-1}=s$. 
This implies the following relations 
\beas
\cO_1(\frac{\dl}{\dl\Jt})(y)  & =  &
s^{-1}\int_C\frac{dx}{2\pi i}\frac{\dl}{\dl\Jt_1(y,x)}
-s^{-1}a'^1_{-1}B_{1/3}, \\
\cO_2(\frac{\dl}{\dl\Jt})(y) & =&
s^{-1}\int_{C''}\frac{dx'}{2\pi i}
\int_{C'}\frac{dy'}{2\pi i}s^{-1}\int_C\frac{dx}{2\pi i}
\frac{\dl}{\dl\Jt_2(y,x,y',x')}
+c^{-1}s^{-1}\tilde{a}\frac{\dl}{\dl\Jt_A(y)}B_{1/3}, 
\eeas
\beq
\cdots , 
\label{O1O2spinflip}
\eeq
with $\tilde{a}$ being some constant,
which are a consequence of the above expansions. 
The first equation of (\ref{O1O2spinflip}) is the 
integrated version of eq. (\ref{5-12-1}).   
Roughly speaking, $\cO_m(\frac{\dl}{\dl\Jt})(y)$ is
obtained from $\frac{\dl}{\dl\Jt_m}$ 
by integrating the domains except $y$,
and the factor $s^{-1}$ is regarded as a sort of finite 
renormalization accompanied with flipping of  a single spin.

Now substituting the expansions
(\ref{expansionJ2-1}) and (\ref{expansionJ2-2}),
and using the formula such as (\ref{4-12-2}),  
we can directly check the validity of 
the formula (\ref{formulaJ2}), and thus justify the 
algebra (\ref{algebraAAB1}). 
 From these results, it is 
expected that the algebra of higher
operators also coincides with
 those of the matrix model 
operators before taking the double-scaling limit. 
In order to prove this for 
the general case, we need to obtain
more precise information 
with respect to the coefficients $a$'s
in the expansions of  generic loop operators. 
We 
would like to emphasize again that,
 since we have started from the matrix model
which is already an 
integral solution of the constraints, the closure of the 
Schwinger-Dyson algebra is guaranteed in our approach. 
The confirmation of the closure of the algebra 
is useful, however, as a consistency check of the 
continuum limit. Since the structure of the 
algebra of the Schwinger-Dyson operators 
is essentially determined by the splitting and 
merging processes of the loops which are closely related 
with those in the 
Hamiltonian, it is natural to 
suppose that the algebra is not affected by the 
scaling limit, in view of our discussions in section 5.

 It is straightforward to see that 
the matrix model operators (\ref{MMoperators})
 form a closed algebra such as 
$$
[ \TmatrixA_A, \TmatrixA_A ] \sim \TmatrixA_A , 
{}~~~  [ \TmatrixB_B, \TmatrixB_B ] \sim  \TmatrixB_B, 
$$
$$
[ \TmatrixA_A, \TmatrixA_n ] \sim  \TmatrixA_n , 
{}~~~ [ \TmatrixB_B, \TmatrixB_n ] \sim  \TmatrixB_n, 
$$
$$
[ \TmatrixA_n, \TmatrixA_m ] \sim  \TmatrixA_{n+m} ,
{}~~~ [\TmatrixB_n, \TmatrixB_m ] \sim  \TmatrixB_{n+m}, 
$$
$$
[ \TmatrixA_A, \TmatrixB_n ] \sim  \TmatrixB_n ,
 ~~~  [ \TmatrixB_B, \TmatrixA_n ] \sim  \TmatrixA_n, 
$$
$$
[ \TmatrixA_A, \TmatrixB_B ] =0 ,  ~~~
[ \TmatrixA_n, \TmatrixB_m ] \sim
\TmatrixA_{n+m}+\TmatrixB_{n+m}, 
$$
\beq
{}~~~(n, m\geq 1). 
\eeq
The corresponding algebra of the continuum generators 
\beq
\prt_y T_A^A(y), ~T_n^A(y;y_1,x_1,\cdots,y_n,x_n),
{}~\prt_x T_B^B(x), 
{}~T_n^B(y_1,x_1,\cdots,y_n,x_n;x)  ~~~(n=1,2,\cdots) 
\label{generators}
\eeq
must have the same structure. 
Half of the generators with the superscript $A(B)$ form a
 subalgebra of the full algebra .\footnote{
The algebra similar to this subalgebra
is presented in ref. \cite{NS}.} 
We note that, as already emphasized in the 
beginning of this section, the generators contained in the 
Hamiltonian $\cH$ 
are only the particular symmetrized combinations of the 
general Schwinger-Dyson operators. 
To ensure the closure of the full algebra, we are lead to  
introduce all of the above general Schwinger-Dyson operators.


\vspace{1cm}

\section{Reduction to the $W_3$ Constraints}
\setcounter{equation}{0}

  In the previous section, we obtained the huge algebra of
the Schwinger-Dyson operators. 
Although this algebra has a very 
complicated structure, we will next demonstrate how  
 the $W_3$ constraints, 
characterizing the $c=1/2$ noncritical string, is naturally 
derived from the integrability condition of the first few 
constraint operators appearing in our  
Hamiltonian. 
This will provide us yet another
consistency check of the preceding 
results. 

   The constraints associated with
the Hamiltonian (\ref{Hamiltonian6}) are 
\bea
-\prt_y T^A_A(y) Z[J] & = & 0,      \label{7-1-1} \\
-\prt_x T^B_B(x) Z[J] & = & 0,    \\
T_1(y,x)Z[J] & = & 0,   \\
T_2(y_1,x_1,y_2,x_2) Z[J]& = & 0, 
\eea
$$
\cdots .
$$
Considering the closure of this constraint algebra 
\bea
\lefteqn{[\prt_yT_A^A(y), T_1(y',x)]} \nn \\
 & = & -\gst^2\prt_y\left\{\frac{1}{(y-y')}\prt_{y'}T_1(y',x)+
\frac{1}{(y-y')^2}(T_1(y,x)+3T_1(y',x))\right.\nn \\
 & & \left.+\frac{-2}{(y-y')^2}(T^A_1(y;y',x)
+T^A_1(y';y,x)+T^B_1(y,x;x)+T^B_1(y',x;x))
\right\}, 
\eea
which follows from (\ref{algebraAAA1}) and
(\ref{algebraAAB1}), we obtain a
 new 
combination of the Schwinger-Dyson operators,  
\beq
(T^A_1(y;y',x)+T^A_1(y';y,x)+T^B_1(y,x;x)+T^B_1(y',x;x))Z[J]=0. 
\eeq
Similarly, from the algebra $[\prt_x T^B_B(x), T_1(y,x')]$, we have 
\beq
(T^B_1(y,x;x')+T^B_1(y,x';x)+T^A_1(y;y,x)+T^A_1(y;y,x'))Z[J]=0. 
\eeq
Let us consider the integrated versions of these:
\bea
\lefteqn{\left[s^{-1}\int_{C''}
\frac{dx}{2\pi i}\int_{C'}\frac{dy'}{2\pi i}
(T^A_1(y;y',x)+T^A_1(y';y,x))\right. }\nn\\
& &\left. +s^{-1}\int_{C''}\frac{dx}{2\pi i}
\int_{C'}\frac{dy'}{2\pi i}T^B_1(y',x;x)
\right]Z[J]=0, 
\label{7-2-1}
\eea
\beq
s^{-1}\int_{C''}\frac{dx'}{2\pi i}
\int_{C'}\frac{dx}{2\pi i}(T^B_1(y,x;x')
+T^B_1(y,x';x))Z[J]=0, 
\label{7-2-2}
\eeq
where the integral contours $C, C'$ and $C''$ 
successively wrap around the negative real axis 
and the singularities in the left half plane.

In order to derive the $W_3$ constraints,
it is important to examine the explicit 
forms of  these integrated operators. 
First,  by using the formulas in the Appendix B, we have
\bea
\lefteqn{\intyC D_z(y,y')
\frac{\dl}{\dl\Jt_I(z,x,y_1,x_1,\cdots)} }\nn\\
 & = & -\frac{\dl}{\dl\Jt_I(y',x,y_1,x_1,\cdots)}
+\sum_{n=0}^{\infty}y'^nc^I_n(x,y_1,x_1,\cdots), 
\label{7-3-1}
\eea
where the second term stands for
the polynomial part with respect to $y'$ of the 
large $y'$ expansion of
$\frac{\dl}{\dl\Jt_I(y',x',y_1,x_1,\cdots)}$.
The meaning of 
eq. (\ref{7-3-1}) is as follows.
In terms of the domain length $l'$ conjugate to $y'$, 
the l.h.s. represents the operator
with the domain length $l'+\ep$, where $\ep$ comes 
from the procedure of changing the contour 
$\intyC\cdots=\lim_{\ep\limit +0}\int_{-i\infty}^{i\infty}
\frac{dy}{2\pi i}\e^{\ep y}\cdots$. 
When $l'\neq 0$, the limit $\ep\limit +0$ is smooth, 
and thus it coincides with 
the first term of  
the r.h.s. However, for the singular terms
supported only at $l'=0$, if any, the l.h.s. gives no 
contribution, since $\lim_{\ep\limit +0}\dl^{(n)}(\ep)=0
{}~(n=0,1,2,\cdots)$ in the prescription 
of the Appendix B.  The second term is needed to
subtract such a contribution. 
As is seen from the expansions (\ref{expansionJA})
$\sim$(\ref{expansionJ2-1}), 
for example, $c^I_n$'s are all zero for $I=A,B,1$, and
$c^2_1=a^2_1$, 
$c^2_0=a^2_0 x+a'^2_0 x_1+a''^2_0 y_1$,
the others vanish for $I=2$. 
 Also, for composite operators,
the above formula can be used. For instance, 
\beq
\intyC D_z(y,y')\frac{\dl}{\dl\Jt_A(z)}\frac{\dl}{\dl\Jt_1(z,x)} 
= -\frac{\dl}{\dl\Jt_A(y')}\frac{\dl}{\dl\Jt_1(y',x)}
+\sum_{n=0}^{\infty}y'^nc^{A1}_n(x).
\eeq
{}From the short-domain expansions,
$c^{A1}_n(x)$'s turn out to be polynomials of $x$ and 
to vanish for $n\geq 4$. Thus, we have 
\beq
\int_{C''}\frac{dx}{2\pi i}\int_{C'}\frac{dy}{2\pi i}
D_z(y,y')\frac{\dl}{\dl\Jt_A(z)}\frac{\dl}{\dl\Jt_1(z,x)} 
= -\frac{\dl}{\dl\Jt_A(y')}
\int_{C''}\frac{dx}{2\pi i}\frac{\dl}{\dl\Jt_1(y',x)}, 
\eeq
and similarly 
\beq
\int_{C''}\frac{dx'}{2\pi i}\int_{C'}\frac{dx}{2\pi i}
D_z(x,x')\frac{\dl}{\dl\Jt_B(z)}\frac{\dl}{\dl\Jt_1(y,z)} 
= -\int_{C''}\frac{dx'}{2\pi i}
\frac{\dl}{\dl\Jt_B(x')} \frac{\dl}{\dl\Jt_1(y,x')}. 
\label{compositeoperator}
\eeq

   After using the above formulas and doing the integral
in the r.h.s. of (\ref{compositeoperator}) 
by substituting the expanded forms
 (\ref{expansionJB}), (\ref{expansionJ1}), 
the integrated operators are written as 
\bea
\lefteqn{s^{-1}\int_{C''}\frac{dx}{2\pi i}
\int_{C'}\frac{dy'}{2\pi i}T^A_1(y;y',x)}\nn\\
& = & -\frac{\dl}{\dl\Jt_A(y)}\cO_1(\frac{\dl}{\dl\Jt})(y)
-c\cO_2(\frac{\dl}{\dl\Jt})(y) \nn\\
&  & +s^{-1}(\tilde{a}-a'^1_{-1})
\frac{\dl}{\dl\Jt_A(y)}B_{1/3} \nn\\
 & & -\gst^2\intyp\Jt_A(y')\prt_{y'}D_z(y,y')
\cO_1(\frac{\dl}{\dl\Jt})(z)\nn\\
& & +(\mbox{terms containing $\Jt_I~~(I\neq A)$}), 
\label{7-4-1}
\eea
\bea
\lefteqn{s^{-1}\int_{C''}\frac{dx'}{2\pi i}
\int_{C'}\frac{dx}{2\pi i}T^B_1(y,x;x')} \nn\\
& = & -s^{-1}a^B_{4/3}a^1_{-7/3}y^4
-s^{-1}(a^B_{4/3}a'^1_{-7/3}+a^B_{-2/3}a^1_{-1/3})Ty^2 \nn\\
& & -s^{-1}(a^B_{4/3}a''^1_{-7/3}+a^1_{2/3})B_{2/3}y \nn\\
& & -s^{-1}(a^B_{4/3}a'''^1_{-7/3}+a^1_{5/3})B_{5/3}
-s^{-1}a^B_{-2/3}a'^1_{-1/3}T^2 \nn\\
 & & -c(1+c^{-1}s^{-1}a^B_{4/3}a^{ ({\rm v})1}_{-7/3})
\cO_2(\frac{\dl}{\dl\Jt})(y) \nn\\
&  & +s^{-1}(\tilde{a}-a^B_{4/3}a^{({\rm iv}) 1}_{-7/3}-a^1_{1/3})
                \frac{\dl}{\dl\Jt_A(y)}B_{1/3} \nn\\
 & & +(\mbox{terms containing $\Jt_I~~(I\neq A)$}). 
\label{7-4-2}
\eea
 The disk parts of  (\ref{7-4-1}) and (\ref{7-4-2})
can be interpreted to represent the continuum versions of 
the Schwinger-Dyson equations 
(\ref{A-2-2}) and (\ref{A-2-3}), respectively. 
In (\ref{7-4-2}), thanks to the short-domain expansions, the integral 
$\int_C\frac{dx'}{2\pi i}\frac{\dl}{\dl\Jt_B(x')}\frac{\dl}{\dl\Jt_1(y,x')}$
whose $y$-dependence was difficult to see in this form turns out to 
be  essentially 
a polynomial of $y$. This property
 is crucial for the derivation of the $W_3$ constraints. 

    Also, from the results (\ref{W2-1}) and
(\ref{W2-2}) in the Appendix D, we 
can assume the following identity holds:
\bea
\lefteqn{s^{-1}\int_{C''}\frac{dx'}{2\pi i}\int_{C'}
\frac{dy'}{2\pi i}s^{-1}\int_C\frac{dx}{2\pi i}
\frac{\dl}{\dl\Jt_2(y,x,y',x')}}\nn\\
 & = & 
s^{-1}\int_{C''}\frac{dx'}{2\pi i}\int_{C'}
\frac{dy'}{2\pi i}s^{-1}\int_C\frac{dx}{2\pi i}
\frac{\dl}{\dl\Jt_2(y',x,y,x')} \nn\\
 & = & 
s^{-1}\int_{C''}\frac{dx'}{2\pi i}\int_{C'}
\frac{dx}{2\pi i}s^{-1}\int_C\frac{dy'}{2\pi i}
\frac{\dl}{\dl\Jt_2(y,x,y',x')} \nn \\
 & = & 
s^{-1}\int_{C''}\frac{dx'}{2\pi i}\int_{C'}
\frac{dx}{2\pi i}s^{-1}\int_C\frac{dy'}{2\pi i}
\frac{\dl}{\dl\Jt_2(y,x',y',x)}. 
\eea
In the Appendix D, we have confirmed these equations 
for some simple cases. 
This means that the results of the  spin-flip processes   
are independent of their orderings.  
 By using these identities for the explicit forms
of $T^A_1$, $T^B_1$, we 
can see the symmetry properties: 
\beas
\lefteqn{\left.s^{-1}\int_{C''}\frac{dx}{2\pi i}
\int_{C'}\frac{dy'}{2\pi i}T^A_1(y;y',x)
Z[J]\right|_{J_I=0~(I\neq A)}}  \\
 & = & 
\left.s^{-1}\int_{C''}\frac{dx}{2\pi i}
\int_{C'}\frac{dy'}{2\pi i}T^A_1(y';y,x)
Z[J]\right|_{J_I=0~(I\neq A)}, 
\eeas
\bea
\lefteqn{\left.s^{-1}\int_{C''}\frac{dx'}{2\pi i}
\int_{C'}\frac{dx}{2\pi i}T^B_1(y,x;x')
Z[J]\right|_{J_I=0~(I\neq A)}} \nn \\
 & = & 
\left.s^{-1}\int_{C''}\frac{dx'}{2\pi i}\int_{C'}
\frac{dx}{2\pi i}T^B_1(y,x';x)
Z[J]\right|_{J_I=0~(I\neq A)}. 
\label{7-6-1}
\eea

  Setting $\Jt_I=0~(I\neq A)$, eqs. (\ref{7-2-1})
and (\ref{7-2-2}) become respectively 
\bea
\lefteqn{[-\frac{\dl}{\dl\Jt_A(y)}
\cO_1(\frac{\dl}{\dl\Jt})(y)+\cO_0'
    -c\cO_2(\frac{\dl}{\dl\Jt})(y)+s^{-1}(\tilde{a}
-a'^1_{-1})\frac{\dl}{\dl\Jt_A(y)}B_{1/3}}\nn\\
 & - & \left.\gst^2\int_{-i\infty}^{i\infty}
\frac{dy'}{2\pi i}\Jt_A(y')\prt_{y'}D_z(y,y')
       \cO_1(\frac{\dl}{\dl\Jt})(z)]Z[J]\right|_{J_I=0~(I\neq A)}=0, 
\label{7-7-1}
\eea
\bea
\lefteqn{[-s^{-1}a^B_{4/3}a^1_{-7/3}y^4
     -s^{-1}(a^B_{4/3}a'^1_{-7/3}
+a^B_{-2/3}a^1_{-1/3})Ty^2}\nn\\
 & - & s^{-1}(a^B_{4/3}a''^1_{-7/3}
+a^1_{2/3})B_{2/3}y
     -s^{-1}(a^B_{4/3}a'''^1_{-7/3}
+a^1_{5/3})B_{5/3}-s^{-1}a^B_{-2/3}a'^1_{-1/3}T^2\nn\\
 & - & c(1+c^{-1}s^{-1}a^B_{4/3}a^{({\rm v}) 1}_{-7/3})
\cO_2(\frac{\dl}{\dl\Jt})(y)\nn\\
 & + &\left. s^{-1}(\tilde{a}-a^B_{4/3}a^{({\rm iv}) 1}_{-7/3}
-a^1_{1/3})
\frac{\dl}{\dl\Jt_A(y)}B_{1/3} ]Z[J]\right|_{J_I=0~(I\neq A)}=0. 
\label{7-7-2}
\eea
We used (\ref{7-4-1}), (\ref{7-4-2}) and (\ref{7-6-1}),
and $\cO_0'$ is the $y$-independent operator 
$$
\cO_0'=\oh s^{-1}\int_{C''}\frac{dx}{2\pi i}\int_{C'}
\frac{dy'}{2\pi i}T^B_1(y',x;x).
$$
Also, from the once-integrated version
of  (\ref{7-1-1}), we have 
\bea
\lefteqn{[-\frac{\dl^2}{\dl\Jt_A(y)^2}
-c\cO_1(\frac{\dl}{\dl\Jt})(y)+\cO''_0}\nn\\
 & - &\left.  \gst^2\int_{-i\infty}^{i\infty}
\frac{dy'}{2\pi i}\Jt_A(y')\prt_{y'}D_z(y,y')
          \frac{\dl}{\dl\Jt_A(z)}]Z[J]\right|_{J_I=0~(I\neq A)}=0, 
\label{7-8-1}
\eea
where $\cO''_0$ is a $y$-independent
operator introduced as an integration constant. 

   The three eqs. ({\ref{7-7-1}), (\ref{7-7-2})
and (\ref{7-8-1}) lead to 
a closed 
equation of the loop operator $\tilde{\Phi}_A(y)$: 
\bea
\lefteqn{[ a_4y^4+a_2Ty^2+a_1B_{2/3}y
+a_0B_{5/3}+a'_0T^2+c\cO_0'}\nn\\
&  &+ \frac{\dl^3}{\dl\Jt_A(y)^3}
-\frac{\dl}{\dl\Jt_A(y)}(\cO''_0+a_A B_{1/3})
       +\oh\gst^2\prt_y^2\frac{\dl}{\dl\Jt_A(y)} \nn \\
 &  & +\gst^2\int_{-i\infty}^{i\infty}
\frac{dy'}{2\pi i}\Jt_A(y')\prt_{y'}D_z(y,y')
              \left(\frac{\dl}{\dl\Jt_A(z)}
\frac{\dl}{\dl\Jt_A(y)}+\frac{\dl^2}{\dl\Jt_A(z)^2}\right)\nn\\
 &  &+\gst^4\int_{-i\infty}^{i\infty}
\frac{dy'}{2\pi i}\int_{-i\infty}^{i\infty}\frac{dy''}{2\pi i}
      \Jt_A(y')\Jt_A(y'')\prt_{y'}\prt_{y''}
D_z(y,y'')D_w(z,y')\frac{\dl}{\dl\Jt_A(w)} ]\nn \\
 &\times & Z[J]|_{J_I=0~(I\neq A)}=0. 
\label{7-8-2}
\eea
The coefficients $a$'s are defined by 
$$
a_4=\frac{cs^{-1}a^B_{4/3}a^1_{-7/3}}{1+c^{-1}s^{-1}a^B_{4/3}
a^{({\rm v}) 1}_{-7/3}}, ~~~
a_2=\frac{cs^{-1}(a^B_{4/3}a'^1_{-7/3}+a^B_{-2/3}a^1_{-1/3})} 
         {1+c^{-1}s^{-1}a^B_{4/3}a^{({\rm v}) 1}_{-7/3}}, 
$$
$$
a_1=\frac{cs^{-1}(a^B_{4/3}a''^1_{-7/3}+a^1_{2/3})}  
      {1+c^{-1}s^{-1}a^B_{4/3}a^{({\rm v}) 1}_{-7/3}},  ~~~
a_0=\frac{cs^{-1}(a^B_{4/3}a'''^1_{-7/3}+a^1_{5/3})}  
      {1+c^{-1}s^{-1}a^B_{4/3}a^{({\rm v}) 1}_{-7/3}},
$$
$$
a'_0=\frac{cs^{-1}a^B_{-2/3}a'^1_{-1/3}}
 {1+c^{-1}s^{-1}a^B_{4/3}a^{({\rm v}) 1}_{-7/3}},~~~
a_A=\frac{cs^{-1}(\tilde{a}-a^B_{4/3}
a^{({\rm iv}) 1}_{-7/3}-a^1_{1/3})}
          {1+c^{-1}s^{-1}a^B_{4/3}
a^{({\rm v}) 1}_{-7/3}}-cs^{-1}(\tilde{a}-a'^1_{-1}). 
$$
Further, by rescaling as 
$$
\frac{\dl}{\dl\Jt_A(y)}\limit \left(\frac{-a_4}{16}\right)^{1/3}
\frac{\dl}{\dl\Jt_A(y)}, ~~~
\Jt_A(y)\limit\left(\frac{-a_4}{16}\right)^{-1/3}\Jt_A(y), 
$$
$$
\gst\limit\left(\frac{-a_4}{16}\right)^{1/3}\gst,
{}~~~T\limit -\frac{a_4}{a_2}T, 
$$
and putting 
$$
a_1B_{2/3}=\frac{a_4}{16}\cO_{\Delta},~~~
a_0B_{5/3}+a'_0T^2+c\cO_0'=\frac{a_4}{16}\cO_1, ~~~
\cO''_0+a_A B_{1/3}=\left(\frac{-a_4}{16}\right)^{2/3}\cO_0, 
$$
eq. (\ref{7-8-2}) takes the form 
\bea
\lefteqn{[ -16y^4+16Ty^2-y\cO_{\Delta}-\cO_1}\nn\\
&  &+ \frac{\dl^3}{\dl\Jt_A(y)^3}-\frac{\dl}{\dl\Jt_A(y)}\cO_0
       +\oh\gst^2\prt_y^2\frac{\dl}{\dl\Jt_A(y)} \nn \\
 &  & +\gst^2\int_{-i\infty}^{i\infty}\frac{dy'}{2\pi i}
\Jt_A(y')\prt_{y'}D_z(y,y')
              \left(\frac{\dl}{\dl\Jt_A(z)}\frac{\dl}{\dl\Jt_A(y)}
+\frac{\dl^2}{\dl\Jt_A(z)^2}\right)\nn\\
 &  &+ \gst^4\int_{-i\infty}^{i\infty}
\frac{dy'}{2\pi i}\int_{-i\infty}^{i\infty}\frac{dy''}{2\pi i}
      \Jt_A(y')\Jt_A(y'')\prt_{y'}\prt_{y''}
D_z(y,y'')D_w(z,y')\frac{\dl}{\dl\Jt_A(w)} ]\nn \\
 &\times & Z[J]\left|_{J_I=0~(I\neq A)}=0. \right. 
\label{7-10-1}
\eea

     Let us now
 confirm that this equation represents the $W_3$ constraints. 
In order to do so, we have to identify 
the disk and cylinder singular parts.
For the disk amplitude $w(y)$, the Schwinger-Dyson equation
obtained from (\ref{7-10-1}) is 
\beq
w(y)^3-\bra\cO_0\ket_0 w(y)-16y^4+16Ty^2
-y\bra\cO_{\Delta}\ket_0-\bra\cO_1\ket_0
=0, 
\eeq
where we consider the solution regular
except on the negative real axis. Note that this 
condition does not determine the solution uniquely.
In fact, we find 
 two solutions: \\
i) $~~~ w(y)=2^{4/3}(y-\sqrt{\frac{T}{6}})(y+3\sqrt{\frac{T}{6}})^{1/3}, $
\beq
\bra\cO_0\ket_0=0,~~~\bra\cO_{\Delta}\ket_0
=128\left(\frac{T}{6}\right)^{3/2}, ~~~
    \bra\cO_1\ket_0= -\frac{4}{3}T^2, 
\eeq
ii) $~~~ w(y)=(y+\sqrt{y^2-T})^{4/3}+(y-\sqrt{y^2-T})^{4/3}, $
\beq
\bra\cO_0\ket_0=3T^{4/3},~~~\bra\cO_{\Delta}\ket_0=0, ~~~
    \bra\cO_1\ket_0= 2T^2. 
\label{7-11-0}
\eeq
It is the solution ii) that reproduces the matrix model result.
By repeating the argument in the Appendix C for deriving 
(\ref{A-8-1}), 
{\bf without} using the $Z_2$ symmetry
$A\leftrightarrow B$, we can see that 
$\bra \cO_{\Delta}\ket_0$ is proportional
to the next leading order  ($O(a^{3})$) of the 
universal part of the quantity $\frac{1}{N}\bra\tr (A-B)\ket_0$. 
This implies that the solution i) spontaneously
breaks the $Z_2$ symmetry. 
Here we only consider the $Z_2$ symmetric solution ii) and 
leave the case i) as a future problem.

   For the cylinder amplitude $w(y,y_1)$, the
Schwinger-Dyson equation is derived from 
the lowest order of $\gst$ in the $\Jt_A(y_1)$
derivative of  (\ref{7-10-1}): 
\bea
\lefteqn{-y\bra\cO_{\Delta}\tilde{\Phi}_A(y_1)\ket_0
-\bra\cO_1\tilde{\Phi}_A(y_1)\ket_0
  -\bra\cO_0\tilde{\Phi}_A(y_1)\ket_0 w(y) }\nn\\
 & + & (3w(y)^2-\bra\cO_0\ket_0)w(y,y_1) \nn \\
 & + & \gst^2\prt_{y_1}\frac{1}{y-y_1}
(2w(y)^2-w(y)w(y_1)-w(y_1)^2)=0. 
\label{7-11-1}
\eea
If we take the solution ii) as the disk amplitude,
note that it satisfies 
\beq
3w(y)^2-\bra\cO_0\ket_0=0~~~~
\mbox{at}~y=0,\pm\sqrt{\frac{T}{2}}. 
\label{zeropoint}
\eeq
The disk amplitude is regular except the region
$y\leq -\sqrt{T}$ on the real axis. 
By assuming the same property for the cylinder amplitude, 
eq. (\ref{7-11-1}) can be  
solved, by using the similar argument as for 
the case (\ref{cylinderSD}). The result is
\bea
w(y,y_1) & = & \gst^2\frac{4}{9}\frac{1}{f(y,y_1)g(y,y_1)}
\frac{(y+\sqrt{y^2-T})^{1/3}}{(y+\sqrt{y^2-T})^{2/3}
+(y-\sqrt{y^2-T})^{2/3}+T^{1/3}} \nn \\
 &  & \times
\frac{(y_1+\sqrt{y_1^2-T})^{1/3}}{(y_1+\sqrt{y_1^2-T})^{2/3}+
(y_1-\sqrt{y_1^2-T})^{2/3}+T^{1/3}} \nn \\
& \times & \left[1+\frac{3T^{1/3}}{f(y,y_1)}
    +\frac{3(y+\sqrt{y^2-T})^{1/3}
(y_1+\sqrt{y_1^2-T})^{1/3}}{g(y,y_1)}\right], 
\label{7-12-1}
\eea
where 
\beas
f(y,y_1) & = & (y+\sqrt{y^2-T})^{1/3}(y_1+\sqrt{y_1^2-T})^{1/3} \\
 & & + (y-\sqrt{y^2-T})^{1/3}(y_1-\sqrt{y_1^2-T})^{1/3} +T^{1/3}, \\
g(y,y_1) & = & (y+\sqrt{y^2-T})^{2/3}+(y_1+\sqrt{y_1^2-T})^{2/3} \\
 & & + (y+\sqrt{y^2-T})^{1/3}(y_1+\sqrt{y_1^2-T})^{1/3} . 
\eeas
{}From (\ref{7-11-0}) and (\ref{7-12-1}), the singular parts can be found as 
\bea
w^{\mbox{sing}}(y) & = & 2^{4/3}(y^{4/3}-\frac{T}{3}y^{-2/3}), 
\label{7-13-1} \\
w^{\mbox{sing}}(y,y_1) & = & \gst^2\frac{1}{9}\frac{1}{(y-y_1)^2} \nn\\
 & &\times [y^{2/3}y_1^{-2/3}+2y^{1/3}y_1^{-1/3}-6+2y^{-1/3}y_1^{1/3}
                      +y^{-2/3}y_1^{2/3}]. 
\label{7-13-2}
\eea

The connected correlation functions are expanded by
the local operator insertions 
$g_{\alp_1,\cdots,\alp_n}$ as 
\bea
\bra\tilde{\Phi}_A(y_1)\ket & = & 
             w^{\mbox{sing}}(y_{1})+g^{(1)}(y_{1}),  \nn   \\
\bra\tilde{\Phi}_A(y_1)\tilde{\Phi}_A(y_2)\ket & = & 
w^{\mbox{sing}}(y_{1},y_{2})+g^{(2)}(y_{1},y_{2}), \nn  \\
\bra\tilde{\Phi}_A(y_1)\cdots\tilde{\Phi}_A(y_K)\ket
& = & g^{(K)}(y_{1},\cdots,y_{K}) ~~~(K\geq 3), \nn\\
g^{(n)}(y_{1},\cdots,y_{n}) & = & \sum_{\alp_{1},\cdots,\alp_{n}}
                  g_{\alp_{1},\cdots,\alp_{n}}
           y_{1}^{-\alp_{1}-1}\cdots y_{n}^{-\alp_{n}-1}, \label{7-13-3}
\eea
where $\alp_i$'s run over the positive integers $+1/3$ and $+2/3$,
 i.e. $1/3,2/3,4/3,5/3,7/3,\cdots.$

    Using (\ref{7-13-1})$\sim$(\ref{7-13-3}),
we expand (\ref{7-10-1}) similarly 
as in the argument of 
the Virasoro constraints in the Appendix B. 
{}From here the analysis is parallel to the reference \cite{GN}, 
 where  the $W_3$ 
constraints were explicitly derived from the two-matrix model 
for the first time. 
So, we show only the results.
For $\cO_0$, $\cO_{\Delta}$, $\cO_1$ insertions, 
\bea
\bra \cO_{0}\ket & = & 3\cdot 2^{4/3}g_{1/3},   \nn \\
\bra \cO_{0}\tilde{\Phi}_A(y_{1})\ket & = &
 \gst^{2}2^{4/3}\frac{1}{3} y_{1}^{-2/3}
     +3\cdot 2^{4/3}\sum_{\alp_{1}}
g_{1/3,\alp_{1}}y_{1}^{-\alp_{1}-1},  \nn \\
\bra \cO_{0}\tilde{\Phi}_A(y_{1})
\cdots\tilde{\Phi}_A(y_{K})\ket & = & 
                  3\cdot 2^{4/3}\sum_{\alp_{1},\cdots,\alp_{K}}
   g_{1/3,\alp_{1},\cdots,\alp_{K}}y_{1}^{-\alp_{1}-1}\cdots 
                   y_{K}^{-\alp_{K}-1},  \nn \\
\bra \cO_{\Delta}\ket & = & 3\cdot 2^{8/3}g_{2/3},   \nn \\
\bra \cO_{\Delta}\tilde{\Phi}_A(y_{1})\ket & = &
\gst^{2}2^{8/3}\frac{2}{3} y_{1}^{-1/3}
     +3\cdot 2^{8/3}\sum_{\alp_{1}}
g_{2/3,\alp_{1}}y_{1}^{-\alp_{1}-1},  \nn \\
\bra \cO_{\Delta}\tilde{\Phi}_A(y_{1})
\cdots\tilde{\Phi}_A(y_{K})\ket & = & 
                  3\cdot 2^{8/3}\sum_{\alp_{1},\cdots,\alp_{K}}
   g_{2/3,\alp_{1},\cdots,\alp_{K}}y_{1}^{-\alp_{1}-1}\cdots 
                   y_{K}^{-\alp_{K}-1},  \nn \\
\bra \cO_{1}\ket & = & \frac{16}{3}T^2
+3\cdot 2^{8/3}g_{5/3},   \nn \\
\bra \cO_{1}\tilde{\Phi}_A(y_{1})\ket & = &
 \gst^{2}2^{8/3}\frac{5}{3} y_{1}^{2/3}
     +3\cdot 2^{8/3}\sum_{\alp_{1}}
g_{5/3,\alp_{1}}y_{1}^{-\alp_{1}-1},  \nn \\
\bra \cO_{1}\tilde{\Phi}_A(y_{1})
\cdots\tilde{\Phi}_A(y_{K})\ket & = & 
                  3\cdot 2^{8/3}\sum_{\alp_{1},\cdots,\alp_{K}}
   g_{5/3,\alp_{1},\cdots,\alp_{K}}y_{1}^{-\alp_{1}-1}\cdots 
                    y_{K}^{-\alp_{K}-1} \nn\\
 & & ~~~~~~~~~~~~~~~~~~~~~~~~~~~~~~~~  (K\geq 2).    
\eea
This means that $\cO_0$, $\cO_{\Delta}$, $\cO_1$ are
expressed in terms of 
the loop operator $\frac{\dl}{\dl\Jt_A}$ 
\beas
\cO_0 & = & 3\cdot 2^{4/3}\int_C
\frac{dy}{2\pi i}y^{1/3}\frac{\dl}{\dl\Jt_A(y)}, \\
\cO_{\Delta} & = & 3\cdot 2^{8/3}\int_C
\frac{dy}{2\pi i}y^{2/3}\frac{\dl}{\dl\Jt_A(y)}, \\
\cO_1 & = & \frac{16}{3}T^2+
      3\cdot 2^{8/3}\int_C\frac{dy}{2\pi i}y^{5/3}\frac{\dl}{\dl\Jt_A(y)}, 
\eeas
as operators acting on $Z[J]|_{J_I=0~(I\neq A)}$. 

    For the other contributions, by introducing the generating function 
\beq
\ln Z(\mu)=\sum_{n_{\alp}=0}^{\infty}\frac{\mu_{1/3}^{n_{1/3}}}{n_{1/3}!}
                               \frac{\mu_{2/3}^{n_{2/3}}}{n_{2/3}!}\cdots
   g_{\underbrace{\mbox{\scriptsize 1/3,$\cdots$,1/3}}_{n_{1/3}},
        \underbrace{\mbox{\scriptsize 2/3,$\cdots$,2/3}}_{n_{2/3}},\cdots},
\eeq
they can be expressed as 
\bea
L_n Z(\mu)=0~~~& & (n\geq -1), \nn\\
W_{n'} Z(\mu)=0~~~& & (n' \geq -2), 
\eea
where 
\bea
L_{-1} & = & 2\cdot 2^{4/3}\frac{\prt}{\prt\mu_{4/3}}
+\gst^2\frac{2}{3}\sum_{\alp}\alp\mu_{\alp}\frac{\prt}{\prt\mu_{\alp-1}} 
 +\gst^4\frac{4}{81}\mu_{2/3}(\mu_{1/3}-\gst^{-2}3\cdot 2^{4/3}T), \nn\\
L_0 & = & 2\cdot 2^{4/3}\left(\frac{\prt}{\prt\mu_{7/3}}
    -\frac{T}{3}\frac{\prt}{\prt\mu_{1/3}}\right)
    +\gst^2\frac{2}{3}\sum_{\alp}\alp\mu_{\alp}\frac{\prt}{\prt\mu_{\alp}} 
  +\gst^2\frac{2}{27}, \nn\\
L_l & = & 2\cdot 2^{4/3}\left(\frac{\prt}{\prt\mu_{l+7/3}}
    -\frac{T}{3}\frac{\prt}{\prt\mu_{l+1/3}}\right)
     +\sum_{\beta+\beta'=l}\frac{\prt^2}{\prt\mu_{\beta}\prt\mu_{\beta'}} 
  +\gst^2\frac{2}{3}\sum_{\alp}\alp\mu_{\alp}\frac{\prt}{\prt\mu_{\alp+l}}
\nn\\
 & & ~~~~~~~~~~~~~~~~~~~~~~~~~~~~~
{}~~~~~~~~~~~~~~~~~~~~~~~~~~~~~(l\geq 1), 
\eea
\bea
W_{-2} & = & 3\cdot 2^{8/3}\left(\frac{\prt}{\prt\mu_{8/3}}
      -\frac{2}{3}T\frac{\prt}{\prt\mu_{2/3}}\right) \nn\\
 & &+\gst^22\cdot 2^{4/3}\sum_{\alp}\alp\mu_{\alp}
\left(\frac{\prt}{\prt\mu_{\alp+1/3}}
        -\frac{T}{3}\frac{\prt}{\prt\mu_{\alp-5/3}}\right) \nn\\
 & & +\gst^2\sum_{\alp}\alp\mu_{\alp}\sum_{\beta+\beta'=\alp-2}
          \frac{\prt^2}{\prt\mu_{\beta}\prt\mu_{\beta'}} \nn\\
 & &+\gst^4\frac{1}{3}\sum_{\alp,\alp'}\alp\alp'\mu_{\alp}\mu_{\alp'}
         \frac{\prt}{\prt\mu_{\alp+\alp'-2}}  \nn \\
 & & +\gst^6\frac{4}{3^5}\mu_{4/3}(\mu_{1/3}
-\gst^{-2}3\cdot 2^{4/3}T)^2
          +\gst^6\frac{8}{3^6}\mu_{2/3}^3,   \nn\\
W_{-1} & = & 3\cdot 2^{8/3}\left(\frac{\prt}{\prt\mu_{11/3}}
      -\frac{2}{3}T\frac{\prt}{\prt\mu_{5/3}}\right)
            +3\cdot 2^{4/3}\frac{\prt^2}{\prt\mu_{2/3}^2} \nn\\
 & &+\gst^22\cdot 2^{4/3}\sum_{\alp}\alp\mu_{\alp}
\left(\frac{\prt}{\prt\mu_{\alp+4/3}}
        -\frac{T}{3}\frac{\prt}{\prt\mu_{\alp-2/3}}\right) \nn\\
 & & +\gst^2\sum_{\alp}\alp\mu_{\alp}
\sum_{\beta+\beta'=\alp-1}
          \frac{\prt^2}{\prt\mu_{\beta}\prt\mu_{\beta'}} \nn\\
 & &+\gst^4\frac{1}{3}\sum_{\alp,\alp'}
\alp\alp'\mu_{\alp}\mu_{\alp'}
         \frac{\prt}{\prt\mu_{\alp+\alp'-1}}  \nn \\
 & & +\gst^6\frac{1}{3^6}(\mu_{1/3}
-\gst^{-2}3\cdot 2^{4/3}T)^3,   \nn\\
W_{m} & = & 3\cdot 2^{8/3}\left(\frac{\prt}{\prt\mu_{m+14/3}}
      -\frac{2}{3}T\frac{\prt}{\prt\mu_{m+8/3}}
+\frac{T^2}{9}\frac{\prt}{\prt\mu_{m+2/3}}\right)
            \nn\\
 & & +3\cdot 2^{4/3}\left(\sum_{\beta+\beta'=m+7/3}
-\frac{T}{3}\sum_{\beta+\beta'=m+1/3}
         \right)\frac{\prt^2}{\prt\mu_{\beta}\prt\mu_{\beta'}} \nn\\ 
 & & +\sum_{\beta+\beta'+\beta''=m}\frac{\prt^3}{\prt\mu_{\beta}
\prt\mu_{\beta'}\prt\mu_{\beta''}}
            \nn\\
 & &+\gst^22\cdot 2^{4/3}\sum_{\alp}\alp\mu_{\alp}
\left(\frac{\prt}{\prt\mu_{\alp+m+7/3}}
        -\frac{T}{3}\frac{\prt}{\prt\mu_{\alp+m+1/3}}\right) \nn\\
 & & +\gst^2\sum_{\alp}\alp\mu_{\alp}\sum_{\beta+\beta'=\alp+m}
          \frac{\prt^2}{\prt\mu_{\beta}\prt\mu_{\beta'}} \nn\\
 & &+\gst^4\frac{1}{3}\sum_{\alp,\alp'}\alp\alp'\mu_{\alp}\mu_{\alp'}
         \frac{\prt}{\prt\mu_{\alp+\alp'+m}}  ~~~~~~~~~~~~(m\geq 0).
\eea
This is nothing but the $W_3$ constraints. 

\vspace{0.5cm}
\section{Conclusion}

Let us first summarize what we have done. 
We have started our paper by discussing the 
nature of PSFTs from the view point of the 
stochastic quantization of the matrix models. 
Then, we have presented detailed derivations 
of the stochastic Hamiltonians in the double-scaling 
limit from the matrix model, 
and investigated the infinite algebras of the Schwinger-Dyson 
operators appearing in the Hamiltonians. 
We have also checked that the algebras contain 
the Virasoro (one-matrix model) and $W_3$ algebras 
(two-matrix model), as they should. 
Proofs of some of the crucial formulas have  
not been completed, because of technical complexity. 
It is therefore desirable to develop more powerful 
methods of treating the double-scaling limit for general 
target spaces.

After these calculations, 
we have to reconsider the questions 
raised in the earlier sections of the present paper. 
Perhaps, one of the most important lessons of our work
is that the structure of the general splitting and merging interactions 
of string fields with arbitrary 
matter configuration is not affected by the mixing of the string-field 
components in taking the scaling limit 
which is defined for a specific background. 
This seems to imply that 
the structure of these terms is completely independent of the 
backgrounds. Recalling the general 
discussion in section 2, we realize that the purely cubic Hamiltonian 
of the matrix model with most general source terms and no 
bare action already 
captures the structure of the continuum Hamiltonian 
in a background-independent way.  
This conforms to  
earlier suggestions \cite{yo}  
and points to an intriguing possibility of formulating 
a background-independent string field 
theory, encompassing critical strings, 
by starting from general matrix integrals 
with infinite number of different matrices. 
For the case $c\le 1$, a related idea has 
already been discussed in ref. \cite{IK3}. 

Before pursuing such possibilities, there remains, however, 
many important issues to be solved. 
Asides the problems mentioned in the Introduction, 
what is  needed to make further progress is a deeper 
understanding on 
how to extract real space-time picture of the string theory 
from matrix models, since matrix models apparently miss 
some important characteristics \cite{pol,JLY} of the string dynamics.

\vspace{1cm}

\noindent
{\large Acknowledgements}

At an early stage of this work, we were benefitted from collaborative 
discussions with A. Tsuchiya and from comments from A. Jevicki. 
The present work was partially supported by the Grant-in-Aid 
for Scientific Research 
(No. 06640378), Grand-in-Aid for Priority Area (No. 06221211) 
 from the Ministry of Education, Science and Culture,  
and also by  US-Japan Collaborative Program from the Japan 
Society of Promotion of Science.

\vspace{2cm}
\appendix
\large
\noindent
\bf Appendix
\renewcommand{\theequation}{\Alph{section}.\arabic{equation}}

\noindent
\section{$k=3$ critical point and the disk amplitude}
\setcounter{equation}{0}

\normalsize
Let us begin from deriving the non-even 
critical potential for $k=3$. It is sufficient 
to recall the well known formulas in the method of orthogonal 
polynomials. For the coefficients $S$ and $R$ in the 
recursion equation for the orthogonal polynomials
$P_n(\lambda)=\lambda^n
+\cdots$,   
$$
\lam P_n(\lam)=P_{n+1}(\lam)+S_n P_n(\lam)+R_n P_{n-1}(\lam),
$$
we have a set of equations \cite{grossmig}
in terms of the potential $V$,
\bea
0 & = & \oint_0\frac{dz}{2\pi i}\frac{1}{z}V'(z+S+\frac{R}{z}),
\label{or1} \\
x & = & \oint_0 \frac{dz}{2\pi i}\frac{N}{\beta}V'(z+S+\frac{R}{z}),
\label{or2}
\eea
in the sphere limit $N\sim \beta \limit\infty$, where 
$x=\frac{n}{\beta}$, $S=S(x)\simeq S_n$, $R=R(x)\simeq R_n$. 
Eqs. (\ref{or1}) and (\ref{or2}) give the relation 
that implicitly determines the 
function $R(x)$, of the form 
$$
x=W(R). 
$$
Since the free energy is determined by $R$ 
as 
\beq
\ln Z \sim \sum_{n=1}^{N-1}(N-n)\ln R_n, 
\label{fe}
\eeq
the following behavior of the $W(R)$, 
\beq
W(R)=1-\mbox{const.}(1-R)^3    \label{W(R)}
\eeq
as $x\limit 1$, $R\limit1$, leads to the $k=3$ criticality of the free energy 
$$
\ln Z\sim (1-\frac{N}{\beta})^{7/3}.
$$
This shows that the minimal order of the $k=3$ critical potential 
is four. 
After lengthy calculations, we find that the potential  
\beq
V(M)=\frac{\beta}{N} (\frac{g_2}{2}M^2
+\frac{g_3}{3}M^3+\frac{1}{20}M^4) 
\eeq 
with (\ref{g3}) and (\ref{g2}) 
satisfies all of the above conditions.  

Next let us derive the disk amplitude in the sphere approximation. 
By using the 
method of ref. \cite{BIPZ}, the disk amplitude in the 
large $N$ limit is given by
\beq
\bra \Phi(\ze)\ket_{0}=\oh V'(\ze)
  +\oh \frac{\beta}{N}(-\frac{1}{5}\ze^{2}+A\ze+B)
        \sqrt{(\ze-b_{+})(\ze-b_{-})},
                                      \label{k=3disk}
\eeq
where 
\beas
A & = & -g_{3}-\frac{1}{10}(b_{+}+b_{-}),  \\
B & = & -g_{2}-\frac{g_{3}}{2}(b_{+}+b_{-})-\frac{1}{20}(b_{+}+b_{-})^2
        -\frac{1}{40}(b_{+}-b_{-})^2.
\eeas
By introducing the variable 
\beq
z=10g_{3}+3(b_{+}+b_{-}),            \label{cut1}
\eeq
the end points of the cut  $b_{-}<\ze<b_{+}$ are determined by the equations: 
\beq
-\frac{1}{4}z^4+36z^2-256z-720+81920z^{-2}=3888\frac{N}{\beta},  
                                     \label{cut2}
\eeq
\beq
(b_{+}-b_{-})^2=-\frac{2}{27}z^2+\frac{80}{9}-\frac{2560}{27}z^{-1}.
                                      \label{cut3}
\eeq
 
The disk amplitude with a microscopic loop can be read off 
from the coefficient of $\ze^{-2}$ in the large $\ze$ expansion of
eq. (\ref{k=3disk}), 
\bea
\lefteqn{\left\bra\frac{1}{N}\tr M\right\ket_{0}=
  \frac{1}{32}\frac{\beta}{N}(b_{+}-b_{-})^2[g_{2}(b_{+}+b_{-})
  +g_{3}(b_{+}+b_{-})^2}\nn\\
 & & +\frac{g_{3}}{8}(b_{+}-b_{-})^2+\frac{3}{20}(b_{+}+b_{-})^3
     +\frac{3}{40}(b_{+}+b_{-})(b_{+}-b_{-})^2].
                                       \label{k=3sdisk}
\eea

In the scaling limit 
\beq
\frac{N}{\beta}=1-a^3t              \label{N/Bappendix}
\eeq
($a$ means a lattice spacing), eq. (\ref{cut2}) is iteratively solved as 
\bea
z & = & -8\left(1+a\frac{3}{4}t^{1/3}+a^2\frac{7}{64}t^{2/3}
                    -a^3\frac{71}{1024}t \right.   \nn \\
 & &       \left. -a^4\frac{8515}{442368}t^{4/3}+O(a^5)\right). 
\eea
Then $b_+$ and $b_-$ are 
\bea
b_+ & = & \frac{-5g_3+2}{3}-a2t^{1/3}-a^2\frac{7}{24}t^{2/3}
                            -a^3\frac{49}{384}t        \nn \\
    & &     -a^4\frac{1205}{165888}t^{4/3}+O(a^5),     
\label{b+}       \\
b_- & = & \frac{-5g_3-10}{3}+a^3\frac{5}{16}t+a^4\frac{15}{256}t^{4/3}
                      +O(a^5).
\label{b-}
\eea
Also, it can be seen that $\ze$ should be tuned to the critical value at $b_+$

\beq
\ze=\ze_*(1+ay), ~~~\ze_*=\frac{-5g_3+2}{3}.
\label{zetaappendix}
\eeq
Substituting (\ref{N/Bappendix})$\sim$(\ref{zetaappendix})
into (\ref{k=3disk}) and 
(\ref{k=3sdisk}), we have 
\bea
\bra \Phi(\ze) \ket_0 & = & \oh V'(\ze)+a^{5/2}w(y)+O(a^{7/2}),  \\
\left \bra \frac{1}{N} \tr M \right \ket_0 & = & -\frac{32+25g_3}{15}
     -a^3\frac{4}{5}t+a^4\frac{3}{4}t^{4/3}+O(a^5), 
\eea
where
\beq
w(y)=-\frac{1}{5}\ze_*^{5/2}(y^2-\oh T^{1/3}y+\frac{3}{8}T^{2/3})
  \sqrt{y+T^{1/3}} 
\eeq
with $T=(2\ze_*^{-1})^3t$.

\vspace{0.8cm} 

\noindent
\section{Expression for the local operator $\cO_0$}
\setcounter{equation}{0}

  We give the proof of eqs. (\ref{O01}) and (\ref{O02}) 
using the Schwinger-Dyson equations.  
We first rewrite the once-integrated Schwinger-Dyson equation 
\beq
T(y)Z[J]=0
\eeq
into the relations among the local operators. Here,  
\beq
T(y)=T_0(y)+\int_0^ydy' \tilde{\rho}(y') +\cO_1, 
\eeq
the $y$-independent operator $\cO_1$
appears as an integration constant. 

    Now to correctly extract the 
local operators, we must identify and
subtract the singular parts in the correlation 
functions 
\beq
\left.\frac{\dl^{K}\ln Z[J]}{\dl\tilde{J}(y_{1})\cdots\dl\tilde{J}(y_{K})}
   \right|_{J=0} =\bra \tilde{\Phi}(y_1)\cdots\tilde{\Phi}(y_K)\ket. 
\eeq
It appears only in the disk and cylinder 
amplitudes. For the disk, from the large $y$ expansion of (\ref{k=3diskPSFT}) 
\beq
w(y)=y^{5/2}+\frac{5}{16}T y^{-1/2}-\frac{15}{128}T^{4/3}y^{-3/2}
+\cdots,         \label{expdisk}
\eeq
we see that the first two terms of (\ref{expdisk})
correspond to the singular part 
\beq
w^{\mbox{sing}}(y)= y^{5/2}+\frac{5}{16}T y^{-1/2}, 
\label{k=3disksing}
\eeq
because it is known that the disk one-point  function of the local operator 
behaves as 
$T^{4/3+\Delta}$ ($\Delta\geq 0$) 
from the analysis of the continuum theory \cite{KPZDDK}. 

   For the cylinder amplitude, we start with
a derivation of the amplitude $w(y,y_1)$ 
from the lowest order of $\gst$ in the Schwinger-Dyson equation 
\beq
\left.\frac{\dl}{\dl\tilde{J}(y_1)}T(y)Z[J]\right|_{J=0}=0, 
\eeq
that is, 
\beq
2w(y)w(y,y_1)+\gst^2\prt_{y_1}D_z(y,y_1)w(z)
-y\bra\cO_0\tilde{\Phi}(y_1)\ket_0-\bra\cO_1\tilde{\Phi}(y_1)\ket_0 =0. 
\label{cylinderSD}
\eeq
It can be easily solved by noting the fact that $w(y)$
given in (\ref{k=3diskPSFT}) 
has single zeroes at $y=\alp, \bar{\alp}$, where $\alp, \bar{\alp}$ are the 
solutions of the quadratic equation 
$$
y^{2}-\oh T^{1/3}y+\frac{3}{8}T^{2/3}=0, 
$$
explicitly 
\beq
\left.\begin{array}{c}\alp \\  \bar{\alp} \end{array} \right\}=
\frac{1\pm i\sqrt{5}}{4} T^{1/3}. 
\eeq
By setting $y=\alp$ or $\bar{\alp}$ in (\ref{cylinderSD}), the first term
vanishes, \footnote{
Here, we assumed that $w(y,y_1)$ does not have
any poles at $y=\alp, \bar{\alp}$. This is justified, since $\tilde{\Phi}(y)$ 
is regular except in the negative real axis as is seen from its definition.} 
then $\bra\cO_0\tilde{\Phi}(y_1)\ket_0$, 
$\bra\cO_1\tilde{\Phi}(y_1)\ket_0$ are determined as 
\bea
\bra\cO_0\tilde{\Phi}(y_1)\ket_0 & = & \gst^2\prt_{y_1}\sqrt{y_1+T^{1/3}}, 
\label{cylinderO0} \\
\bra\cO_1\tilde{\Phi}(y_1)\ket_0 & = & \gst^2 \prt_{y_1}(y_1-\oh T^{1/3})
    \sqrt{y_1+T^{1/3}}.    
\label{cylinderO1}
\eea
Substituting these into (\ref{cylinderSD}), we obtain 
\beq
w(y,y_1)=\frac{\gst^2}{4}\frac{1}{\sqrt{y+T^{1/3}}\sqrt{y_1+T^{1/3}}}
   \frac{1}{(\sqrt{y+T^{1/3}}+\sqrt{y_1+T^{1/3}})^2}. 
\eeq
   Since the cylinder amplitude of the local operators behaves as 
$T^{1/3+\Delta_1+\Delta_2}$ ($\Delta_1, \Delta_2\geq 0$), the 
singular part is identified to be 
\beq
w^{\mbox{sing}}(y,y_{1}) = \gst^{2}\frac{1}{4}
    \frac{1}{(y-y_{1})^{2}}\left(\sqrt{\frac{y}{y_{1}}}-2+
        \sqrt{\frac{y_{1}}{y}}\right) .
\label{k=3cylindersing}
\eeq

    Thus, the connected correlation functions are expressed as 
\bea
\bra\tilde{\Phi}(y_1)\ket & = & 
             w^{\mbox{sing}}(y_{1})+g^{(1)}(y_{1}),  \nn   \\
\bra\tilde{\Phi}(y_1)\tilde{\Phi}(y_2)\ket & = & 
w^{\mbox{sing}}(y_{1},y_{2})+g^{(2)}(y_{1},y_{2}), \nn  \\
\bra\tilde{\Phi}(y_1)\cdots\tilde{\Phi}(y_K)\ket
& = & g^{(K)}(y_{1},\cdots,y_{K}) ~~~(K\geq 3),
\eea
where $g^{(n)}(y_{1},\cdots,y_{n})$ is the part interpreted as
local operator insertions, and it is expanded by the correlators among 
local operators $g_{\alp_1,\cdots,\alp_n}$ 
\beq
g^{(n)}(y_{1},\cdots,y_{n})=\sum_{\alp_{1},\cdots,\alp_{n}}
                  g_{\alp_{1},\cdots,\alp_{n}}
           y_{1}^{-\alp_{1}-1}\cdots y_{n}^{-\alp_{n}-1}, \label{k=3localop}
\eeq
where $\alp_{i}$'s run over the positive half odd integers
$1/2,3/2,5/2,\cdots$. 
Using (\ref{k=3disksing}), (\ref{k=3cylindersing})$\sim$(\ref{k=3localop}), 
we expand the Schwinger-Dyson equations 
\beq
\left. \frac{\dl^K}{\dl\tilde{J}(y_1)
\cdots\dl\tilde{J}(y_K)}T(y)Z[J]\right|_{J=0}
 =0~~~(K=0,1,2, \cdots), 
\label{k=3KpointSD}
\eeq
and  perform the similar analysis as in ref. \cite{FKN}. 
  For example, the result for $K=0$ is 
\beas
\lefteqn{-y\bra\cO_0\ket-\bra\cO_1\ket+\left(\frac{5}{16}T\right)^2y^{-1}
+\gst^2\frac{1}{16}y^{-2}} \\
& & 
+2\sum_{\alp}g_{\alp}(y^{-\alp+3/2}+\frac{5}{16}T y^{-\alp-3/2})
+\sum_{\alp, \alp'}(g_{\alp}g_{\alp'}+g_{\alp, \alp'})y^{-\alp-\alp'-2}=0.
\eeas
Here, we use the Greek indices $\alp, \alp', \beta, \beta'$ 
for the positive half odd integers. 
{}From the first two powers of  the large $y$,
$\bra\cO_0\ket$, $\bra\cO_1\ket$ 
are determined as
$$
\bra\cO_0\ket=2g_{1/2},~~~\bra\cO_1\ket=2g_{3/2}.
$$
   Performing similar analysis for $K\geq 1$ in (\ref{k=3KpointSD}), we 
obtain for $\cO_0$, $\cO_1$ insertions 
\bea
\bra \cO_{0}\ket & = & 2g_{1/2},   \nn \\
\bra \cO_{0}\tilde{\Phi}(y_{1})\ket & = & \gst^{2}\oh y_{1}^{-1/2}
     +2\sum_{\alp_{1}}g_{1/2,\alp_{1}}y_{1}^{-\alp_{1}-1},  \nn \\
\bra \cO_{0}\tilde{\Phi}(y_{1})\cdots\tilde{\Phi}(y_{K})\ket & = & 
                  2\sum_{\alp_{1},\cdots,\alp_{K}}
   g_{1/2,\alp_{1},\cdots,\alp_{K}}y_{1}^{-\alp_{1}-1}\cdots 
                   y_{K}^{-\alp_{K}-1},  \nn \\
\bra \cO_{1}\ket & = & 2g_{3/2},   \nn \\
\bra \cO_{1}\tilde{\Phi}(y_{1})\ket & = & \gst^{2}\frac{3}{2} y_{1}^{1/2}
     +2\sum_{\alp_{1}}g_{3/2,\alp_{1}}y_{1}^{-\alp_{1}-1},  \nn \\
\bra \cO_{1}\tilde{\Phi}(y_{1})\cdots\tilde{\Phi}(y_{K})\ket & = & 
                  2\sum_{\alp_{1},\cdots,\alp_{K}}
   g_{3/2,\alp_{1},\cdots,\alp_{K}}y_{1}^{-\alp_{1}-1}\cdots 
                    y_{K}^{-\alp_{K}-1} ~~~  (K\geq 2).    
\eea
This shows that 
$\bra \cO_{0}\Phi(y_{1})\ket$ and 
$\bra \cO_{1}\Phi(y_{1})\ket$ have
the singular parts $\gst^{2}\oh y_{1}^{-1/2}$
 and $\gst^2\frac{3}{2}y_1^{1/2}$, respectively.

The other powers of $y$  give the Virasoro constraints. 
For $K=0$, we have  
\beas
\lefteqn{2(g_{l+7/2}+\frac{5}{16}T g_{l+1/2})
  +\sum_{\beta+\beta'=l}(g_{\beta}g_{\beta'}+g_{\beta, \beta'}) } \\
& & +\gst^2\frac{1}{16}\dl_{l,0}+\left(\frac{5}{16}T\right)^2\dl_{l, -1}
=0 ~~~(l=-1,0,1,\cdots), 
\eeas
where $g$ with negative indices is understood as zero. 
For general $K$, by introducing the generating function 
\beq
\ln Z(\mu)=\sum_{n_{\alp}=0}^{\infty}\frac{\mu_{1/2}^{n_{1/2}}}{n_{1/2}!}
                               \frac{\mu_{3/2}^{n_{3/2}}}{n_{3/2}!}\cdots
   g_{\underbrace{\mbox{\scriptsize 1/2,$\cdots$,1/2}}_{n_{1/2}},
        \underbrace{\mbox{\scriptsize 3/2,$\cdots$,3/2}}_{n_{3/2}},\cdots},
\eeq
the relations 
can be expressed in the form
$$
  L_{n}Z(\mu)=0  ~~~(n\geq -1),
$$
\bea
L_{-1} & = & 2\frac{\prt}{\prt\mu_{5/2}}
    +\gst^{2}\sum_{\alp}\alp\mu_{\alp}\frac{\prt}{\prt\mu_{\alp-1}}
    +(\frac{5}{16}T+\gst^{2}\frac{1}{4}\mu_{1/2})^{2},           \nn \\
L_{0} & = & 2(\frac{\prt}{\prt\mu_{7/2}}
               +\frac{5}{16}T\frac{\prt}{\prt\mu_{1/2}})
    +\gst^{2}\sum_{\alp}\alp\mu_{\alp}\frac{\prt}{\prt\mu_{\alp}}
    +\gst^{2}\frac{1}{16},           \nn \\
L_{l} & = & 2(\frac{\prt}{\prt\mu_{l+7/2}}
               +\frac{5}{16}T\frac{\prt}{\prt\mu_{l+1/2}})
    +\sum_{\beta+\beta'=l}\frac{\prt^{2}}{\prt\mu_{\beta}\prt\mu_{\beta'}} 
                                       \nn \\
    &   &  +\gst^{2}\sum_{\alp}\alp\mu_{\alp}\frac{\prt}{\prt\mu_{\alp+l}}
                         ~~~(l\geq 1).   \label{k=3Virasoro}
\eea
Note the operation of 
$\cO_{0}$ ($\cO_{1}$) on $Z(\mu)$ is
expressed as the local operator insertion 
$2\frac{\prt}{\prt\mu_{1/2}}$ ($2\frac{\prt}{\prt\mu_{3/2}}$). 

    Next, we notice that as an operator acting on $Z(\mu)$, 
$\frac{\dl}{\dl\Jty}$ is expanded by the local operator
$\frac{\prt}{\prt\mu_{\alp}}$:  
$$
\frac{\dl}{\dl\Jty}=\sum_{\alp}\frac{\prt}{\prt\mu_{\alp}}y^{-\alp-1} 
{}~~~\mbox{as acting on }Z(\mu).
$$
Furthermore, since the partition function $Z[J]$
is related to the $Z(\mu)$ through the 
rescaling 
\beq
Z[J] = \exp[ \int\frac{dy}{2\pi i}\Jty w^{\mbox{sing}}(y)
           +\oh\int\frac{dy_{1}}{2\pi i}\int\frac{dy_{2}}{2\pi i}
             \tilde{J}(y_{1})\tilde{J}(y_{2})w^{\mbox{sing}}(y_{1},y_{2})]
           Z(\mu),         \label{k=3rescaleZ}          
\eeq
we see that 
\beq
\frac{\dl}{\dl\Jty}=w^{\mbox{sing}}(y)
       +\int\frac{dy_1}{2\pi i}\tilde{J}(y_1)w^{\mbox{sing}}(y,y_1)
       +\sum_{\alp}\frac{\prt}{\prt\mu_{\alp}}y^{-\alp-1}   
\label{4-11-0}
\eeq
as acting on $Z[J]$. 

   After these preparations, we can now prove the
eqs. (\ref{O01}) and (\ref{O02}). 
First, we consider (\ref{O01}). The integral along
the contour $C$ is defined by 
the analytic continuation using the Beta function. 
For example, 
\beas
\int_C\frac{dy}{2\pi i}y^A & = &
 \int_{-\infty}^0\frac{dy}{2\pi i}\,(y-i0)^A
                           +\int_0^{-\infty}\frac{dy}{2\pi i}\,(y+i0)^A  \\
& = & -\frac{\sin\pi A}{\pi}\int_0^{\infty}dy \,y^A
         =-\frac{\sin\pi A}{\pi}B(A+1,-A-1)   \\
& = & -\frac{1}{A+1}\frac{1}{\Gamma(0)}.          
\eeas
For $A\neq -1$ this is zero, and for $A=-1$,
because the integrand has no cut, 
the contour can be deformed as encircling the origin 
$$
\int_C\frac{dy}{2\pi i}y^{-1}=\oint_0\frac{dy}{2\pi i}y^{-1}=1. 
$$
We summarize these into 
\beq
\intyC y^A=\dl_{A,-1}.               \label{4-11-2}
\eeq
{}From this result, in the case that the pole $y=-y_1$
exists inside the contour, 
we obtain
\bea
\intyC\frac{y^A}{y+y_1} & = &
\sum_{n=0}^{\infty}(-y_1)^n \intyC y^{A-n-1}    
\nn \\
 & = & \left\{ \begin{array}{ll} (-y_1)^A    & A=0,1,2,\cdots \\
                                                           0
&\mbox{otherwise} .
\end{array}
\right.     
\label{4-12-1}
\eea
Also, when the pole $y=y_1$ is outside the contour, 
\beq
\intyC \frac{y^A}{y-y_1}=
\left\{ \begin{array}{ll}  0  & A=0,1,2,\cdots \\
                              -y_1^A   &\mbox{otherwise} .   \end{array}
\right.     
\label{4-12-2}
\eeq
By using these formulas (\ref{4-11-2}) and (\ref{4-12-2}),
we can see that 
$$
2\intyC y^{1/2}\sum_{\alp}\frac{\prt}{\prt\mu_{\alp}}y^{-\alp-1} = 
          2\frac{\prt}{\prt\mu_{1/2}} , 
$$
$$
2\intyC y^{1/2}w^{\mbox{sing}}(y)= 0 = \bra \cO_{0}\ket^{\mbox{sing}},
$$
$$
2\intyC y^{1/2}w^{\mbox{sing}}(y,y_{1}) = \gst^{2}\oh y_{1}^{-1/2} 
                   = \bra \cO_{0}\Phi(y_{1})\ket^{\mbox{sing}},  
$$
which just mean that $\cO_0$ is written as in (\ref{O01}).

  Next, we verify the second equality (\ref{O02}). For this  
purpose, 
it is sufficient to show that 
\beq
\lim_{\ep\limit+0}\inty \e^{\ep y}y^A=\intyC \,y^A       \label{4-13-2}
\eeq
for arbitrary $A$, because the formulas such as (\ref{4-12-1}) and 
(\ref{4-12-2}) can be derived from (\ref{4-11-2}). 

    We consider the following integral in the case $A\notin\Z$
\beas
\inty y^A & = & \frac{1}{\pi}\cos\frac{\pi A}{2}\int_0^{\infty}dx \, x^A 
 = \frac{1}{\pi}\cos\frac{\pi A}{2}B(A+1,-A-1) \\
 & = & \frac{1}{\Gamma (0)}\frac{1}{A+1}\frac{1}{2\sin\frac{\pi A}{2}}=0.
\eeas
This implies that for $A\notin\Z$, 
\beq
\lim_{\ep\limit+0}\inty \e^{\ep y}y^A=\lim_{\ep\limit+0}\sum_{n=0}^{\infty}
\frac{\ep^n}{n!}\inty y^{n+A}=0. 
\eeq
Also, when $A=-1,-2,-3,\cdots$, the contour can be deformed to the circle 
enclosing the origin, 
\beas
\lim_{\ep\limit+0}\inty\e^{\ep y}y^A & = & 
         \lim_{\ep\limit+0}\oint_0 \frac{dy}{2\pi i}\e^{\ep y}y^A \\
 & = & \lim_{\ep\limit+0}\frac{\ep^{|A|-1}}{(|A|-1)!}=\dl_{A, -1}. 
\eeas
Further, for $A=0,1,2,\cdots$, the derivative of the $\dl$-function 
at $y=\ep$ appears. In our prescription,
since $\dl^{(A)}(\ep)=0$ for finite $\ep$, 
the limit $\ep\limit+0$ is also zero
$$
\lim_{\ep\limit+0}\inty\e^{\ep y}y^A=\lim_{\ep\limit+0}\dl^{(A)}(\ep)=0. 
$$
These results are summarized into 
\beq
\lim_{\ep\limit+0}\inty\e^{\ep y}y^A=\dl_{A, -1}. 
\eeq
Comparing this with (\ref{4-11-2}), we conclude that eq. (\ref{O02}) holds. 
Similarly, it is easy to see that the operator $\cO_1$ is expressed as 
\bea
\cO_{1} & = & 2\intyC y^{3/2}\frac{\dl}{\dl\Jty},  \label{O11}\\
               & = &  2\lim_{\ep\limit +0}\inty \e^{\ep
y}y^{3/2}\frac{\dl}{\dl\Jty}.
\label{O12}   
\eea

\noindent
\section{Disk Amplitudes in the Two-Matrix Model}
\renewcommand{\thesubsection}{\Alph{section}.\arabic{subsection}}
\setcounter{subsection}{0}
\setcounter{equation}{0}
 
 Here,  we obtain various disk amplitudes
(genus zero one-point functions) in 
the two-matrix model by using
the continuum limit of the Schwinger-Dyson 
equations which give the relations among them.  
Some of the disk amplitudes before taking the continuum limit
have been obtained by Staudacher \cite{St}. We will 
extend his results considerably
and give detailed forms of the continuum disk amplitudes 
that have not appeared in the literatures.

  We introduce the following notations
for the disk amplitudes (some of which 
are borrowed from \cite{St}):
\beas
W_{n} & = & \left\bra\oN\tr A^{n}\right\ket_{0}, \\
W_{n,m}^{(2)} & = & \left\bra\oN\tr A^{n}B^{m}\right\ket_{0}, \\
W(\ze) & = &\left\bra\oN\tr\frac{1}{\ze-A}\right\ket_{0}, \\
W(\sg) & = &\left\bra\oN\tr\frac{1}{\sg-B}\right\ket_{0}, \\
W_{j}(\ze) & = & \left\bra\oN\tr\frac{1}{\ze-A}B^{j}\right\ket_{0}, \\
W^{(2)}(\ze,\sg) & = & \left\bra\oN\tr \frac{1}{\ze-A}\frac{1}{\sg-B}
                     \right\ket_{0},  \\
W_{j}(\ze_{1};\ze_{2},\sg_{2},\cdots,\ze_k,\sg_k) & = & 
         \left\bra\oN\tr\frac{1}{\ze_{1}-A}B^{j}
  \frac{1}{\ze_{2}-A}\frac{1}{\sg_{2}-B} \cdots 
   \frac{1}{\ze_k-A}\frac{1}{\sg_k-B}\right\ket_{0}, \\
W^{(2k)}(\ze_{1},\sg_{1},\cdots,\ze_{k},\sg_{k}) & = & \left\bra\oN\tr
  \frac{1}{\ze_{1}-A}\frac{1}{\sg_{1}-B}\cdots
  \frac{1}{\ze_{k}-A}\frac{1}{\sg_{k}-B}\right\ket_{0}~~~(k=1,2,\cdots).
\eeas

\subsection{$W(\ze)$}

  Let us first start from the disk amplitude with the simplest spin 
configuration on the loop. (The spins on the loop are all $A$-.)  It is 
obtained by combining the following three Schwinger-Dyson equations:
\bea
(\ze-g \ze^{2})W(\ze) & = & cW_{1}(\ze)+W(\ze)^{2}+1-g(\ze+W_{1}), 
                                   \label{A-2-1} \\
(\ze-g \ze^{2})W_{1}(\ze) & = & cW_{2}(\ze)+W(\ze)W_{1}(\ze)+W_{1}
               -g(\ze W_{1}+W_{1,1}^{(2)}),       \label{A-2-2} \\
W_{1}(\ze)-gW_{2}(\ze) & = & c\ze W(\ze)-c.  \label{A-2-3}
\eea
 
 We can eliminate $W_{1}(\ze)$ and 
$W_{2}(\ze)$, and have a cubic equation of $W(\ze)$:
\beq
W(\ze)^{3}+a_{1}W(\ze)^{2}+a_{2}W(\ze)+a_{3}=0,       \label{A-3-1}
\eeq
\beas
a_{1} & = & \frac{c}{g}-2(\ze-g \ze^{2}),\\
a_{2} & = & (\ze-g \ze^{2})^{2}-\frac{c}{g}(\ze-g \ze^{2})+
           (\frac{c^{3}}{g}-g)\ze+1-gW_{1},\\
a_{3} & = & (-1+gW_{1}+g\ze)(\ze-g\ze^{2})+(1-3c+cg\ze)W_{1} \\
     &  &       -g^{2}W_{3}-g+\frac{c}{g}(1-c^{2})-c\ze,
\eeas
where in order to eliminate $W_{2}$ and $W_{1,1}^{(2)}$ we used 
$$
W_{1}-gW_{2}=cW_{1},~~~W_{2}-gW_{3}=cW_{1,1}^{(2)}+1.
$$
The expressions of  $W_{1}$ and $W_{3}$ 
are evaluated by the orthogonal polynomial method 
\cite{Me,BK2} as follows:
\bea
W_{1} & = & \frac{1}{64g^{3}}[3\rho^{4}-6c\rho^{3}-2(1-2c)\rho^{2}
        -2c(1-2c)^{2}\rho^{-1} \nn \\
    &  &  +32g^{2}-(1-2c+4c^{2})(1-2c)],    \label{A-4-1} \\
W_{3} & = &
\frac{1}{16\cdot 64g^{5}}[-16(\rho^{6}-1)+90c(\rho^{5}-1) \nn \\
  &  &  + \left(80 (1-2c)-\frac{531}{4}c^{2}\right)(\rho^{4}-1) \nn \\
  &  &   +(-64-94c+380c^{2}+60c^{3})(\rho^{3}-1) \nn \\
  &  &  +(-48+336c-333c^{2}-150c^{3}-54c^{4})(\rho^{2}-1) \nn \\
  &  &   +2(1-2c)(32-41c-14c^{2}-66c^{3})(\rho-1) \nn \\
   &  & -6c(1-2c)^{2}(7-14c-2c^{2})(\rho^{-1}-1) \nn \\
   &  &  +c(1-2c)^{2}(16-21c-6c^{2}-6c^{3})(\rho^{-2}-1) \nn \\
   &  &  -4c^{3}(1-2c)^{3}(\rho^{-3}-1)
             -\frac{3}{4}c^{2}(1-2c)^{4}(\rho^{-4}-1)],   \label{A-4-2}
\eea
where $\rho$ is implicitly determined by
\bea
g^{2} & = & -\frac{1}{32}[4\rho^{3}-9c\rho^{2}
-4(1-2c)\rho+2c(1-2c+2c^{2})
               \nn \\
  &  & -c(1-2c)^{2}\rho^{-2}].        \label{A-5-1}
\eea
In the 
continuum limit, expanding $g$ and $\rho$ about the critical points 
$$
g_{*}=\sqrt{10c_{*}^3},~~~
\rho_{*}=3c_*~~~(c_*=\frac{-1+2\sqrt{7}}{27}),
$$ 
eq. (\ref{A-5-1}) can be solved iteratively:
\bea
\rho & = & \rho_{*}+a^{2/3}\frac{2}{3}\rho_{*}(5t)^{1/3}+
           a^{4/3}\frac{5}{36}\rho_{*}(5t)^{2/3} \nn \\
  &  & -a^{2}\frac{35}{288}\rho_{*}t
     -a^{8/3}\frac{8557}{311040}\rho_{*}(5t)^{4/3}
       -a^{10/3}\frac{3523}{746496}\rho_{*}(5t)^{5/3} \nn \\
 &  & +a^{4}\frac{21205}{442368}\rho_{*}t^{2}+O(a^{14/3})  \label{A-5-2}
\eea
where $g$ is expanded as $g=g_{*}(1-a^{2}t)$.

  Substituting this into eqs. (\ref{A-4-1}) and (\ref{A-4-2}),
we have $W_{1}$ and 
$W_{3}$ in the expanded form:
\bea
W_{1} & = & W_{1}^{\non}+\hat{W}_{1},  \label{A-6-1} \\
W_{1}^{\non}  & = & \frac{-8\rho_*^{4}+3(2g_{*})^{2}}{3(2g_{*})^{3}}+
        a^{2}\frac{-136\rho_*^{4}+27(2g_{*})^{2}}{27(2g_{*})^{3}}t, \nn \\
\hat{W}_{1} & = &
a^{8/3}\frac{8\rho_{*}^{4}}{27(2g_{*})^{3}}(5t)^{4/3}+
        a^{10/3}\frac{4\rho_{*}^{4}}{81(2g_{*})^{3}}(5t)^{5/3} \nn \\
   &  & +a^{4}\frac{-8527\rho_{*}^{4}
+972(2g_{*})^{2}}{972(2g_{*})^{3}}t^{2}+
        O(a^{14/3}),  \nn 
\eea
\bea
W_{3} & = & W_{3}^{\non}+\hat{W}_{3}, \label{A-6-2} \\
W_{3}^{\non} & = &
\frac{32(420-839\rho_{*})\rho_{*}^{5}}{729(2g_{*})^{5}}+
  a^{2}\frac{160(252-611\rho_{*})\rho_{*}^{5}}
{729(2g_{*})^{5}}t,  \nn \\
\hat{W}_{3} & = &
a^{8/3}\frac{320\rho_{*}^{6}}{81(2g_{*})^{5}}(5t)^{4/3}+
    a^{10/3}\frac{160\rho_{*}^{6}}{243(2g_{*})^{5}}(5t)^{5/3} \nn \\
  &  &  +a^{4}\frac{70(1152-3593\rho_{*})\rho_{*}^{5}}
{729(2g_{*})^{5}}t^{2}
        +O(a^{14/3}),   \nn
\eea
where we denoted the non-universal pieces
by $W_{1}^{\non},W_{3}^{\non}$ 
and the universal ones which give the continuum limit by $\hat{W}_{1},
\hat{W}_{3}$.

     Now, we shall evaluate $W(\ze)$
 in the continuum limit. Shifting $W(\ze)$ as 
\beq
W(\ze)=-\frac{a_{1}}{3}+\hat{W}(\ze),    \label{A-7-1}
\eeq
eq. (\ref{A-3-1}) becomes 
\beq
\hat{W}(\ze)^{3}-\frac{1}{3}A_{2}\hat{W}(\ze)-\frac{1}{27}A_{1}=0
 \label{A-7-2}
\eeq
where 
$$ 
A_{1}=9a_{1}a_{2}-2a_{1}^{3}-27a_{3},~~~A_{2}=a_{1}^{2}-3a_{2}.
$$ 
Then the critical point of $\ze$ denoted by $P_{*}$ is determined by 
\beq
A_{1}|_{*}=A_{2}|_{*}=0,       \label{A-7-3}
\eeq
where $|_{*}$ means that $g,W_{1}$ and $W_{3}$ are
set to the critical values.
 It turns out that eq. (\ref{A-7-3}) gives a cubic equation of $P_{*}$, 
and its solution is threefold: $P_{*}=\frac{1+3c_{*}}{2g_{*}}$.

  After substituting $\ze=P_{*}(1+ay)$ into (\ref{A-7-2})
and expanding with respect to 
$a$, (\ref{A-7-2}) becomes  
\beq
\hat{W}(\ze)^{3}-
\frac{a^{8/3}cs^{8/3}}{40\cdot 2^{2/3}}T^{4/3}\hat{W}(\ze)-
  \frac{a^4c^{3/2}s^{4}}{160\sqrt{10}}(16y^{4}-16Ty^{2}+2T^{2})
+O(a^{13/3})=0,  
                                \label{A-8-1}
\eeq
where $c$ is fixed to be the critical value $c_*$, $s$
is the irrational number $s=2+\sqrt{7}$, 
and the rescaled variable $T=\frac{20}{s^{2}}t $ is introduced. 

 The solution of (\ref{A-8-1}) is 
\bea
\hat{W}(\ze) & = &a^{4/3} \frac{c^{1/2}s^{4/3}}{\sqrt{10}\cdot 2^{4/3}}
 [(y+\sqrt{y^{2}-T})^{4/3}+(y-\sqrt{y^{2}-T})^{4/3}]+O(a^{5/3})
   \nn\\
& \equiv & a^{4/3}\frac{c^{1/2}s^{4/3}}
{\sqrt{10}\cdot 2^{4/3}} w(y)+O(a^{5/3}), 
 \label{A-8-2} 
\eea
which gives the universal part of the disk amplitude.

  Also, 
the non-universal part 
$W^{\non}(\ze)$ is 
\beq
W^{\non}(\ze) = -\frac{a_{1}}{3}=-\frac{c}{3g}+\frac{2}{3}(\ze-g\ze^{2}). 
                            \label{A-8-4} 
\eeq

\subsection{$W_{1}(\ze)$, $W_{2}(\ze)$}

  The amplitude $W_{1}(\ze)$ ($W_{2}(\ze)$)
represents the configuration that 
the spins on the loop all align $A$-
except a small $B$-domain consisting of a
single spin (two spins).

   From (\ref{A-2-1}), 
\bea
W_{1}(\ze) & = & \frac{1}{c}[(\ze-g\ze^{2})W^{\non}(\ze)
-W^{\non}(\ze)^{2}
           -1+g(\ze+W_{1}^{\non})] \nn \\
 &  & +\frac{1}{c}[\ze-g\ze^{2}-2W^{\non}(\ze)]\hat{W}(\ze) \nn \\
 &  & +\frac{1}{c}[-\hat{W}(\ze)^{2}+g\hat{W}_{1}].   \label{A-9-1}
\eea
We identify the universal and non-universal parts as follows. 
If there are polynomials of $y$ and $T$, 
they are non-universal.
Also, if there are amplitudes, with the spin configurations 
simpler than that of $W_1(\ze)$, multiplied by polynomials of $y$ 
and $T$, they are non-universal.
After these identifications, the remaining terms are 
universal. 
    By using this rule, the universal and non-universal parts, denoted by 
$\hat{W}_1(\ze)$ and $W_1^{\non}(\ze)$ respectively, are determined as 
\bea
W_{1}(\ze) & = & W_{1}^{\non}(\ze)+\hat{W}_{1}(\ze),  \label{A-10-2} \\
W_{1}^{\non}(\ze) & = & 
 \frac{1}{c}[(\ze-g\ze^{2})W^{\non}(\ze)-W^{\non}(\ze)^{2}
           -1+g(\ze+W_{1}^{\non})] \nn \\
 &  & +\frac{1}{c}[\ze-g\ze^{2}-2W^{\non}(\ze)]\hat{W}(\ze),
\label{A-10-3}  \\
\hat{W}_{1}(\ze) & = & \frac{1}{c}[-\hat{W}(\ze)^{2}+g\hat{W}_{1}] \nn \\
 & = & a^{8/3}\frac{s^{8/3}}{40\cdot 2^{2/3}}
  [-(y+\sqrt{y^{2}-T})^{8/3}-(y-\sqrt{y^{2}-T})^{8/3}+T^{4/3}]
  +O(a^{3})  \nn \\
 & \equiv & a^{8/3}\frac{s^{8/3}}{40\cdot 2^{2/3}}w_{1}(y)+O(a^{3}). 
                         \label{A-10-1} 
\eea

  By a similar manipulation for eq. (\ref{A-2-2}) using 
$$
\hat{W}(\ze)^{2}=-c\hat{W}_{1}(\ze)+g\hat{W}_{1}, 
$$
we have
\beas
W_{2}(\ze) & = & W_{2}^{\non}(\ze)+\hat{W}_{2}(\ze),  \\
W_{2}^{\non}(\ze) & = & \left(\frac{1}{3g}+\frac{1}{3c}(\ze-g\ze^{2})\right)
 \left[-\frac{1}{c}+\frac{g}{c}\ze-\frac{c}{9g^{2}}+\frac{1}{9g}(\ze-g\ze^{2})
    \right.  \\
 &  &  \left. +\frac{2}{9c}(\ze-g\ze^{2})^{2}+\frac{g}{c}W_{1}^{\non}\right]
       \\
 &  & -\frac{g}{c}\left(\frac{2}{3g}-\frac{1}{3c}(\ze-g\ze^{2})\right)
       \hat{W}_{1} 
       -\frac{1}{c}(2-\frac{1}{c}-g\ze)W_{1}-\frac{g^{2}}{c^{2}}W_{3}-
       \frac{g}{c^{2}} \\
 &  & +\left[\frac{1}{3g^{2}}-\frac{1}{3c^{2}}(\ze-g\ze^{2})^{2}+
     \frac{1}{c^{2}}(1-g\ze)-\frac{g}{c^{2}}W_{1}^{\non}\right]\hat{W}(\ze)
             \\
 &  & +\frac{1}{g}\hat{W}_{1}(\ze)+O(a^{13/3}), \\
\hat{W}_{2}(\ze) & = & -\frac{1}{c}\hat{W}(\ze)\hat{W}_{1}(\ze)
+O(a^{13/3})\\
 &=  &  a^{4}\frac{s^{4}}{160\sqrt{10c}}(16y^{4}-16Ty^{2}+2T^{2})
+O(a^{13/3})\\
 & \equiv & a^{4}\frac{s^{4}}{160\sqrt{10c}}w_2(y)+O(a^{13/3}).
\eeas

\subsection{$W^{(2)}(\ze,\sg)$, $W^{(4)}(\ze_{1},\sg_{1},\ze_{2},\sg_{2})$}

  To discuss the higher disk amplitudes, the recursion equation for 
$W^{(2k)}$ given by Staudacher  (eq. (20) in ref. \cite{St}) is useful: 
\bea
\lefteqn{W^{(2k)}(P_1,Q_1,\cdots,P_k,Q_k)=
\frac{1}{P_1-gP_1^2-cQ_1-W(P_1)}}\nn\\
 & \times & \{ D_Q(Q_1,Q_k)
\left( \frac{g}{c}[Q-gQ^2-W(Q_1)-W(Q_k)]-1+gP_1\right)
                          \nn\\
 & & ~~\times W^{(2k-2)}(P_2,Q_2,\cdots,P_k,Q) \nn\\
 & & -\frac{g^2}{c}D_P(P_2,P_k)
W^{(2k-4)}(P,Q_2,\cdots,P_{k-1},Q_{k-1}) \nn\\
 & & +\frac{g}{c}\sum_{l=2}^{k-1}[D_Q(Q_1,Q_l)
W^{(2l-2)}(P_2,Q_2,\cdots,P_l,Q)] \nn\\
 & & ~~~~~~\times [D_Q(Q_l,Q_k)
W^{(2k-2l)}(P_{l+1}, Q_{l+1},\cdots,P_k,Q)] \nn\\
 & & +cD_P(P_1,P_2)W^{(2k-2)}(P,Q_2,\cdots, P_k,Q_k) \nn\\
 & & -\sum_{l=2}^k W^{(2l-2)}(P_l,Q_1,\cdots,P_{l-1},Q_{l-1}) \nn\\
 & & ~~~~~ \times D_P(P_1,P_l)
 W^{(2k+2-2l)}(P,Q_l,\cdots,P_k,Q_k)\}. 
\label{W(2k)Staudacher}
\eea
Let us consider the continuum limit
for the cases $k=1$ and $k=2$: 
\beq
W^{(2)}(\ze,\sg)=\frac{(1-g\ze)W(\sg)-cW(\ze)-gW_{1}(\sg)}
       {\ze-g\ze^{2}-c\sg-W(\ze)},        \label{A-12-1}
\eeq
\bea
\lefteqn{W^{(4)}(\ze_{1},\sg_{1},\ze_{2},\sg_{2})=\frac{1}
 {\ze_{2}-g\ze_{2}^{2}-c\sg_{2}-W(\ze_{2})}} \nn \\
 &  & \times\{ (c-W^{(2)}(\ze_{1},\sg_{2}))D_{\ze}(\ze_{1},\ze_{2})
       W^{(2)}(\ze,\sg_{1}) \nn \\
 &  & ~~+D_{\sg}(\sg_{1},\sg_{2})[\frac{g}{c}(\sg-g\sg^{2}
-W(\sg_{1})-W(\sg_{2}))
       -1+g\ze_{2}]W^{(2)}(\ze_{1},\sg) \nn \\
 &  & ~~+\frac{g^{2}}{c}W(\ze_{1})\}.    \label{A-16-1}
\eea

 For $k=1$, putting $\ze=P_{*}(1+ay)$, $\sg=P_{*}(1+ax)$
and expanding with respect to $a$, 
(\ref{A-12-1}) becomes  
\bea
W^{(2)}(\ze,\sg) & = & W^{(2)\,\non}(\ze,\sg)+\hat{W}^{(2)}(\ze,\sg),  
                           \label{A-12-2} \\
W^{(2)\,\non}(\ze,\sg) & = & c(1-as(y+x))+\sqrt{10c}
     (\hat{W}(\ze)+\hat{W}(\sg)),   \label{A-12-3} \\
\hat{W}^{(2)}(\ze,\sg) & = &  a^{-1}\frac{10}{s}\frac{1}{y+x}
   (-\hat{W}(\ze)\hat{W}(\sg)-\hat{W}(\ze)^2+c\hat{W}_1(\sg)) \nn \\
&  & +a^{-2}\frac{10^{3/2}}{c^{1/2}s^{2}}
          \frac{\hat{W}(\ze)^{2}\hat{W}(\sg)
         +\hat{W}(\ze)\hat{W}(\sg)^{2}}{(y+x)^{2}} \nn \\
 &  & +a^{2}\frac{cs^{2}}{120}\frac{1}{(y+x)^{2}}[160(y^{4}+x^{4})
        +80(yx^{3}+y^{3}x)-40y^{2}x^{2} \nn \\
 &  & \mbox{ } +12(s-10)T(y^{2}+x^{2})+24(s-5)Tyx
-120T(y^{2}+x^{2}+yx) \nn \\
 &  & \mbox{ } +12sT(y+x)^{2}+15T^{2}] \nn \\
 &  & +O(a^{7/3})  \nn \\
 & \equiv & a^{5/3}\frac{cs^{5/3}}{4\cdot 2^{2/3}}
      w^{(2)}(y,x) +O(a^2),   \label{A-12-4}
\eea
where 
\beq
w^{(2)}(y,x)=\frac{-w(y)^{2}-w(y)w(x)-w(x)^{2}+3T^{4/3}}{y+x}, 
\label{w(2)}
\eeq
and we calculated $\hat{W}^{(2)}(\ze,\sg)$ up to the next leading order 
since it will be necessary to obtain
 $W^{(4)}(\ze_{1},\sg_{1},\ze_{2},\sg_{2})$ 
below.

  Repeating the same procedure for $k=2$, 
we have 
\beq
W^{(4)}(\ze_{1},\sg_{1},\ze_{2},\sg_{2})
= 
W^{(4)\,\non}(\ze_{1},\sg_{1},\ze_{2},\sg_{2})
+\hat{W}^{(4)}(\ze_{1},\sg_{1},\ze_{2},\sg_{2}),
              \label{A-17-1}
\eeq
\bea
W^{(4)\,\non}(\ze_{1},\sg_{1},\ze_{2},\sg_{2})
& = & 10 c^2 -10c[D_{\ze}(\ze_{1},\ze_{2})
  \hat{W}(\ze)+D_{\sg}(\sg_{1},\sg_{2})\hat{W}(\sg)] \nn \\
& & -\sqrt{10c}[D_{\ze}(\ze_{1},\ze_{2})
   (\hat{W}^{(2)}(\ze,\sg_{1})+\hat{W}^{(2)}(\ze,\sg_{2})) \nn \\
& &   +D_{\sg}(\sg_{1},\sg_{2})(\hat{W}^{(2)}(\ze_{1},\sg)+
        \hat{W}^{(2)}(\ze_{2},\sg))],  \label{A-17-2}
\eea
\bea
\lefteqn{\hat{W}^{(4)}(\ze_{1},\sg_{1},\ze_{2},\sg_{2})=a 
 \frac{5c^{2}s}{8}w^{(4)}(y_{1},x_{1},y_{2},x_{2})+O(a^{4/3})} 
         \nn \\
 & = & a\frac{5c^{2}s}{8}\frac{1}{(y_{1}-y_{2})(x_{1}-x_{2})}
    \left[-\frac{8}{3}(y_{1}-y_{2})(x_{1}-x_{2})(y_{1}+y_{2}+x_{1}+x_{2})
     \right. \nn \\
 &  & -\oh(w(y_{1})+w(x_{1})+2w(y_{2})
+2w(x_{2}))w^{(2)}(y_{1},x_{1}) \nn \\
 &  & +\oh(w(y_{1})+w(x_{2})+2w(y_{2})
+2w(x_{1}))w^{(2)}(y_{1},x_{2}) \nn \\
 &  & +\oh(w(y_{2})+w(x_{1})+2w(y_{1})
+2w(x_{2}))w^{(2)}(y_{2},x_{1}) \nn \\
 &  & \left.-\oh(w(y_{2})+w(x_{2})+2w(y_{1})
+2w(x_{1}))w^{(2)}(y_{2},x_{2})
           \right] \nn \\
 &  & +O(a^{4/3}).     \label{A-17-3}
\eea
We note that the above results have the
symmetry under the cyclic 
permutation of variables 
$$
\ze_{1}\limit\sg_{2},~~~\sg_{2}\limit\ze_{2},
{}~~~\ze_{2}\limit\sg_{1},
 ~~~\sg_{1}\limit\ze_{1},
$$
which should be satisfied by the definition of the amplitude. 
In general, however, in the expressions such as 
eq. (\ref{A-16-1}) the symmetry is totally obscure. 
Obtaining amplitudes with correct symmetry property 
thus constitutes a quite nontrivial consistency check for the 
continuum results.

\subsection{$W_{1}(\ze_{1};\ze_{2},\sg_{2})$}

  From the following Schwinger-Dyson equation 
$$
0=\intAB\sum_{\alp=1}^{N^{2}}\frac{\prt}{\prt A_{\alp}}\left(\tr\left(
    \frac{1}{\ze_{1}-A}t^{\alp}
    \frac{1}{\ze_{2}-A}\frac{1}{\sg_{2}-B}\right)\e^{-S}\right),
$$ 
we have 
\bea
W_{1}(\ze_{1};\ze_{2},\sg_{2}) & = &
-\frac{1}{c}[D_{\ze}(\ze_{1},\ze_{2})
  (\ze-g\ze^{2}-W(\ze_{1})-W(\ze_{2}))W^{(2)}(\ze,\sg_{2}) \nn \\
  &  & \mbox{ } +gW(\sg_{2})].   \label{A-14-1}
\eea
By a similar calculation as before, we find 
\beq
W_{1}(\ze_{1};\ze_{2},\sg_{2})=W_{1}^{\non}(\ze_{1};\ze_{2},\sg_{2})+
       \hat{W}_{1}(\ze_{1};\ze_{2},\sg_{2}), \label{A-14-2} 
\eeq
\bea
\lefteqn{W^{\non}(\ze_{1};\ze_{2},\sg_{2})=-D_{\ze}(\ze_{1},\ze_{2})
  (\ze-g\ze^{2})(1-\frac{s}{P_{*}}(\ze+\sg_{2}-2P_{*}))} \nn \\
 &  & +\frac{s}{P_{*}}\left(\frac{2c}{3g}-\frac{2}{3}(\ze_{1}-g\ze_{1}^{2})
      -\frac{2}{3}(\ze_{2}-g\ze_{2}^{2})\right)+g\left(\frac{1}{3g}-
      \frac{2}{3c}(\sg_{2}-g\sg_{2}^{2})^{2}\right) \nn \\
 &  & -\frac{s}{P_{*}}(\hat{W}(\ze_{1})+\hat{W}(\ze_{2}))
         -\frac{g}{c}\hat{W}(\sg_{2}) \nn \\
 &  & -\sqrt{\frac{10}{c}}\left(
          \frac{2c}{3g}-\frac{2}{3}(\ze_{1}-g\ze_{1}^{2})
      -\frac{2}{3}(\ze_{2}-g\ze_{2}^{2})\right)D_{\ze}(\ze_{1},\ze_{2}) 
      \hat{W}(\ze) \nn \\
 &  & -\sqrt{\frac{10}{c}}D_{\ze}(\ze_{1},\ze_{2}) 
      ((\ze-g\ze^{2})(\hat{W}(\ze)+\hat{W}(\sg_{2}))) \nn \\
 &  & -\sqrt{10c}D_{\ze}(\ze_{1},\ze_{2})\hat{W}_{1}(\ze)-
       \frac{1}{c}D_{\ze}(\ze_{1},\ze_{2})((\ze-g\ze^{2})
           \hat{W}^{(2)}(\ze,\sg_{2})) \nn \\
 &  & -\frac{1}{c}\left(\frac{2c}{3g}-\frac{2}{3}(\ze_{1}-g\ze_{1}^{2})
      -\frac{2}{3}(\ze_{2}-g\ze_{2}^{2})\right)D_{\ze}(\ze_{1},\ze_{2}) 
      \hat{W}^{(2)}(\ze,\sg_{2}), \label{A-14-3}
\eea
\bea
\hat{W}_{1}(\ze_{1};\ze_{2},\sg_{2}) & = & \frac{1}{c}
  (\hat{W}(\ze_{1})+\hat{W}(\ze_{2}))D_{\ze}(\ze_{1},\ze_{2})
    \hat{W}^{(2)}(\ze,\sg_{2}) \nn \\
 & = & a^{2}\frac{cs^{2}}{16}(w(y_{1})+w(y_{2}))D_{y}(y_{1},y_{2})
      w^{(2)}(y,x_{2}) \nn \\
 &  & +O(a^{7/3}).     \label{A-15-1}
\eea

 These results show the following scaling behavior 
for the general disk amplitudes \\
$W^{(2k)}(\ze_{1},\sg_{1},\cdots,\ze_{k},\sg_{k})$: 
\beq
\hat{W}^{(2k)}(\ze_{1},\sg_{1},\cdots,\ze_{k},\sg_{k})
=a^{\frac{7}{3}-\frac{2}{3}k}w^{(2k)}(y_{1},x_{1},\cdots,y_{k},x_{k}), 
\label{w(2k)}
\eeq
which is consistent with the analysis
of the boundary conformal field theory 
\cite{IIKMNS}.  An argument for this is 
as follows: The gravitationally 
dressed spin operator exists at the boundary of domains and 
its dimension is $[y]^{2/3}$. This is derived by considering 
the gravitational dressing  of the spin operator, 
whose dimension is 
$[y]^{1/2}$, 
in the boundary conformal field theory in flat space \cite{Cardy}. 
In eq. (\ref{w(2k)}), increasing $k$ by
one unit corresponds to adding the two domains. 
Clearly, the boundaries of domains are also increased  by two.  Then, the 
dimension of $w^{(2k)}$ is changed by a factor 
$$
[y]^{2\cdot \frac{2}{3}+2\cdot (-1)}=[y]^{-\frac{2}{3}},
$$
where $2\cdot\frac{2}{3}$ comes from
the dressed spin operators at the two boundaries, and $2\cdot (-1)$ 
from the two domains. This coincides with (\ref{w(2k)}).

\vspace{0.5cm}

\section{Continuum Spin-Flip Operator }
\setcounter{equation}{0}

  In the matrix model before taking the scaling limit,  
a domain consisting of only a single flipped spin can be 
obtained as an integral of a general domain, 
\beq
\frac{1}{N}\tr\left(\frac{1}{\ze-A}B\cdots\right)=
\ointsg\sg\frac{1}{N}\tr\left(\frac{1}{\ze-A}
\frac{1}{\sg-B}\cdots\right).
               \label{B-1-1}
\eeq
Let us construct the continuum version of this operation. 
We can do this by deriving the relation  
between the universal parts of the both sides in (\ref{B-1-1}). 

  First, let us consider $\hat{W}^{(2)}(\ze,\sg)$
and $\hat{W}_{1}(\ze)$.
Comparing eqs. (\ref{A-10-1}) and (\ref{w(2)}), we have  
\beq
\int_{C}\frac{dx}{2\pi i}w^{(2)}(y,x)=-w(y)^{2}+T^{4/3}
=w_{1}(y)-2T^{4/3},
                      \label{B-2-1}
\eeq
where the contour $C$ surrounds
the negative real axis and the pole $x=-y$. 
The calculation can be performed by
using the formula (\ref{4-12-1}) after 
expanding the numerator of  $w^{(2)}(y,x)$
with respect to the large $x$. 
In such a calculation, we assume that the
unintegrated variable $y$ is 
outside the contour, and $-y$ is inside. 
By including the overall factors, it is rewritten as
\beq
s^{-1}\hat{\oint}\frac{d\sg}{2\pi i}\sg
\hat{W}^{(2)}(\ze,\sg)=\hat{W}_{1}(\ze)-a^{8/3}
 \frac{2^{1/3}s^{8/3}}{40}T^{4/3} +O(a^{3})  \label{B-2-2}
\eeq
where the integral symbol $\hat{\oint}\frac{d\sg}{2\pi i}\sg$
 is used in the sense of 
$$
\hat{\oint}\frac{d\sg}{2\pi i}\sg=P_{*}^{2}a\int_{C}\frac{dx}{2 \pi i}.
$$

  Next, for $\hat{W}^{(4)}(\ze_{1},\sg_{1},\ze_{2},\sg_{2})$ and 
$\hat{W}_{1}(\ze_{1};\ze_{2},\sg_{2})$, we use the formulas:
\bea
\int_{C}\frac{dx_{1}}{2\pi i}\frac{1}{x_{1}-x_{2}}
  \frac{x_{1}^{\alp}}{x_{1}+y_{1}} & = &
-\frac{x_{2}^{\alp}}{y_{1}+x_{2}}
  ~~~(\alp\notin\mbox{\boldmath $Z$}), \label{B-3-2} \\
\int_{C}\frac{dx_{1}}{2\pi i}\frac{1}{x_{1}-x_{2}}
  \frac{x_{1}^{n}}{x_{1}+y_{1}} & = &
 -\frac{y_{1}^{n}}{y_{1}+x_{2}}
  ~~~(n=0,1,2,\cdots), \label{B-3-3} 
\eea
which are derived from the formulas in the Appendix B,
where we regard again that 
the unintegrated variables 
$x_2, y_1$ are outside the contour, and $-x_2, -y_1$ are inside. 
After some calculations, we have 
\beq
\int_{C}\frac{dx_{1}}{2\pi i}w^{(4)}(y_{1},x_{1},y_{2},x_{2})=
 (w(y_{1})+w(y_{2}))D_{y}(y_{1},y_{2})w^{(2)}(y,x_{2}), \label{B-4-2}
\eeq
\beq
s^{-1}\hat{\oint}\frac{d\sg_{1}}{2\pi i}\sg_{1}
 \hat{W}^{(4)}(\ze_{1},\sg_{1},\ze_{2},\sg_{2})=
\hat{W}_{1}(\ze_{1};\ze_{2},\sg_{2})
  +O(a^{7/3}) .   \label{B-4-3}
\eeq 

     We can use this method also for a domain consisting of 
two flipped spins. For a preparation,
we shall compute the amplitudes 
$\hat{W}_2(\ze_1; \ze_2, \sg_2)$ and
$\hat{W}_1(\ze_1;\ze_2,\sg_2,\ze_3,\sg_3)$. 
{}From the analysis of the Schwinger-Dyson equations
similar in the Appendix C, 
we obtain 
$$
\hat{W}_2(\ze_1;\ze_2,\sg_2) = 
a^{10/3}\frac{s^{10/3}\sqrt{10c}}{320\cdot 2^{1/3}}
w_2(y_1;y_2,x_2),  
$$
\bea
\lefteqn{w_2(y_1;y_2,x_2)}\nn \\
& =  & 
(-w(y_1)^2-w(y_1)w(y_2)-w(y_2)^2
+3T^{4/3})D_y(y_1,y_2)w^{(2)}(y,x_2) ,
\eea
\bea
\hat{W}_1(\ze_1;\ze_2,\sg_2,\ze_3,\sg_3) & = & 
a^{4/3}\frac{5c^2s^{4/3}}{16\cdot 2^{1/3}}
w_1(y_1;y_2,x_2,y_3,x_3),   \nn \\
w_1(y_1;y_2,x_2,y_3,x_3)  & = &
 (w(y_1)+w(y_2))D_y(y_1,y_2)w^{(4)}(y,x_2,y_3,x_3)\nn \\
 & & -D_y(y_2,y_3)w^{(2)}(y,x_2) D_y(y_3,y_1)w^{(2)}(y,x_3). 
\eea
Then it is easy to see that the following formulas hold:
\bea
\int_{C_1}\frac{dx_2}{2\pi i}\int_C
\frac{dy_2}{2\pi i}w_1(y_1;y_2,x_2) & = & 
\int_{C_1}\frac{dx_2}{2\pi i}\int_C
\frac{dy_2}{2\pi i}w_1(y_2;y_1,x_2)  \nn \\
& = &\int_{C_1}\frac{dx_2}{2\pi i}\int_C
\frac{dx_1}{2\pi i}w_1(y_1,x_1;x_2)  \nn \\
& = &\int_{C_1}\frac{dx_2}{2\pi i}\int_C
\frac{dx_1}{2\pi i}w_1(y_1,x_2;x_1) \nn \\
& = & 
w_2(y_1)+2T^{4/3}w(y_1),
\label{D-1}
\eea
where the contour $C_1$ wraps around the contour $C$.  
Moreover, after a straightforward but lengthy
calculation, we can show that 
\bea
\lefteqn{\int_{C_1}\frac{dx_2}{2\pi i}\int_C
\frac{dy_2}{2\pi i}w_1(y_1;y_2,x_2,y_3,x_3)}\nn\\
 &  =  & \int_{C_1}\frac{dx_2}{2\pi i}\int_C
\frac{dy_2}{2\pi i}w_1(y_2;y_3,x_3,y_1,x_2)\nn \\
& = &
\int_{C_1}\frac{dx_2}{2\pi i}\int_C
\frac{dx_1}{2\pi i}w_1(y_1,x_1;x_2,y_3,x_3)\nn \\
& = & \int_{C_1}\frac{dx_2}{2\pi i}\int_C
\frac{dx_1}{2\pi i}w_1(y_1,x_2;x_1,y_3,x_3)\nn \\
& = & 
w_2(y_1;y_3,x_3)-2T^{4/3}D_y(y_1,y_3)w^{(2)}(y,x_3).
\label{D-3}
\eea
Now by taking the overall factors into account,
(\ref{D-1}) and (\ref{D-3}) are rewritten as 
\bea
\lefteqn{s^{-1}\hat{\oint}\frac{d\sg_2}{2\pi i}\sg_2
\left(\hat{\oint}\frac{d\ze_2}{2\pi i}
\hat{W}_1(\ze_1;\ze_2,\sg_2)\right)}\nn\\ 
& = & 
s^{-1}\hat{\oint}\frac{d\sg_2}{2\pi i}\sg_2
\left(\hat{\oint}\frac{d\ze_2}{2\pi i}
\hat{W}_1(\ze_2;\ze_1,\sg_2)\right) \nn \\
& = & 
s^{-1}\hat{\oint}\frac{d\sg_2}{2\pi i}\sg_2
\left(\hat{\oint}\frac{d\sg_1}{2\pi i}
\hat{W}_1(\ze_1,\sg_1;\sg_2)\right)\nn \\
& = & 
s^{-1}\hat{\oint}\frac{d\sg_2}{2\pi i}\sg_2
\left(\hat{\oint}\frac{d\sg_1}{2\pi i}
\hat{W}_1(\ze_1,\sg_2;\sg_1)\right)\nn \\
& = &
 \hat{W}_2(\ze_1)
+a^{8/3}\frac{2^{1/3}s^{8/3}}{40}T^{4/3}
\hat{W}(\ze_1)+O(a^{13/3}), 
\label{W2-1}
\eea
\bea
\lefteqn{s^{-1}\hat{\oint}\frac{d\sg_2}{2\pi i}
\sg_2\left(\hat{\oint}\frac{d\ze_2}{2\pi i}
\hat{W}_1(\ze_1;\ze_2,\sg_2,\ze_3,\sg_3)\right)}\nn\\
& = &
s^{-1}\hat{\oint}\frac{d\sg_2}{2\pi i}\sg_2
\left(\hat{\oint}\frac{d\ze_2}{2\pi i}
\hat{W}_1(\ze_2;\ze_3,\sg_3,\ze_1,\sg_2)\right) \nn \\
& = & 
s^{-1}\hat{\oint}\frac{d\sg_2}{2\pi i}\sg_2
\left(\hat{\oint}\frac{d\sg_1}{2\pi i}
\hat{W}_1(\ze_1,\sg_1;\sg_2,\ze_3,\sg_3)\right) \nn \\
& = & 
s^{-1}\hat{\oint}\frac{d\sg_2}{2\pi i}\sg_2
\left(\hat{\oint}\frac{d\sg_1}{2\pi i}
\hat{W}_1(\ze_1,\sg_2;\sg_1,\ze_3,\sg_3)\right) \nn \\
 & = &
 \hat{W}_2(\ze_1;\ze_3,\sg_3)
-a^{8/3}\frac{2^{1/3}s^{8/3}}{40}T^{4/3}D_{\ze}(\ze_1,\ze_3)
\hat{W}^{(2)}(\ze,\sg_3)+O(a^{11/3}), 
\label{W2-2}
\eea
where the integral symbols are used in the sense of 
$$
\hat{\oint}\frac{d\sg_2}{2\pi i}\sg_2 =
 P_*^2a\int_{C_1}\frac{dx_2}{2\pi i}, ~~~
\hat{\oint}\frac{d\ze_2}{2\pi i} = P_*a\int_C\frac{dy_2}{2\pi i}, ~~~
\hat{\oint}\frac{d\sg_1}{2\pi i} = P_*a\int_C\frac{dx_1}{2\pi i}.
$$
These formulas show us that 
 the domain consisting of two flipped spins is 
constructed by shrinking the domain between 
two microscopic domains 
consisting of a single flipped spin. 
Indeed, by substituting eq. (\ref{B-4-3}),
 the first line of (\ref{W2-1}) becomes
$$
s^{-1}\hat{\oint}\frac{d\sg_2}{2\pi i}\sg_2
\left(\hat{\oint}\frac{d\ze_2}{2\pi i}\left(
s^{-1}\hat{\oint}\frac{d\sg_{1}}{2\pi i}\sg_{1}
 \hat{W}^{(4)}(\ze_{1},\sg_{1},\ze_{2},\sg_{2})\right)\right). 
$$
Thus the symbol 
 $s^{-1}\hat{\oint}\frac{d\sg_i}{2\pi i}\sg_i $ is the single-spin 
flip operator, while $\hat{\oint}\frac{d\ze_2}{2\pi i}$ 
shrinks the domain $\ze_2$ to nothing in conformity with 
the original matrix-model operation. 

\vspace{1cm}
 

\section{Commutativity of the Mixing Matrix
 with Splitting and Merging Processes}
\setcounter{equation}{0}

  Here, we present the calculations which leads to eq. (\ref{splitmix})
$$
\left(\MderJh\vee\MderJh\right)_{I}=
\left(\cM\left(\derJh\vee\derJh\right)
       \right)_{I}            
$$
for the first several components $I=A,B,1,2$, 
and eq. (\ref{mergemix}) 
$$
\left(\wedge\MderJh\right)_{I,J} = \sum_{K,L}\cM_{IK}\cM_{JL}
\left(\wedge\derJh\right)_{K,L}
$$
for $(I,J)=(A,A), (B,B), (A,1), (B,1), (A,2), (B,2), (1,1).$ 

  First, we consider about eq. (\ref{splitmix}).
For $I=A, B$, it is trivial from 
the definition of $\cM$ and $\vee$ 
\bea
\MderJh_{A} &= & \frac{\dl}{\dl \Jh_{A}(\ze)}, \nn \\
\MderJh_{B} &= & \frac{\dl}{\dl \Jh_{B}(\sg)}, \nn \\
\left(\derJh\vee\derJh\right)_{A} & = & -\prt_{\ze}\left(
  \frac{\dl^{2}}{\dl\Jh_{A}(\ze)^{2}}\right),  \label{C-1-2} \\
\left(\derJh\vee\derJh\right)_{B} & = & -\prt_{\sg}\left(
  \frac{\dl^{2}}{\dl\Jh_{B}(\sg)^{2}}\right).  
\eea
For $I=1$, using 
\bea
\MderJh_{1} & = & \sqrt{10c}\left(\frac{\dl}{\dl\Jh_{A}(\ze)}+
        \frac{\dl}{\dl\Jh_{B}(\sg)}\right)+
       \frac{\dl}{\dl\Jh_{1}(\ze,\sg)}, \label{C-1-3} \\
\spJ_{1} & = & -2\left(\frac{\dl}{\dl J_{A}(\ze)}\prt_{\ze}
      \frac{\dl}{\dl J_{1}(\ze,\sg)}+
         \frac{\dl}{\dl J_{B}(\sg)}\prt_{\sg}
      \frac{\dl}{\dl J_{1}(\ze,\sg)}\right), 
   \label{C-1-4}
\eea
we have
\bea
\lefteqn{\left(\MderJh\vee\MderJh\right)_{1}} \nn \\
&= &-2\left(\MderJh_{A}\prt_{\ze}\MderJh_{1}
+\MderJh_{B}\prt_{\sg}\MderJh_{1}
       \right) \nn \\
& = & -2\sqrt{10c}\left(
  \frac{\dl}{\dl\Jh_{A}(\ze)}\prt_{\ze}\frac{\dl}{\dl\Jh_{A}(\ze)}+
\frac{\dl}{\dl\Jh_{B}(\sg)}\prt_{\sg}
    \frac{\dl}{\dl\Jh_{B}(\sg)}\right) \nn \\
& & -2\left(\frac{\dl}{\dl\Jh_{A}(\ze)}\prt_{\ze}
      \frac{\dl}{\dl\Jh_{1}(\ze,\sg)}+
         \frac{\dl}{\dl\Jh_{B}(\sg)}\prt_{\sg}
      \frac{\dl}{\dl \Jh_{1}(\ze,\sg)}\right).   \label{C-1-5}
\eea
On the other hand,
$$
\left(\cM\left(\derJh\vee\derJh\right)\right)_{1}=
\sqrt{10c}\left(
 \left(\derJh\vee\derJh\right)_{A}
+\left(\derJh\vee\derJh\right)_{B}\right)
 +\left(\derJh\vee\derJh\right)_{1},
$$
which is nothing but the r.h.s. of (\ref{C-1-5}). 
Similarly, we can show the validity of
the formula for $I=2$ by noticing 
the following identities
\beas
\lefteqn{\sum_{j=1}^{2}\frac{\dl}{\dl \Jh_{A}(\ze_{j})}\prt_{\ze_{j}}
  D_{\ze}(\ze_{1},\ze_{2})\frac{\dl}{\dl\Jh_{1}(\ze,\sg_{1})}}\\
&= & D_{\ze}(\ze_{1},\ze_{2})
\frac{\dl}{\dl \Jh_{A}(\ze)}\prt_{\ze}\frac{\dl}{\dl\Jh_{1}(\ze,\sg_{1})}
 -D_{\ze}(\ze_{1},\ze_{2})\frac{\dl}{\dl J_{A}(\ze)}
     D_{\ze}(\ze_{1},\ze_{2})\frac{\dl}{\dl J_{1}(\ze,\sg_{1})},
\eeas
\bea
\sum_{j=1}^{2}\frac{\dl}{\dl \Jh_{A}(\ze_{j})}\prt_{\ze_{j}}
  D_{\ze}(\ze_{1},\ze_{2})\frac{\dl}{\dl\Jh_{A}(\ze)}
 +\left( D_{\ze}(\ze_{1},\ze_{2})\frac{\dl}{\dl\Jh_{A}(\ze)}\right)^{2}
=D_{\ze}(\ze_{1},\ze_{2})\frac{\dl}{\dl\Jh_{A}(\ze)}\prt_{\ze}
  \frac{\dl}{\dl\Jh_{A}(\ze)}.
\eea

     Next, we consider eq. (\ref{mergemix}).
Note that $\cM$ takes the upper-triangular 
form: $\cM_{IJ}=0$ for $I<J$. 
For $(I,J)=(A,A)$, 
\bea
\left(\wedge\MderJh\right)_{A,A}(\ze;\ze') & = & 
\left(\wedge\derJh\right)_{A,A}(\ze;\ze') \nn \\
& = & \sum_{K,L}\cM_{AK}\cM_{AL}
\left(\wedge\derJh\right)_{K,L}(\ze;\ze'), 
\eea
because $\cM_{AL}=\dl_{A,L}$. 
For $(I,J)=(A,1)$, 
\bea
\lefteqn{\left(\wedge\MderJh\right)_{A,1}(\ze';\ze_1,\sg_1) =
-\prt_{\ze'}\prt_{\ze_1}D_z(\ze_1,\ze')\MderJh_1(z,\sg_1)} \nn \\
 & = & -\sqrt{10c}\prt_{\ze'}\prt_{\ze_1}
D_z(\ze_1,\ze')\frac{\dl}{\dl \hat{J}_A(z)}
        -\prt_{\ze'}\prt_{\ze_1}D_z(\ze_1,\ze')
\frac{\dl}{\dl \hat{J}_1(z,\sg_1)}.
\label{E-2-1}
\eea
On the other hand, 
\bea
\lefteqn{ \sum_{K,L}\cM_{AK}\cM_{1L}
\left(\wedge\derJh\right)_{K,L}(\ze';\ze_1,\sg_1)}
\nn \\
 & = & \sqrt{10c}\left(\wedge\derJh\right)_{A,A}(\ze';\ze_1)+
      \left(\wedge\derJh\right)_{A,1}(\ze';\ze_1,\sg_1), 
\eea
which is nothing but the r.h.s. of (\ref{E-2-1}).

    Similarly, for $(I,J)=(A,2), (1,1)$, checking the formula is  
straightforward by utilizing the identities such as 
\bea
\lefteqn{D_z(\ze_1,\ze_2)\prt_zD_w(\ze',z)\frac{\dl}{\dl\hat{J}_A(w)}}\nn \\
& = & \prt_{\ze_1}D_z(\ze',\ze_1)D_w(z,\ze_2)\frac{\dl}{\dl\hat{J}_A(w)}
  +\prt_{\ze_2}D_z(\ze',\ze_2)D_w(z,\ze_1)\frac{\dl}{\dl\hat{J}_A(w)}, 
\eea
\beq
D_z(\ze_1,\ze_1')D_w(\ze_1,\ze_1')D_{\ze}(z,w)\frac{\dl}{\dl\hat{J}_A(\ze)}
=\prt_{\ze_1}\prt_{\ze'_1}D_z(\ze_1,\ze_1')\frac{\dl}{\dl\hat{J}_A(z)}. 
\eeq
The validity for $(I,J)=(B,1), (B,2)$ is obvious from the symmetry
of $\cM$ with respect to $A\leftrightarrow B$.

\newpage
 


\begin{thebibliography}{99}
\bibitem{Kaku-Kikkawa} M. Kaku and K. Kikkawa,
Phys. Rev. D10 (1974) 1110;1823. \\
E. Cremmer and J. L. Gervais, Nucl. Phys. B90 (1975) 707. 
\bibitem{Witten} E. Witten, Nucl. Phys. B268 (1986) 253. \\
H. Hata, K. Itoh, T. Kugo, H. Kunitomo and K. Ogawa, 
Phys. Rev. D34 (1986) 2360; D35 (1987) 1318;1356. \\
 For recent developments, see, e.g., B. Zwiebach, Nucl. Phys. 
B390 (1993) 33.
\bibitem{KKMW} H. Kawai, N. Kawamoto, T. Mogami and Y. Watabiki,
Phys. Lett. B306 (1993) 19.
\bibitem{IK} N. Ishibashi and H. Kawai, Phys. Lett. B314 (1993) 190.
\bibitem{IIKMNS} M. Ikehara, N. Ishibashi, H. Kawai, T. Mogami, 
R. Nakayama and  N. Sasakura, Phys. Rev. D50 (1994)7467. \\
M. Ikehara, Phys. Lett. B348 (1995) 365. 
\bibitem{JR} A. Jevicki and J. Rodrigues, Nucl. Phys. B421 (1994) 278.
\bibitem{Banks-Martinec} See, e.g.,
 T. Banks and E. Martinec, Nucl. Phys. B294 (1987) 733,
and references therein. 
\bibitem{IK2} N. Ishibashi and H. Kawai, Phys. Lett. B322 (1994) 67.
\bibitem{NS} R. Nakayama and T. Suzuki, Phys. Lett. B354 (1995) 69.
\bibitem{Parisi-Wu} G. Parisi and Y. Wu, Sci. Sin. 24 (1981) 483. 
\bibitem{BIPZ} E. Brezin, C. Itzykson, G. Parisi and J. Zuber, 
Comm. Math. Phys. 59 (1978) 35.
\bibitem{Watabiki} Y. Watabiki, Nucl. Phys. B441 (1995) 119. 
\bibitem{Kleb} S. S. Gubser and I. R. Klebanov,
Nucl. Phys. B414 (1994) 827.
\\A. Tsuchiya, unpublished. 
\bibitem{FKN} M. Fukuma, H. Kawai and R. Nakayama,
Int. J. Mod. Phys. A6 (1991) 1385.
\\ R. Dijkgraaf, E. Verlinde and H. Verlinde, Nucl. Phys. B348 (1991) 435. 
\bibitem{IK3} N. Ishibashi and H. Kawai, 
Phys. Lett. B352 (1995) 75. \\
M. Ikehara, preprint KEK-TH-434, hep-th/9504094. 
\bibitem{KPZDDK} V. Knitzhnik, A. Polyakov and A. Zamolodchikov, 
Mod. Phys. Lett. A3 (1988) 819. \\
F. David, Mod. Phys. Lett. A3 (1988) 1651.\\
J. Distler and H. Kawai, Nucl. Phys. B321 (1989) 509.
\bibitem{Me} M. L. Mehta, Comm. Math. Phys. 79 (1981) 327.
\bibitem{BK2} D. Boulatov and V. Kazakov, Phys. Lett. B186 (1987) 379.
\bibitem{GN} E. Gava and K. S. Narain, Phys. Lett. B263 (1991) 213. 
\bibitem{St} M. Staudacher, Phys. Lett. B305 (1993) 332. 
\bibitem{yo} T. Yoneya, {\it in} Proc. Seventh Workshop
on Grand Unification/ICOBAN'86,
ed. J. Arafune (World Scientific, Singapore, 1986).\\
H. Hata, K. Itoh, T. Kugo, H. Kunitomo and K. Ogawa,
Phys. Lett. B175 (1986) 138.
\\ G. T. Horowitz, J. Lykken, R. Rohm and A. Strominger, 
Phys. Rev. Lett. 57 (1986) 283.
\bibitem{pol} M. Natsuume and J. Polchinski, 
Nucl. Phys. B424 (1994) 137. \\
J. Polchinski, Phys. Rev. Lett. 74 (1995) 638. 
\bibitem{JLY} A. Jevicki, M. Li and T. Yoneya,
Nucl. Phys. B448 (1995) 
277. 
\bibitem{grossmig} E. Brezin and V. Kazakov,
Phys. Lett. B236 (1990) 144. \\
M. Douglas and S. Shenker, Nucl. Phys. B335 (1990) 635. \\
D. Gross and A. A. Migdal, Phys. Rev. Lett. 64 (1990) 127;
Nucl. Phys. B340 (1990) 333.
\bibitem{Cardy} J. L. Cardy, Nucl. Phys. B275 [FS17] (1986) 200. 
\end{thebibliography}
\end{document}